\documentclass[fleqn,usenatbib]{mnras}

\usepackage[T1]{fontenc}

\DeclareRobustCommand{\VAN}[3]{#2}
\let\VANthebibliography\thebibliography
\def\thebibliography{\DeclareRobustCommand{\VAN}[3]{##3}\VANthebibliography}

\usepackage{graphicx}	
\usepackage{amssymb}  
\usepackage{amsmath}	
\usepackage{bm}
\usepackage{mathtools}
\usepackage{xspace}
\usepackage{graphicx}
\usepackage{graphbox}
\usepackage{bbm}
\usepackage{soul}
\usepackage{float}
\usepackage{booktabs}
\usepackage{tabularx}
\usepackage{multirow}
\usepackage{multicol}
\usepackage{braket}
\usepackage{natbib}
\defcitealias{Koposov11}{K11}
\defcitealias{Jenkins21}{J21}
\defcitealias{Longeard22}{L22}
\defcitealias{Walker23}{W23}
\defcitealias{Ding25}{D25}
\defcitealias{Sandford26}{S26}
\defcitealias{Yang25}{Y25}

\newcolumntype{I}{@{\extracolsep{\fill}}c}

\newcommand{\gnn}{\textsc{GraphNPE}\xspace}

\newcommand{\bootes}{Bo\"{o}tes~I\xspace}
\newcommand{\sfive}{\ensuremath{S^5}\xspace}
\newcommand{\sfivecomb}{\ensuremath{S^5_\mathrm{comb}}\xspace}
\newcommand{\feh}{\ensuremath{\mathrm{[Fe/H]}}\xspace}

\newcommand{\sigmalos}{\ensuremath{\sigma_\mathrm{los}}\xspace}
\newcommand{\kappalos}{\ensuremath{\kappa_\mathrm{los}}\xspace}
\newcommand{\vlos}{\ensuremath{v_\mathrm{los}}\xspace}
\newcommand{\rhalf}{\ensuremath{R_\mathrm{1/2}}\xspace}
\newcommand{\rhoinner}{\ensuremath{\rho_{150}}\xspace}
\newcommand{\betaprime}{\ensuremath{\beta'}\xspace}
\newcommand{\gprime}{\ensuremath{g'}\xspace}
\newcommand{\vrfour}{\ensuremath{\braket{v_r^4}}\xspace}
\newcommand{\vtfour}{\ensuremath{\braket{v_t^4}}\xspace}
\newcommand{\vrvtfour}{\ensuremath{\braket{v_r^2v_t^2}}\xspace}

\newcommand{\mvir}{\ensuremath{M_{200}}\xspace}

\newcommand{\mviram}{\ensuremath{M_{200}^\mathrm{am}}\xspace}
\newcommand{\gammain}{\ensuremath{\gamma_\mathrm{in}}\xspace}

\newcommand{\modot}{\ensuremath{\mathrm{M_\odot}}\xspace}
\newcommand{\kms}{\ensuremath{\mathrm{km \, s^{-1}}}\xspace}
\newcommand{\pc}{\ensuremath{\mathrm{pc}}\xspace}
\newcommand{\kpc}{\ensuremath{\mathrm{kpc}}\xspace}

\title[SBI on Draco and Boötes I]{Dark Matter in Draco and Boötes I: Hints of a Core in an Ultra-Faint Dwarf from Simulation-Based Inference}

\author[T. Nguyen et al.]{Tri Nguyen,$^{1,2}$\thanks{E-mail: trivtnguyen@northwestern.edu}
Lina Necib,$^{3, 4}$
Ting S. Li,$^{5}$
Justin Read,$^{6}$
Andrés Bañares-Hernández,$^{7, 8}$
\newauthor
Claude-André Faucher-Giguère,$^{1, 2, 9}$
Kohei Hayashi,$^{10, 11, 12}$
Kevin McKinnon,$^{5}$
Andrew B. Pace,$^{13}$
\newauthor
Nathan R. Sandford$^{5}$ and 
Hao Yang$^{14, 15}$
\\
$^{1}$Center for Interdisciplinary Exploration and Research in Astrophysics, Northwestern University, 1800 Sherman Ave, Evanston, IL 60201\\
$^{2}$NSF-Simons AI Institute for the Sky, 172 E. Chestnut St., Chicago, IL 60611, USA\\
$^{3}$Kavli Institute for Astrophysics and Space Research, Massachusetts Institute of Technology, Cambridge, MA 02139, USA\\
$^{4}$NSF AI Institute for Artificial Intelligence and Fundamental Interactions, Cambridge, MA 02139, USA\\
$^{5}$Department of Astronomy and Astrophysics, University of Toronto, 50 St. George Street, Toronto ON, M5S 3H4, Canada\\
$^{6}$Department of Physics, University of Surrey, Guildford, GU2 7XH, UK\\
$^{7}$Instituto de Astrofísica de Canarias, La Laguna, Tenerife, E-38200, Spain \\
$^{8}$Departamento de Astrofísica, Universidad de La Laguna\\
$^{9}$Department of Physics \& Astronomy, Northwestern University, 2145 Sheridan Road, Tech F165, Evanston, IL 60208, USA\\
$^{10}$National Institute of Technology, Sendai College, 4-16-1 Ayashi-Chuo, Sendai, Japan\\
$^{11}$Astronomical Institute, Tohoku University, 6-3 Aoba, Sendai, Japan\\
$^{12}$ICRR, The University of Tokyo, 5-1-5 Kashiwanoha, Kashiwa, Japan\\
$^{13}$Department of Astronomy, University of Virginia, 530 McCormick Road, Charlottesville, VA 22904, USA\\
$^{14}$State Key Laboratory of Dark Matter Physics, Tsung-Dao Lee Institute \& School of Physics and Astronomy, \\ Shanghai Jiao Tong University, Shanghai 201210, China\\
$^{15}$Key Laboratory for Particle Astrophysics and Cosmology (MOE) \& Shanghai Key Laboratory for Particle Physics and Cosmology, \\ Shanghai 200240, China \\
}
\date{Accepted XXX. Received YYY; in original form ZZZ}
\pubyear{\the\year{}}

\begin{document}
\label{firstpage}
\pagerange{\pageref{firstpage}--\pageref{lastpage}}
\maketitle

\begin{abstract}
The density profiles of dwarf spheroidal galaxies are among the most sensitive probes of dark matter physics, yet extracting them from noisy stellar kinematics remains a fundamental obstacle. 
We present GraphNPE, a simulation-based inference method for dynamical mass modeling that incorporates measurement uncertainties and spectroscopic selection functions in the forward model. 
Using mock data, we show that methods relying solely on line-of-sight velocity dispersion are biased toward cuspy density profiles, even in the absence of the mass-anisotropy degeneracy. By accessing higher-order velocity moments, particularly line-of-sight kurtosis, GraphNPE breaks key degeneracies and recovers density profiles with substantially less bias. 
We apply GraphNPE to Draco and Bo\"otes I using MMT/Hectochelle and DESI for Draco, and the S5 survey for Bo\"otes I. For each, we report density profiles and dark matter $J$- and $D$-factors. For Draco, GraphNPE yields consistent results across datasets, marginally preferring a cuspy inner profile ($\rho_{150} \sim 1.6-1.9 \times 10^8\,\mathrm{M}_\odot\,\mathrm{kpc}^{-3}$) in agreement with literature. 
On DESI, however, second-order Jeans modeling fits the dispersion but fails to reproduce the kurtosis, demonstrating higher-order moments are essential. For Bo\"otes I, limited statistical power prevents definitive determination of the inner slope. GraphNPE recovers $\rho_{150} = 0.36^{+0.15}_{-0.11} \times 10^8\,\mathrm{M}_\odot\,\mathrm{kpc}^{-3}$, significantly lower than literature and consistent with a cored inner profile. This places Bo\"otes I among the lowest density dwarfs at comparable stellar masses.
\end{abstract}

\begin{keywords}
dark matter - galaxies: dwarf - galaxies: structure - stars: kinematics and dynamics - software: machine learning 
\end{keywords}

\section{Introduction}
\label{section:intro}

The particle nature of dark matter (DM) remains one of the most fundamental open questions in physics, with viable candidates spanning a wide range in mass and interaction properties \citep[e.g.][]{2005PhR...405..279B, 2018Natur.562...51B}.
Alternative DM models, including self-interacting and dissipative DM, predict deviations from collisionless cold DM that are most pronounced on the smallest galactic scales \citep[e.g.][]{2018PhR...730....1T, 2017ARA&A..55..343B, 2000PhRvL..84.3760S, 2019ApJ...878L..32N, 2021ApJ...917....7N, 2021JCAP...08..062N}.
Among the most powerful probes of DM physics are dwarf spheroidal galaxies, located at the faint end of the luminosity function.
These systems are dominated by DM, with mass-to-light ratios often exceeding $100~\mathrm{M}_\odot/\mathrm{L}_\odot$, making them ideal laboratories for studying DM properties through their gravitational effects on stellar kinematics \citep[e.g.][]{2007ApJ...670..313S, 2009ApJ...704.1274W, 2019ARA&A..57..375S}.

The mass density profile of dwarf galaxies provides a direct link between observed stellar dynamics and the underlying DM physics, with the inner slope being particularly sensitive to the collisional dynamics of DM.
Cold DM models are known to follow ``cuspy'' inner density profiles \citep{1997ApJ...490..493N}, with mass density inversely proportional to the radius from the galaxy's center, $\rho \propto r^{-1}$.
Self-interactions between DM particles can create constant-density cores, $\rho \propto r^{0} = \mathrm{constant}$ \citep{2000PhRvL..84.3760S, 2013MNRAS.430...81R, 2013MNRAS.430..105P, 2017MNRAS.472.2945R, 2018PhR...730....1T}, which over time can undergo gravothermal collapse, leading to cuspy profiles \citep{2002ApJ...568..475B, 2020PhRvD.101f3009N}.
Dissipative DM models can further accelerate this process and create even stronger cusps \citep{2021MNRAS.506.4421S, 2024ApJ...966..131S, 2024ApJ...967...21G, 2024arXiv240815317R}.

Density profiles also affect the sensitivity of indirect DM detection experiments, which search for signatures of non-gravitational DM interactions.
Annihilation and decay of DM particles, for example, can produce $\gamma$-rays, which can be detected by telescopes such as Fermi-LAT \citep{2011PhRvL.107x1302A, 2011PhRvL.107x1303G, 2014PhRvD..89d2001A, 2015PhRvL.115w1301A, 2024AAS...24331503A} and Cherenkov Telescope Array \citep{2011ExA....32..193A, 2013APh....43..189D, 2015JCAP...03..055S}.
For annihilating and decaying DM, the predicted flux depends on the astrophysical $J$-factor and $D$-factor (see e.g. \citet{2011JCAP...03..051C}), 
\begin{align}
J = \int_{\mathrm{l.o.s.}} \int_{\Delta\Omega} \rho^2(r) \, \mathrm{d}\Omega \, \mathrm{d}l, \\
D = \int_{\mathrm{l.o.s.}} \int_{\Delta\Omega} \rho(r) \, \mathrm{d}\Omega \, \mathrm{d}l,
\end{align}
which integrates the squared density and density along the line of sight within a solid angle $\Delta\Omega$, respectively.
The $J$-factor is proportional to $\rho^2(r)$ and sensitive to the inner density slope, though this sensitivity is diminished by integration along the line of sight, which averages over the 3D density structure and reduces radial discriminating power \citep[e.g.][]{2011MNRAS.418.1526C, 2011ApJ...733L..46W, 2015MNRAS.446.3002B, 2020JCAP...09..004A}.

DM inferences from density profile measurements face several systematic uncertainties.
Stellar feedback, gas outflows, and/or dynamical friction from sinking gas, stellar or even DM subhalo clumps can modify the inner density profiles and create degeneracies with DM models \citep{ElZant2001, Nipoti2004, Nipoti2015, Goerdt2010, 2014MNRAS.437..415D, 2014Natur.506..171P, 2014MNRAS.441.2986D, 2015MNRAS.454.2981C, 2017ApJ...835..193E, 2019MNRAS.490..962F, 2019MNRAS.484.1401R, 2020MNRAS.497.2393L, Orkney2021, 2024MNRAS.535.1015D}.
In case of DM indirect detection, astrophysical backgrounds can also mimic signatures of DM interactions in indirect detection searches \citep{2011JCAP...03..010A, 2013PhRvD..88h3521G, 2016PhRvL.116e1103L, 2016PhRvL.116e1102B, 2018NatAs...2..387M, 2019JCAP...09..042M, 2022NatAs...6..703G}.
In dwarf galaxies, extragalactic and foreground Milky Way emission can contribute significant background \citep{2020JCAP...09..004A}, while internal backgrounds are typically limited to specific cases, such as globular clusters that may be stripped dwarf nuclei \citep{Gray2025} or gas-rich, massive dwarfs such as Sagittarius \citep[e.g.][]{2022NatAs...6.1317C} or the Small Magellanic Cloud \citep{Caputo2016}.

Satellites around the Milky Way are particularly valuable targets because they largely circumvent these issues: their quenched star formation and relatively low baryon fractions minimize the impact of baryonic feedback \citep{2012ApJ...759L..42P, 2014ApJ...794L...3W, 2016MNRAS.459.2573R, 2021ApJ...913...53P, 2024ApJ...976..118G}.
Furthermore, their proximity enables detailed spectroscopic observations, allowing for precise measurements of stellar kinematics that trace the underlying gravitational potential.

The traditional approach for inferring mass density profiles employs the Jeans equations \citep{1915MNRAS..76...70J, 2015MNRAS.446.3002B}.
Under the assumptions of dynamical equilibrium, the Jeans equations relate the velocity dispersion (i.e. the second velocity moment) to the underlying gravitational potential and the velocity anisotropy of the tracer population.
For a given density profile and anisotropy model, the predicted line-of-sight velocity (LOSV) dispersion profile $\sigmalos(r)$ can be computed numerically and compared to the observed stellar kinematics.
The mass parameters are then inferred by constructing a Gaussian likelihood and performing Bayesian inference, typically using Markov Chain Monte Carlo (MCMC) or nested sampling methods \citep{2008ApJ...678..614S}.
Jeans methods are fast and straightforward to implement, and have been extended beyond spherical symmetry to triaxial systems \citep{Hayashi2020, Hayashi23} and tidally disrupted systems \citep{2018MNRAS.481..860R, 2024MNRAS.535.1015D}.

A major challenge of Jeans methods that rely solely on velocity dispersions is the \textit{mass-anisotropy degeneracy}: different combinations of the enclosed mass $M(<r)$ and velocity anisotropy $\beta(r)$ can produce identical LOSV dispersion profiles \citep[e.g.,][]{1990AJ.....99.1548M, 2002MNRAS.330..778W, 2003MNRAS.343..401L, 2009MNRAS.395...76D, 2010MNRAS.406.1220W}.
This degeneracy can be broken by incorporating multiple tracer populations with different spatial or kinematic distributions \citep{2011ApJ...742...20W, 2012MNRAS.419..184A, 2016MNRAS.463.1117Z, 2017MNRAS.471.4541R}, proper motions when available \citep{2007ApJ...657L...1S, 2017MNRAS.471.4541R, 2018ApJ...860...56S, 2020A&A...633A..36M, 2024ApJ...970....1V, 2025A&A...693A.104B, 2026ApJ...998..206V}, or higher-order velocity moments \citep{2009MNRAS.394L.102L, 2013MNRAS.429.3079M, 2013MNRAS.432.3361R, 2017MNRAS.471.4541R, 2025ApJ...982..167W, 2026A&A...705A.212B}.
Alternatively, one can directly model the phase-space distribution function $f(\mathbf{x}, \mathbf{v})$ by assuming a functional form (or family of forms) and fitting all moments simultaneously, as in Schwarzschild modeling \citep[e.g.,][]{2013A&A...558A..35B, 2014ApJ...791L...3B, 2019MNRAS.482.5241K} or distribution function mapping methods \citep{2002MNRAS.330..778W, 2018MNRAS.480..927P, 2021MNRAS.501..978R, 2025A&A...700A..77P, 2026arXiv260424855P}.

Recently, \citet{gnn1} developed an alternative approach using Neural Posterior Estimation (NPE) to infer density profiles directly from kinematic samples drawn from a distribution function.
This framework, dubbed \gnn, is part of a broader class of simulation-based inference (SBI) methods \citep{2020PNAS..11730055C}, which have seen increasing adoption across diverse astrophysics applications including galaxy cluster mass estimation \citep[e.g.,][]{2022NatAs...6..936H, 2022NatAs...6.1325D}, gravitational wave parameter inference \citep[e.g.,][]{2020PhRvD.102j4057G}, line intensity mapping \citep{2026JCAP...02..008S}, and stellar stream modeling \citep{2021MNRAS.507.1999H, 2023MNRAS.525.3662A, 2025ApJ...987...96M, 2025arXiv251207960N}.
\citet{gnn2} then applied \gnn to satellites orbiting Milky Way-mass hosts from the FIRE-2 simulations \citep{2018MNRAS.480..800H} and demonstrated that it can recover density profiles and halo structural parameters including the pre-infall virial mass and maximum circular velocity.
Crucially, \citet{gnn2} showed that \gnn remains robust even when applied to satellites undergoing tidal stripping, a systematic not explicitly modeled during training, demonstrating the framework's ability to generalize beyond idealized equilibrium systems.

In this work, we extend the \gnn framework to bridge the gap between idealized simulations and real observations by incorporating heterogeneous velocity measurement uncertainties and spectroscopic selection functions.
While \citet{gnn1} explored the model's performance under different velocity uncertainties ($0.1, \, 2.0, \, 5.0 \; \mathrm{km/s}$), they assumed uniform uncertainties across all stars and did not explicitly model uncertainties within the inference pipeline.
Similarly, spectroscopic surveys of dwarf galaxies are subject to selection effects that vary spatially: magnitude limits reduce completeness in the outer regions, while crowding limits sampling in dense central regions, both of which can bias density profile inference if not properly accounted for.

We apply this updated framework to two Milky Way dwarf spheroidal galaxies selected to bracket the luminosity regime where the core--cusp problem is most informative: Draco and \bootes.
Draco is one of the faintest classical dwarf spheroidals, with a well-measured velocity dispersion profile \citep{2009ApJ...704.1274W, 2018ApJ...860...66M} and a long history at the center of the core--cusp debate \citep[e.g.][]{2018MNRAS.481..860R, Hayashi2020, Yang25}.
\bootes, in turn, is one of the brightest ultra-faint dwarfs (UFDs) and now has one of the largest spectroscopic member samples among UFDs \citep{Sandford26}, making it a uniquely tractable target in a regime where Jeans modeling is typically limited by small sample sizes and large measurement uncertainties \citep{2006ApJ...647L.111B, 2006ApJ...650L..51M, 2007MNRAS.380..281M, 2007ApJ...670..313S, Koposov11}.

For each galaxy, we perform independent analyses on spectroscopic samples from multiple datasets to ensure robustness against measurement systematics and selection effects.
For Draco, we use stellar kinematics from MMT/Hectochelle \citep{Walker23} and complementary measurements from the Dark Energy Spectroscopic Instrument (DESI; \citealt{Ding25}).
For \bootes, we employ kinematic measurements from the Southern Stellar Stream Spectroscopic Survey \citep[$S^5$;][]{2019MNRAS.490.3508L} as presented in \citet{Sandford26}, supplemented by archival spectroscopy from MMT/Hectochelle \citep{Walker23} and the Very Large Telescope (VLT; \citealt{Koposov11, Jenkins21}).

For each dwarf, we first validate the framework on mock kinematic datasets constructed to match the observational properties of the real data, including empirical selection functions and star-by-star measurement uncertainties drawn from the spectroscopic catalogs.
We then apply \gnn to the observational data to infer the DM density and velocity anisotropy profiles, inner DM densities, and $J$- and $D$-factors for each system.
For both Draco and \bootes, we analyze each spectroscopic dataset independently to assess consistency across instruments.
For \bootes, we additionally analyze the combined dataset to leverage the full kinematic information available.

This paper is structured as follows.
Section~\ref{section:method} describes the \gnn framework, including modifications to incorporate measurement uncertainties and selection effects, and the second-order Jeans modeling used for comparison.
Section~\ref{section:data} describes the spectroscopic datasets for Draco and \bootes.
Section~\ref{section:mock} validates the framework on mock datasets and compares the performance of \gnn and Jeans modeling.
Section~\ref{section:result} presents the main results, including the inferred density and velocity anisotropy profiles (Section~\ref{section:result_profiles}), inner densities (Section~\ref{section:result_inner_density}), and $J$- and $D$-factors (Section~\ref{section:result_DJ}).
Section~\ref{section:bootes_core} examines the inferred inner DM distribution of \bootes in the context of baryonic feedback, tidal processing by the Milky Way, and alternative dark matter models.
Section~\ref{section:discussion} discusses the broader implications of our results, including the role of higher-order moments in the Draco DESI inference (Section~\ref{section:discussion_draco}) and potential sources of systematic uncertainty (Section~\ref{section:systematics}).
Finally, Section~\ref{section:summary} summarizes our conclusions.

\section{Methodology}
\label{section:method}

\subsection{Spherical Second-order Jeans Modeling}
\label{section:jeans}

We first provide a brief summary of the second-order Jeans modeling used as a baseline to compare with \gnn in Section~\ref{section:result}.
We also discuss how velocity measurement uncertainties and selection effects impact the parameter inference of the mass profile.
As previously noted, Jeans-based methods can incorporate higher-order velocity moments in addition to the LOSV dispersions \citep[e.g.][]{2009MNRAS.394L.102L, 2013MNRAS.429.3079M, 2013MNRAS.432.3361R, 2017MNRAS.471.4541R, 2025ApJ...982..167W, 2026A&A...705A.212B}.
Thus, throughout this work, we will refer to our baseline Jeans model as ``second-order Jeans'' to distinguish it from more complex, Jeans-based methods. 

The collisionless Boltzmann equation (CBE) describes the phase-space distribution function (DF) $f(\mathbf{x}, \mathbf{v}, t)$ of stars:
\begin{equation}
    \label{eq:cbe}
    \frac{\partial f}{\partial t} + \mathbf{v} \cdot \nabla_{\mathbf{x}} f - \nabla_{\mathbf{x}} \Phi \cdot \nabla_{\mathbf{v}} f = 0,
\end{equation}
where $\mathbf{x}$ and $\mathbf{v}$ are the 3D position and velocity, $\Phi$ is the gravitational potential, and $t$ is time.
The positions $\mathbf{x}$ and velocities $\mathbf{v}$ are in the frame of the system.

Taking the first moment of the CBE in spherical coordinates and assuming steady state yields the spherical Jeans equation (for a full derivation, see \citealt{1980MNRAS.190..873B, 2008gady.book.....B}):
\begin{equation}
    \label{eq:jeans}
    \frac{1}{\nu} \left[\frac{\partial}{\partial r} (\nu \sigma_r^2) + \frac{2 \beta(r)}{r} (\nu \sigma_r^2)\right] =  -\frac{GM(<r)}{r^2},
\end{equation}
where $G$ is the gravitational constant,  $\sigma_r$ is the radial velocity dispersion, $\nu(r)$ is the tracer density profile, $\beta(r)$ is the velocity anisotropy profile, and $M(<r)$ is the enclosed mass profile.
Throughout this work, we also adopt the $\braket{\cdot}$ notation to denote an expectation over the velocity distribution.
Under this notation, the velocity dispersion is thus $\sigma^2 \equiv \braket{v^2} - \braket{v}^2$, where $\braket{v}$ is the systemic velocity of the system.

The velocity anisotropy is defined as:
\begin{equation}
    \label{eq:beta_ani}
    \beta(r) \equiv 1 - \frac{\sigma^2_t}{2 \sigma^2_r},
\end{equation}
where $\sigma_t$ is the tangential velocity dispersion.
By this definition, $\beta = -\infty$, $0$, and $1$ correspond to tangentially biased, isotropic, and radially biased velocity distributions, respectively.

To highlight the mass-anisotropy degeneracy inherent in the Jeans equations, we can rearrange Eq.~\ref{eq:jeans} to express the enclosed mass as:
\begin{equation}
    \label{eq:mass_from_jeans}
    M(<r) = -\frac{r^2 \sigma_r^2}{G} \left[\frac{\mathrm{d} \ln(\nu \sigma_r^2)}{\mathrm{d} r} + \frac{2 \beta(r)}{r}\right].
\end{equation}
For systems with similar velocity dispersion, a more radially anisotropic orbit distribution ($\beta \to 1$) corresponds to a lower enclosed mass, while a more tangentially anisotropic distribution ($\beta \to -\infty$) requires higher mass.
When only LOSV is available, degeneracies arise because the projected velocity dispersion $\sigma_{\rm los}(R)$ provides limited constraints on the 3D radial profiles of both $M(<r)$ and $\beta(r)$.

The solution to the Jeans equation for the radial velocity dispersion is:
\begin{equation}
    \label{eq:veldisp_3d}
    \sigma_r^2(r) = \frac{1}{\nu(r) g(r)} \int_r^\infty g(s) \frac{G M(< s) \nu(s)} {s^2} \mathrm{d}s,
\end{equation}
where the function $g(r)$ is
\begin{equation}
    \label{eq:gint}
    g(r) \equiv \exp \left( 2 \int \frac{\beta(s)}{s} \mathrm{d}s \right).
\end{equation}
Projecting along the line of sight via the Abel transform \citep{Abel1826} gives the observable LOSV dispersion:
\begin{equation}
    \label{eq:veldisp_los}
    \sigma_{\rm los}^2(R) = \frac{2}{\Sigma_\star(R)} \int_R^\infty \left( 1 - \beta(r) \frac{R^2}{r^2} \right) \frac{\nu(r) \sigma_r^2(r) r}{\sqrt{r^2 - R^2}} \mathrm{d} r,
\end{equation}
where $R$ is the projected radius and $\Sigma_\star(R)$ is the projected tracer surface density.

Following \citet{2008ApJ...678..614S}, for a system with $N$ independent tracers, we define a Gaussian likelihood:
\begin{equation}
    \label{eq:loglike_jeans}
    \mathcal{L}_\mathrm{Jeans} = \prod_{i=1}^{N} 
    \frac{(2\pi)^{-1/2}}{\sqrt{\sigma_{\rm los}^2(R_i) + \Delta_i^2}}
    \exp\left[
    -\frac{1}{2} \frac{(v_{{\rm los}, i} - V_{\rm sys})^2}{\sigma_{\rm los}^2(R_i) + \Delta_i^2}
    \right],
\end{equation}
where $(R_i, v_{{\rm los}, i}, \Delta_i)$ are the observed projected radius, LOSV, and velocity uncertainty of the $i$-th tracer, and $V_{\rm sys}$ is the systemic LOSV of the galaxy.
In Jeans modeling, the parameters of interest are typically those of the mass, velocity anisotropy, and tracer density profiles, while $V_{\rm sys}$ is treated as a nuisance parameter.
Bayesian inference is then performed by combining this likelihood with priors on the model parameters to obtain posterior distributions via MCMC or nested sampling.

From Eq.~\ref{eq:loglike_jeans}, we see that the measurement uncertainties add in quadrature with the intrinsic velocity dispersion, $\sigma_{\rm los}^2(R_i) + \Delta_i^2$.
Underestimating (and overestimating) the uncertainties thus inflates (and suppresses) the inferred intrinsic velocity dispersion and subsequently the mass.
The Gaussian likelihood also assumes normally distributed measurement errors.
In practice, $\Delta_i$ is estimated by the spectroscopic pipeline from spectral line fitting and thus may deviate from Gaussianity for low signal-to-noise spectra; we do not address, however, non-Gaussian uncertainties in this work.

On the other hand, spectroscopic selection functions bias the observed tracer surface density $\Sigma_\star(R)$, which enters the velocity dispersion calculations (Eqs.~\ref{eq:veldisp_3d} and~\ref{eq:veldisp_los}).
We can account for this by first fitting $\Sigma_\star(R)$ using photometric data, which is independent of spectroscopic selection and typically more complete.
We then perform a joint inference of all profiles, while either fixing the surface density to its photometric best-fit or allowing it to vary within the fit uncertainties.
This assumes approximately constant mass-to-light ratio as a function of radius, which is a reasonable approximation given the old, metal-poor stellar populations of both Draco \citep{Kirby2011, Ding25} and \bootes \citep{Longeard22, Sandford26}.
Allowing the parameters of the surface density to vary during the joint fit might also help account for potential deviations from this assumption.

\subsection{GraphNPE and Forward Model}

We now describe the formalism of \gnn (Graph Neural Posterior Estimation) introduced in \citet{gnn1, gnn2}.
We first review the NPE framework, then describe the forward model for generating synthetic dwarf galaxy kinematics, and finally discuss the incorporation of observational effects.

\subsubsection{GraphNPE}
\label{section:npe}

The \gnn model is built on NPE, an SBI technique that directly approximates the Bayesian posterior distribution $p(\bm{\theta} | \bm{d})$ without requiring an explicit likelihood function.
Instead, NPE learns the posterior from samples $(\bm{d}, \bm{\theta})$ drawn from the joint distribution $p(\bm{d}, \bm{\theta})$:
\begin{align}
    \bm{\theta} &\sim p(\bm{\theta}), \\
    \bm{d} &\sim p(\bm{d} | \bm{\theta}),
\end{align}
where $\bm{\theta}$ is the parameter vector, $\bm{d}$ is the data, $p(\bm{\theta})$ is the prior, and $p(\bm{d} | \bm{\theta})$ is the simulator (forward model).

The NPE model $q_\phi(\bm{\theta} | \bm{d})$ with learnable weights $\phi$ approximates the true posterior by minimizing the forward Kullback-Leibler (KL) divergence:
\begin{equation}
    \label{eq:kld}
    D_{\rm KL}[p(\bm{\theta} | \bm{d}) \| q_\phi(\bm{\theta} | \bm{d})] = \int p(\bm{\theta} | \bm{d}) \ln \frac{p(\bm{\theta} | \bm{d})}{q_\phi(\bm{\theta} | \bm{d})} \, \mathrm{d}\bm{\theta}.
\end{equation}
In practice, the approximate posterior $q_\phi$ is often modeled using neural density estimators such as normalizing flows \citep{2019arXiv191202762P}, which can represent complex, high-dimensional distributions through invertible transformations of a simple base distribution.

The forward KL divergence is mode-covering: it severely penalizes the model for failing to place probability mass where the true posterior is non-zero (i.e. when $p(\bm{\theta} | \bm{d}) > 0$ and $q_\phi(\bm{\theta} | \bm{d}) = 0$).
This encourages the model to be more conservative and thus better suited for scientific applications, although, in practice, one should always perform calibration tests \citep[e.g.][]{Harrison2015, Talts2018, 2023PMLR..20219256L}.

Eq.~\ref{eq:kld} can be reduced to minimizing the expected negative log-density of the approximate posterior:
\begin{align}
    \label{eq:loss_fn}
    l_\mathrm{NPE}(\phi) &= -\mathbb{E}_{\bm{\theta} \sim p(\bm{\theta}), \bm{d} \sim p(\bm{d} | \bm{\theta})} \left[ \log q_\phi(\bm{\theta} | \bm{d}) \right] \nonumber \\
    &\approx -\frac{1}{M} \sum_{i=1}^{M} \log q_\phi(\bm{\theta}_i | \bm{d}_i), 
\end{align}
where the approximation is a Monte Carlo estimate over a training set of $M$ simulation pairs $\{(\bm{\theta}_i, \bm{d}_i)\}_{i=1}^M$.
Once trained, the model provides \textit{amortized} inference, i.e. given any new observation $\bm{d}_{\rm obs}$, the posterior $q_\phi(\bm{\theta} | \bm{d}_{\rm obs})$ can be evaluated directly without retraining or running additional simulations.

In our previous work \citep{gnn1, gnn2}, the data vector $\bm{d}$ consists of the positions and LOSV  of all observed tracers in a system, i.e., $\bm{d} = \{(x_i, y_i, v_{\mathrm{los}, i})\}_{i=1}^N$ for $N$ tracers, where $(x_i, y_i)$ are the 2D projected coordinates of tracer $i$-th.
Since this has much higher dimensionality than the parameter vector $\bm{\theta}$, we first convert $\bm{d}$ into a graph representation $\mathcal{G}$ with edges $\mathcal{E}$ and node features $\mathcal{N}$.
The graph edges are computed by connecting each star to its $k$-nearest neighbors based on their distances in the $x$-$y$ plane.
The node features are $\mathcal{H}_i = (R_i, v_{\mathrm{los}, i})$, where $R_i = \sqrt{x_i^2 + y_i^2}$ is the projected radius.
We then employ GNN layers to encode $\mathcal{G}$ into a low-dimensional latent representation, which serves as the conditioning context for a normalizing flow that models the posterior distribution.
We discuss updates to the graph construction procedure to incorporate measurement uncertainties and selection functions in Section~\ref{section:method_new}.

Both the GNN encoder and the flow are trained end-to-end by minimizing the loss in Eq.~\ref{eq:loss_fn}, allowing the GNN encoder to extract features most informative for parameter inference.
Further details on the model architecture and training procedure are provided in Appendix~\ref{app:training}.

\subsubsection{Forward model}
\label{section:forward_model}

In NPE, the forward model (or simulator) $p(\bm{d} | \bm{\theta})$ generates synthetic observations from parameters.
NPE only requires that the forward model can be sampled from, in contrast to likelihood-based approaches that require an explicit, tractable likelihood. 
Thus, NPE is particularly powerful when the data-generating process can be simulated (e.g. $N$-body simulation, mock observation), but the likelihood is intractable due to complex selection effects or high-dimensional correlations.
The forward model encodes all physical assumptions and observational systematics, and the quality of posterior inference depends directly on how faithfully it captures the true underlying physics.
Below, we describe the forward model used to generate synthetic dwarf galaxy kinematics for training \gnn.

The forward model is largely the same as the version used in \cite{gnn2}, with some minor updates.
First, we model the DM density profiles as the Zhao profile \citep{1996MNRAS.278..488Z}:
\begin{equation}
    \label{eq:zhao}
    \rho_\mathrm{dm}^\mathrm{Zhao}(r) = \rho_s \left(\frac{r}{r_s}\right)^{-\gamma}
    \left(1 + \left(\frac{r}{r_s}\right)^\alpha\right)^{-(\beta-\gamma)/\alpha},
\end{equation}
where $\rho_s$, $r_s$, $\alpha$, $\beta$, and $\gamma$ are the characteristic density, scale radius, transition sharpness, outer slope, and inner slope, respectively.
Setting $(\alpha, \beta) = (1, 3)$ recovers the generalized Navarro-Frenk-White (NFW; \citealt{1997ApJ...490..493N}) profile.
The inner slope $\gamma = 0$ and $\gamma = 1$ denote cored and cuspy profiles, respectively.

The tracer mass density is modeled as a Plummer profile \citep{1911MNRAS..71..460P}:
\begin{equation}
    \label{eq:plummer_3D}
    \nu(r) = \frac{3M_\star}{4 \pi r_\star^3} \left(1 + \frac{r^2}{r_\star^2} \right)^{-5/2},
\end{equation}
with surface density:
\begin{equation}
    \label{eq:plummer_2D}
    \Sigma_\star(R) = \frac{M_\star}{\pi r_\star^2} \left(1 + \frac{R^2}{r_\star^2} \right)^{-2},
\end{equation}
where $r_\star$ is the Plummer scale radius and $M_\star$ is the total stellar mass.
The 3D and projected half-mass radii are related to the Plummer scale radius by $r_\mathrm{1/2} = 1.305 \, r_\star$ and $\rhalf = r_\star$, respectively.
As discussed in Section~\ref{section:jeans}, we assume the stellar mass distribution follows the light distribution (constant mass-to-light ratio), such that the Plummer scale radius $r_\star$ can be constrained from photometric measurements of the half-light radius and used directly in dynamical modeling.

The velocity anisotropy follows the Cuddeford-Osipkov-Merritt (COM) profile \citep{1979PAZh....5...77O, 1985AJ.....90.1027M, 1991MNRAS.253..414C}:
\begin{equation}
    \label{eq:velani_COM}
    \beta^\mathrm{COM}(r) = \frac{\beta_0 + (r/r_a)^2}{1 + (r/r_a)^2},
\end{equation}
where $\beta_0$ is the central anisotropy and $r_a$ is the anisotropy scale radius.
This profile transitions from $\beta(r \ll r_a) \approx \beta_0$ in the inner regions to $\beta(r \gg r_a) \approx 1$ (radially biased) in the outer regions.
When $\beta_0=0$, the COM profile reduces to the standard Osipkov-Merritt (OM) profile.

\begin{table*}
\centering
\caption{Summary of model parameters and their ranges of the training dataset. 
The notations $\mathcal{U}$ and $\log\mathcal{U}$ represent the uniform and log-uniform distribution.
The NPE model predicts the posterior distribution of all parameters except the Plummer scale radius $r_\star$, which is used as a conditioning variable.
}
\renewcommand{\arraystretch}{1.3}
\begin{tabular}{llll}
\hline\hline
Parameter & Name & Equation & Prior \\
\hline
$\alpha$ & Transition sharpness & Zhao (Eq.~\ref{eq:zhao}) & $\mathcal{U}[0.5, 3]$ \\
$\beta$ & DM outer slope & Zhao (Eq.~\ref{eq:zhao}) & $\mathcal{U}[1, 10]$ \\
$\gamma$ & DM inner slope & Zhao (Eq.~\ref{eq:zhao}) & $\mathcal{U}[-1, 2]$ \\
$\rho_s / (\mathrm{M}_\odot~\mathrm{kpc}^{-3})$ & DM density normalization & Zhao (Eq.~\ref{eq:zhao}) & $\log\mathcal{U}[10^3, 10^{10}]$ \\
$r_s / \mathrm{kpc}$ & DM scale radius & Zhao (Eq.~\ref{eq:zhao}) & $\log\mathcal{U}[10^{-2}, 10^{2}]$ \\
\hline
$r_\star / r_{\rm dm}$ & Stellar density scale radius & Plummer (Eqs.~\ref{eq:plummer_3D}, \ref{eq:plummer_2D}) & $\log\mathcal{U}[10^{-3}, 1]$ \\
\hline
$r_a / r_\star$ & Anisotropy scale radius & COM (Eq.~\ref{eq:velani_COM}) & $\log\mathcal{U}[10^{-1}, 10^3]$ \\
$\beta_0$ & Central anisotropy & COM (Eq.~\ref{eq:velani_COM}) & $\mathcal{U}[-0.5, 1]$ \\
\hline
\end{tabular}
\label{tab:parameters}
\end{table*}

The forward model parameters and their prior ranges are shown in Table~\ref{tab:parameters}.
In total, \gnn has seven free parameters $\bm{\theta}=(\alpha, \beta, \gamma, \rho_s, r_s, \beta_0, r_a)$ that are inferred from the data.
The total stellar mass $M_\star$ is assumed to be negligible compared to the DM halo mass and only sets the normalization of the tracer density; it does not enter the dynamical parameter inference in either Jeans modeling or \gnn.
Compared to \citet{gnn1, gnn2}, this includes two additional DM profile shape parameters $\alpha$ and $\beta$, while the Plummer scale radius $r_\star$ is no longer inferred but instead used as conditioning context for the normalizing flow to account for spectroscopic selection effects (further discussed in Section~\ref{section:method_new}).
In addition, we significantly extend the prior ranges of $r_\star$ and $r_a$ to $[10^{-3}, 1] \, r_s$ and $[10^{-1}, 10^3] \, r_\star$, respectively, compared to \citet{gnn2}.
Lastly, following our past studies, we let the prior of the inner slope go down to $\gamma = -1$ to avoid running into the prior edge for density profiles with pronounced cores at $\gamma=0$, while avoiding more negative, unphysical values. 
Results with $\gamma > 0$ are reported in Appendix~\ref{app:results_gm0} for easier comparisons with literature values. 

We generate mock galaxies using \textsc{Agama} \citep{2019MNRAS.482.1525V}, an action-based library for dynamical modeling.
Given a set of parameters $\bm{\theta}$ describing the underlying gravitational potential, \textsc{Agama} uses the \texttt{df::QuasiSphericalCOM} class with a generalized Eddington inversion formula \citep{1916MNRAS..76..572E} to compute the anisotropic DF $f(E, L)$ from the density profile $\rho(r)$ and potential $\Phi(r)$, where $E$ and $L$ are the specific orbital energy and angular momentum, respectively.
Mock stars are then drawn from $f(E, L)$ via adaptive rejection sampling and converted to 6D phase space $(\mathbf{x}, \mathbf{v})$.

For each mock galaxy in the training set, the number of tracers $N$ is sampled from a Poisson distribution with mean of $100$ tracers, i.e., $N \sim \mathrm{Poisson}(100)$.
This introduces small variability in sample sizes during training, which helps the model generalize across galaxies with different numbers of spectroscopic members.
Using the graph construction procedure described in Section~\ref{section:method_new}, we find the model to be robust to tracer counts well outside this range, as also demonstrated in Section~\ref{section:mock} and Section~\ref{section:result}.

A limitation of this approach is that the generalized Eddington inversion becomes computationally unstable for $\beta_0 < -0.5$, as it requires evaluating higher-order derivatives $\mathrm{d}^2\rho/\mathrm{d}\Phi^2$.
We therefore restrict the central anisotropy prior to $\beta_0 \in [-0.5, 1]$.
This restriction is physically well-motivated, however.
Observations of stellar tracers in Sculptor and Draco show tangential anisotropy typically in the range $\beta_0 \approx -0.2$ to $0$ \citep{2016MNRAS.463.1117Z, 2024ApJ...970....1V}, though this can depend on the metallicity of the stellar populations.
Similarly, \citet{2017MNRAS.472.4786G} show that dwarf galaxies in the FIRE-2 cosmological simulations are near-isotropic to mildly tangentially anisotropic in their central regions. 
This is consistent with the FIRE-2 samples analyzed in \citet{gnn2}, which include dwarf galaxies with stellar masses $M_\star \gtrsim 10^5 \, \mathrm{M}_\odot$ under both cold and self-interacting DM scenarios.

To conclude, each mock galaxy is generated by sampling $N$ tracers independently from the DF $f(\mathbf{x}, \mathbf{v}; \bm{\theta})$, yielding a likelihood of the form
\begin{equation}
    \label{eq:loglike_npe}
    \mathcal{L}_\mathrm{NPE} = \prod_{i=1}^{N} f(\mathbf{x}_i, \mathbf{v}_i; \bm{\theta}).
\end{equation}
This likelihood is equivalent to that of DF mapping methods \citep{2002MNRAS.330..778W, 2018MNRAS.480..927P, 2021MNRAS.501..978R}.
GraphNPE offers key advantages over traditional DF methods: amortized inference enabling rapid posterior evaluation without retraining, natural handling of incomplete phase-space data (see Section~\ref{section:method_new}), and straightforward extension to non-equilibrium systems (we refer readers to Section 7.2 of \citet{gnn2} for more details).

\subsubsection{Incorporating measurement uncertainties and selection functions}
\label{section:method_new}
 
As discussed in Section~\ref{section:jeans}, velocity measurement uncertainties and spectroscopic selection functions can bias dynamical inference if not properly accounted for.
Here, we describe how these observational effects are incorporated into the \gnn forward model.
We first describe how the DF is transformed to account for these effects at the likelihood level, before describing the practical implementation in the forward model.
In brief, all transformations are applied conditional on the Plummer radius $r_\star$, which encodes the projected stellar surface density.
This conditioning enables the model to infer the selection function implicitly by comparing the observed spatial distribution to the expected distribution.

The idealized likelihood in Eq.~\ref{eq:loglike_npe} assumes full 6D phase-space observations, which, in practice, is rarely available.
First, the DF is projected onto the observable space $(x, y, v_\mathrm{los})$ by marginalizing over the unobserved coordinates $(z, v_x, v_y)$:
\begin{equation}
    \label{eq:df_proj}
    f_\mathrm{proj}(x, y, v_\mathrm{los}; \bm{\theta}) = \int f(\mathbf{x}, \mathbf{v}; \bm{\theta})\,
    \mathrm{d}z\, \mathrm{d}v_x\, \mathrm{d}v_y.
\end{equation}
Assuming spherical symmetry, $f_\mathrm{proj}$ depends on $(x, y)$ only through the projected radius $R = \sqrt{x^2 + y^2}$.

Assuming the selection function, denoted as $\mathcal{S}(R)$, is only a function of the projected radii $R$, the DF is then transformed as:
\begin{equation}
    \label{eq:df_sel}
    f_\mathrm{sel}(x, y, v_\mathrm{los}; \bm{\theta}) \propto \mathcal{S}(R)\, f_\mathrm{proj}(x, y, v_\mathrm{los}; \bm{\theta}),
\end{equation}
where the normalization ensures $\int \mathcal{S}(R)\, \Sigma_\star(R; \bm{\theta})\, \mathrm{d}R = 1$, with $\Sigma_\star(R; \bm{\theta})$ the projected surface density.

The DF is then convolved with a per-star uncertainty kernel $\mathcal{K}$ to account for measurement uncertainties in $v_\mathrm{los}$:
\begin{equation}
    \label{eq:df_conv}
    \tilde{f}(x, y, v_\mathrm{los}; \bm{\theta}, \Delta) = \int f_\mathrm{sel}(x, y, v_\mathrm{los}';
    \bm{\theta})\, \mathcal{K}(v_\mathrm{los}', v_\mathrm{los}; \Delta)\, \mathrm{d}v_\mathrm{los}',
\end{equation}
where $\Delta$ is the measurement uncertainty.
Note that, throughout this work, we assume the uncertainties on the positions are negligible.

Finally, this yields the transformed likelihood:
\begin{equation}
    \label{eq:loglike_npe2}
    \mathcal{L}_\mathrm{NPE} = \prod_{i=1}^{N} \tilde{f}(x_i, y_i, v_{\mathrm{los},i}; \bm{\theta}, \Delta_i).
\end{equation}
While the integrals in Eqs.~\ref{eq:df_proj} and~\ref{eq:df_conv} are generally costly to evaluate directly, generating samples from the transformed DF $\tilde{f}(x, y, v_\mathrm{los})$ is straightforward.
Given a set of samples $\{(\mathbf{x}_i, \mathbf{v}_i)\}$ drawn from the original DF $f(\mathbf{x}, \mathbf{v})$ via \textsc{Agama}, each transformation is applied sequentially as follows.
First, the projection onto $(x, y, v_\mathrm{los})$ is trivially achieved by retaining only the $(x_i, y_i, v_{\mathrm{los},i})$ components of each sample.
We then apply the selection function $\mathcal{S}(R)$ by subsampling stars with a probability $p_i \propto \mathcal{S}(R_i)$.
Finally, measurement uncertainties are incorporated by perturbing each $v_{\mathrm{los},i}$ with a noise sample drawn from the kernel $\mathcal{K}(\cdot\,; \Delta_i)$.

We now discuss the procedure for including $\mathcal{S}$ and $\mathcal{K}$ in the training process.
Importantly, we do not marginalize over the spaces of possible selection functions or velocity uncertainty kernels.
Rather, we assume specific functions $\mathcal{S}(R)$ and $\mathcal{K}(\cdot)$ exist for each galaxy.
The model learns to account for them implicitly: $\mathcal{S}$ through conditioning on $r_\star$, which encodes the true projected stellar surface density $\Sigma_\star(R; \bm{\theta}, r_\star)$, and $\mathcal{K}$ through the LOSV uncertainties $\Delta_i$ provided directly as input features.
To ensure the model generalizes across a broad range of observing conditions without retraining, we define a \textit{hyperprior} over both $\mathcal{S}$ and $\mathcal{K}$ during training.


We model $\mathcal{K}$ as a Gaussian kernel, $\mathcal{K}(v_\mathrm{los}', v_\mathrm{los}; \Delta_i) = \mathcal{N}(v_\mathrm{los}'; v_\mathrm{los}, \Delta_i^2)$.
We draw the per-star uncertainty $\Delta_i$ from a Jeffreys prior, which is a non-informative prior distribution and proportional to the square root of the determinant of the Fisher information matrix \citep{1946RSPSA.186..453J}.
For the scale parameter of a Gaussian, the Jeffreys prior takes the form:
\begin{equation}
    \label{eq:jeffreys_prior}
    p(\Delta_i) = \frac{1}{\Delta_i \ln(\Delta_\mathrm{max}/\Delta_\mathrm{min})},
\end{equation}
where $\Delta_\mathrm{min}$ and $\Delta_\mathrm{max}$ bound the support of the prior.
We place further hyperpriors on these bounds:
\begin{align}
    \label{eq:hyperprior_delta}
    \Delta_\mathrm{min} &\sim \mathcal{U}(0.01, 0.1) \; \mathrm{km\,s^{-1}}, \\
    \Delta_\mathrm{max} &\sim \mathcal{U}(5, 30) \; \mathrm{km\,s^{-1}},
\end{align}
which should cover the full range of velocity uncertainties reported in existing spectroscopic catalogs of Milky Way dwarfs.

For a mock galaxy with $N$ tracers, we first sample $\Delta_\mathrm{min}$ and $\Delta_\mathrm{max}$ from the hyperprior once, then draw $N$ samples from $p(\Delta_i)$.
This procedure has two main limitations.
As noted in Section~\ref{section:jeans}, the uncertainties are obtained from spectroscopic fitting pipeline and may not be Gaussian for low-signal-to-noise spectra. 
One can in principle account for non-Gaussian uncertainties by adopting a Monte Carlo procedure in \cite{2023ApJ...952L..10W}.
However, this requires marginalizing over the uncertainty distribution \textit{during training}, which can be computationally expensive for large $N$.
Additionally, our procedure does not account for any dependence of $\Delta_i$ on the physical properties of the tracer stars within the galaxy, such as their position, magnitude, or color.
We leave both of these limitations as directions for future work.

We model the selection function $\mathcal{S}$ as a radial dropout scheme, where a fraction $f_\mathrm{drop} \sim \mathcal{U}(0, 0.5)$ of stars are removed according to one of three modes:
\begin{enumerate}
    \item $\mathcal{S}_\mathrm{outer}$ removes the outermost $f_\mathrm{drop}$ fraction of stars mimicking magnitude-limited incompleteness at large $R$; 
    \item $\mathcal{S}_\mathrm{inner}$ removes the innermost $f_\mathrm{drop}$ fraction mimicking crowding incompleteness at small $R$; \
    \item $\mathcal{S}_\mathrm{rand}$ removes a random $f_\mathrm{drop}$ fraction regardless of position.
\end{enumerate}
During training, one mode is sampled per mock galaxy with mixture probabilities $(0.4, 0.4, 0.2)$ for $(\mathcal{S}_\mathrm{outer}, \mathcal{S}_\mathrm{inner}, \mathcal{S}_\mathrm{rand})$, respectively.
The mixture probabilities are obtained through experimentation, though we note that the results are not sensitive to the exact choice, as long as $\mathcal{S}_\mathrm{outer}$ and $\mathcal{S}_\mathrm{inner}$ are not negligible relative to $\mathcal{S}_\mathrm{rand}$.

Observationally, the selection function is not homogeneous but instead decreases towards both the inner and outer edges of the galaxy.
However, we find that the model is not sensitive to the exact functional form of $\mathcal{S}$ during training.
This insensitivity likely arises because the model is conditioned on the Plummer radius $r_\star$ (see Section~\ref{section:npe}), which encodes the true projected surface density $\Sigma_\star(R)$.
In this sense, the model already possesses sufficient information to infer the underlying stellar distribution and identify deviations from it caused by selection effects.
The dropout scheme does not prescribe the selection function to the model but rather encourages it to develop this capability during training.

We note a few updates to the graph construction procedure relative to \citet{gnn1, gnn2}.
The node features now include LOSV uncertainties: $\mathcal{H}_i = (R_i, v_{\mathrm{los}, i}, \Delta_i)$.
Additionally, rather than fixing the number of nearest neighbors to $k=20$ as in our previous work, we set $k$ to 20\% of the total number of tracers in each mock galaxy.
We find this adaptive scheme is critical for model performance: the selection function can reduce the tracer count by up to 50\%, and a fixed $k=20$ would result in overly dense graphs for small samples or excessively sparse graphs for large samples.
Setting $k$ proportional to $N$ ensures a similar graph connectivity across the full range of sample sizes encountered during training.

Finally, we note that all transformations (i.e. projection, uncertainty kernel, selection function, and adaptive graph construction) are applied \textit{on-the-fly} during training rather than pre-computed.
This significantly improves sample efficiency: even for a fixed parameter, the model is unlikely to encounter the same graph twice.

\subsubsection{Inference on observational data}

We briefly outline the inference procedure on observational data.
Since the Plummer scale radius $r_\star$ is used as conditioning context for the model, we first obtain $r_\star$ and its uncertainty by fitting a Plummer profile to the photometric light profile.
This can be obtained through direct fitting to photometry data, though in this work we use measurements reported in the Local Volume Database (LVDB; \citealt{2024arXiv241107424P}).
We assume a Gaussian posterior for $r_\star$ centered on the reported value with the reported uncertainty, and draw $N_{\mathrm{MC}} = 100$ samples of $r_\star$ from this posterior.
For each sampled $r_\star$, we condition the trained \gnn model on the observed stellar kinematics and generate $N_{\mathrm{post}} = 100$ posterior samples.
This yields a total of $10^4$ samples that account for both the uncertainty in $r_\star$ and the posterior uncertainty in the mass profile parameters.
The graph is constructed from the observed data without applying any of the training-time transformations.

\section{Data}
\label{section:data}

\subsection{Spectroscopic Datasets and Membership}
\label{section:dataset}

\begin{table*}
\centering
\caption{Summary of spectroscopic datasets for Draco and \bootes.}
\label{tab:datasets}
\renewcommand{\arraystretch}{1.3}
\begin{tabular}{llccccccccc}
\hline
Galaxy & Dataset & Reference & $N_{\mathrm{mem}}$ & \multicolumn{2}{c}{$R_{\mathrm{min}}$} & \multicolumn{2}{c}{$R_{\mathrm{max}}$} & $\sigma^\mathrm{global}_{\mathrm{los}}$ & $\langle \Delta_\mathrm{los} \rangle$ \\
 &  &  & $(p_\mathrm{mem} > 0.8)$ & arcmin & kpc ($R_{1/2}$) & arcmin & kpc ($R_{1/2}$) & km\,s$^{-1}$ & km\,s$^{-1}$ \\
\hline
Draco & MMT/Hectochelle & \citetalias{Walker23} & 510 & 0.16 & 0.004 (0.02) & 121.7 & 2.858 (12.5) & $8.96^{+0.30}_{-0.28}$ & 1.25 \\
& DESI & \citetalias{Ding25} & 157 & 1.10 & 0.026 (0.11) & 80.7 & 1.910 (8.33) & $9.35^{+0.59}_{-0.53}$ & 1.64 \\
\hline
\bootes & VLT/FLAMES & \citetalias{Koposov11}, \citetalias{Jenkins21} & 56 & 0.59 & 0.011 (0.06) & 13.0 & 0.251 (1.31) & $4.30^{+0.67}_{-0.56}$ & 3.11 \\
& Arx. AAT & \citetalias{Longeard22} & 42 & 1.67 & 0.032 (0.17) & 51.4 & 0.991 (5.16) & $3.83^{+0.70}_{-0.58}$ & 2.24 \\
& MMT/Hectochelle & \citetalias{Walker23} & 37 & 1.67 & 0.032 (0.17) & 31.4 & 0.606 (3.15) & $3.64^{+0.51}_{-0.42}$ & 1.01 \\
& $S^5$ AAT & \citetalias{Sandford26} & 56 & 1.67 & 0.032 (0.17) & 59.5 & 1.149 (5.98) &  $3.22^{+0.54}_{-0.47}$ & 3.36 \\
& Combined & \citetalias{Sandford26} & 115 & 0.59 & 0.011 (0.06) & 59.5 & 1.149 (5.98) &  $3.98^{+0.59}_{-0.39}$ & 2.56 \\
\hline
\end{tabular}
\begin{flushleft}
\textit{Notes.} $N_{\mathrm{mem}}$ is the number of member stars with membership probability $p_{\mathrm{mem}} > 0.8$ used in the analysis. $R_{\mathrm{min}}$ and $R_{\mathrm{max}}$ indicate the minimum and maximum projected distance from the galaxy center in angular and physical units; parenthetical values in the kpc columns are in units of $R_{1/2}$. $\sigma^\mathrm{global}_{\mathrm{los}}$ is the global LOS velocity dispersion with $1\sigma$ uncertainties. $\langle \Delta_\mathrm{los} \rangle$ is the mean velocity measurement uncertainty. We adopt distances and half-light radii of $d = 81.6 \, \kpc$ and $R_{1/2} = 0.229 \, \kpc \; (9.76')$ for Draco \citep{2024AJ....167..247B, 2018ApJ...860...66M}, and $d = 66.4 \, \kpc$ and $R_{1/2} = 0.192 \, \kpc \; (9.97')$ for \bootes \citep{2006ApJ...653L.109D, 2018ApJ...860...66M}. \\
\textit{References.} \citetalias{Walker23}: \citet{Walker23}; \citetalias{Ding25}: \citet{Ding25}; \citetalias{Koposov11}: \citet{Koposov11}; \citetalias{Jenkins21}: \citet{Jenkins21}; \citetalias{Longeard22}: \citet{Longeard22}; \citetalias{Sandford26}: \citet{Sandford26}.
\end{flushleft}
\end{table*}

We summarize the spectroscopic datasets and the membership selection used to constrain the mass profiles of Draco and \bootes.
For Draco, we analyze two homogeneous samples, each obtained with a single instrument and reduction pipeline: MMT/Hectochelle observations from \cite{Walker23} (hereafter \citetalias{Walker23}) and DESI observations from \cite{Ding25} (hereafter \citetalias{Ding25}).
The two Draco samples have distinct radial distributions -- the MMT sample is more concentrated toward the galaxy center, while the DESI sample extends further into the outskirts.
For \bootes, we analyze the two samples compiled in \cite{Sandford26} (hereafter \citetalias{Sandford26}): a homogeneous \sfive AAT sample, which minimizes systematics associated with combining heterogeneous observations, and a \sfivecomb sample that extends \sfive with published VLT and MMT measurements and with archival AAT spectroscopy re-processed through the \sfive pipeline, maximizing kinematic sample size and radial coverage.
Table~\ref{tab:datasets} summarizes the sample sizes, radial coverage, global line-of-sight velocity dispersions, and mean per-star velocity uncertainties of all datasets considered here.
Fig.~\ref{fig:velocity_dist} shows the LOSV, binned velocity dispersion profiles, and radial distributions of tracers in these samples.

\subsubsection{Draco}

For Draco, we use stellar kinematics from two large spectroscopic surveys: MMT/Hectochelle observations from \citetalias{Walker23} and more recent DESI observations from \citetalias{Ding25}.
Both datasets provide heliocentric radial velocities and \feh measurements.
Of the datasets considered in this work, \citetalias{Walker23} has the smallest mean per-star velocity uncertainty ($\sim 1.25\,\kms$; see Table~\ref{tab:datasets}).

The DESI sample combines Survey Validation and Year-1 observations obtained as part of the DESI Milky Way Survey \citep{2023ApJ...947...37C}.
\citetalias{Ding25} select candidate Draco members using color--magnitude and data-quality cuts, and assign membership probabilities using a two-component Gaussian mixture model (GMM) applied to heliocentric LOSV, \feh, and \emph{Gaia} DR3 proper motions.
We adopt the \citetalias{Ding25} membership catalog with a cut of $p_\mathrm{mem} > 0.8$, which yields 157 Draco members spanning projected radii of $\sim 0.11-8.33\,\rhalf$.

The \citetalias{Walker23} catalog does not itself provide a membership classification for Draco.
To obtain a self-consistent membership list for the MMT sample, we use the classifications from \citetalias{Ding25}, who apply the same GMM procedure used for their DESI sample to the \citetalias{Walker23} catalog and report Draco membership probabilities for the MMT stars (see their Section~5.2).
Applying the same $p_\mathrm{mem} > 0.8$ cut used for DESI yields 510 MMT member stars spanning $\sim 0.02-12.5\,\rhalf$.

We use the DESI and MMT datasets because they are the two largest single-instrument samples currently available for Draco.
Analyzing them separately, rather than constructing a combined catalog, avoids introducing systematics associated with combining heterogeneous instruments, reduction pipelines, and selection functions.
Moreover, as shown in Fig.~\ref{fig:velocity_dist}, the two samples have distinct radial distributions: the MMT sample is more concentrated toward the galaxy center, while the DESI sample extends further into the outskirts and is relatively sparse in the innermost regions.
This allows us to probe how \gnn and Jeans modeling respond to different tracer distributions.

\subsubsection{\bootes}

\bootes currently has the largest spectroscopic member sample among UFDs, and \citetalias{Sandford26} present the most up-to-date compilation of this sample.
A notable feature of \citetalias{Sandford26} is that they leverage the 16-year baseline spanned by their combined spectroscopic data to identify binary candidates among \bootes members.
Following their analysis, we exclude all stars flagged as binaries by \citetalias{Sandford26} from the samples used in this work, as unresolved binary motion can inflate the observed velocity dispersion and bias the inferred DM density profile \citep[e.g.][]{2017AJ....153..254S, 2023ApJ...956...91W}.
We also exclude RR~Lyrae variables, which can also inflate the observed velocity dispersion due to stellar pulsations.
\citetalias{Sandford26} provide two datasets for \bootes, both of which we use here.

The first dataset, which we refer to as \sfive, is a homogeneous sample drawn from new observations taken as part of the Southern Stellar Stream Spectroscopic Survey \citep[\sfive;][]{2019MNRAS.490.3508L} using the Two-degree Field \citep{2002MNRAS.334..673L} fiber-fed AAOmega spectrograph \citep{2006SPIE.6269E..0GS} on the Anglo-Australian Telescope (AAT).
Membership probabilities are assigned using a two-component GMM applied to heliocentric velocity, \feh, and \emph{Gaia} DR3 proper motions.
After excluding identified binary candidates and RR~Lyrae variables and applying $p_\mathrm{mem} > 0.8$, the \sfive sample contains 56 member stars spanning $\sim 0.17-5.98\,\rhalf$.

The second dataset, which we refer to as \sfivecomb, combines the \sfive AAT observations with three additional data sources: (i) published VLT/FLAMES measurements from \cite{Jenkins21}, based on a re-analysis of the multi-epoch spectroscopy originally presented in \cite{Koposov11}; (ii) published MMT/Hectochelle measurements from \citetalias{Walker23}; and (iii) archival AAT/AAOmega spectroscopy originally published by \citet{2010ApJ...723.1632N} and \cite{Longeard22}.
For the VLT/FLAMES and MMT/Hectochelle data, \citetalias{Sandford26} adopt the velocity and \feh measurements as published in \cite{Jenkins21} and \citetalias{Walker23}.
For the archival AAT spectra, \citetalias{Sandford26} instead re-reduces and re-analyzes the raw data using the same pipeline applied to the new \sfive observations, rather than adopting the measurements published in \citet{2010ApJ...723.1632N} and \citet{Longeard22} (see \citetalias{Sandford26} for details).
Although \sfivecomb is assembled from multiple instruments, reduction pipelines, and spectral resolutions, \citetalias{Sandford26} apply zero-point corrections to homogenize the combined catalog, measuring and correcting velocity and metallicity offsets between datasets using stars observed in more than one sample.
\citetalias{Sandford26} then apply the same GMM procedure to this homogenized catalog to derive membership probabilities in a manner consistent with the \sfive sample.
After excluding binaries and RR~Lyrae variables and applying the same $p_\mathrm{mem} > 0.8$ cut, the \sfivecomb sample contains 115 member stars spanning $\sim 0.06-5.98\,\rhalf$.

As shown in Fig.~\ref{fig:velocity_dist}, the main increase in sample size from \sfive to \sfivecomb occurs within the half-light radius $\rhalf$, and this inner-region gain is contributed mainly by the VLT/FLAMES observations: the smaller field of view of VLT/FLAMES concentrates its coverage in the central region, while the large aperture of the VLT reaches fainter stars than the AAT, allowing denser sampling of \bootes's inner regions.

We use both datasets for \bootes for complementary reasons: \sfive is the cleanest test case, based on data from a single instrument and reduction pipeline, while \sfivecomb represents the largest kinematic sample currently available for any ultra-faint dwarf and provides the broadest radial coverage for density-profile inference.



\subsection{Velocity Dispersion Profiles}
\label{section:vdisp_profile}

\begin{figure*}
    \centering
    \includegraphics[width=0.98\linewidth]{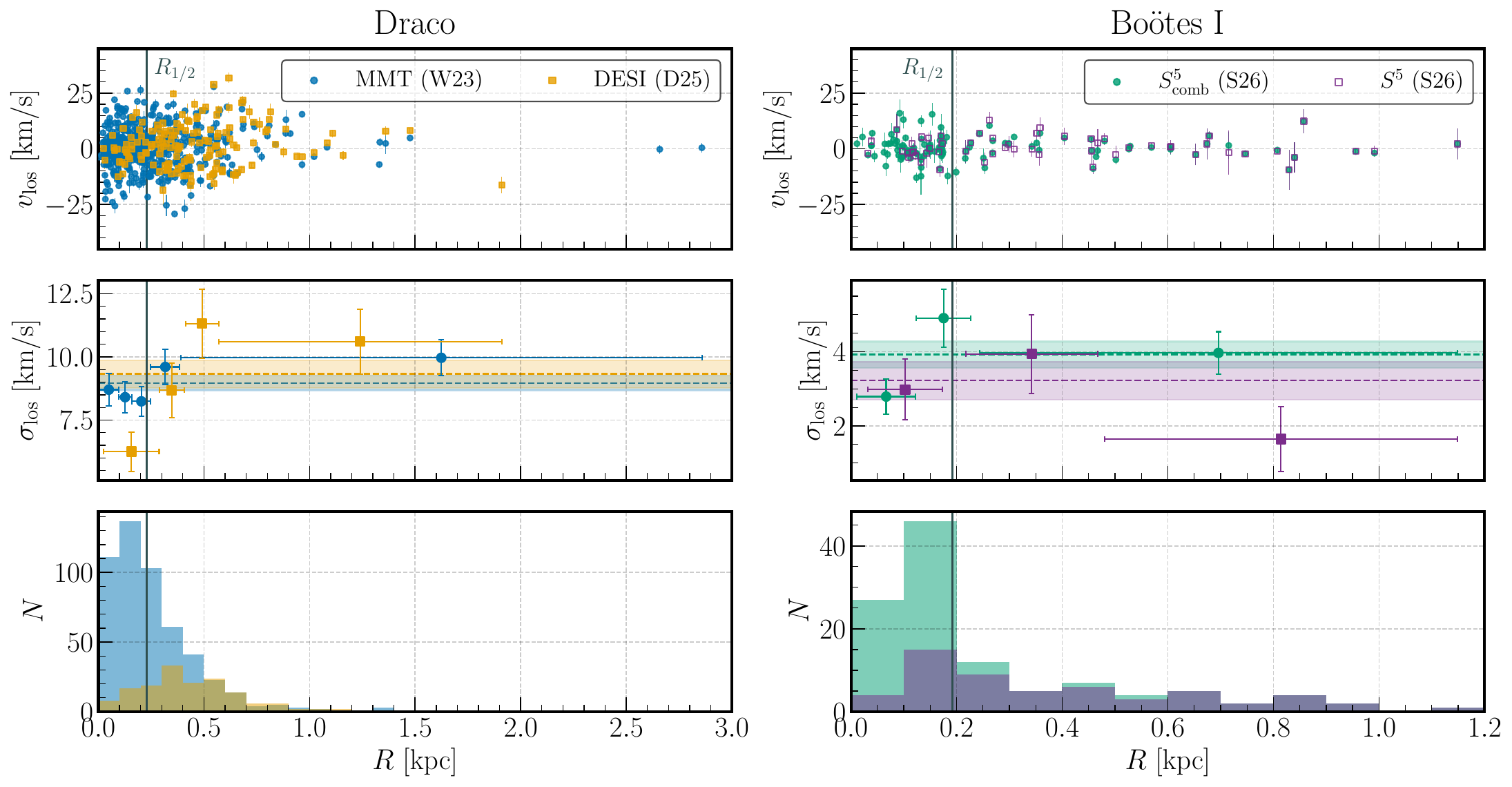}
    \caption{
    The LOSV (top), binned dispersion profiles (middle), and radial distributions of tracers (bottom) for Draco (left) and \bootes (right).
    For Draco, the MMT (\citetalias{Walker23}, blue circles) and DESI (\citetalias{Ding25}, orange squares) datasets are shown.
    For \bootes, the \sfivecomb (\citetalias{Sandford26}, teal circles) and \sfive (\citetalias{Sandford26}, purple squares) datasets are shown.
    Gray vertical lines denote the half-light radius $\rhalf$.
    In the middle panels, data points show the binned velocity dispersions obtained by fitting a Gaussian model to stars in each radial bin via MCMC, properly accounting for per-star measurement uncertainties.
    Vertical error bars denote the $1\sigma$ uncertainty in the dispersion, while horizontal error bars denote the center and width of the radial bins.
    Horizontal dashed lines and shaded bands show the median and 68\% percentile of the global velocity dispersion.
    }
    \label{fig:velocity_dist}
\end{figure*}

We compute the binned LOSV dispersion profiles for visualization and comparison with literature measurements.
Although the binned profiles are not used directly in our analysis (both Jeans modeling and \gnn operate on unbinned stellar kinematics), they provide a useful visual summary of the kinematic data.

For each tracer, we compute the projected distance from the galaxy center, adopting heliocentric distances of $d = 81.6 \, \mathrm{kpc}$ for Draco \citep{2024AJ....167..247B} and $d = 66.4 \, \mathrm{kpc}$ for \bootes \citep{2006ApJ...653L.109D}.
We then divide the stars into radial bins with equal numbers of tracers per bin to compute binned velocity dispersion profiles using the MCMC procedure described in Appendix~\ref{app:kurtosis_fit}.
We also compute the global velocity dispersion for each dataset using the same procedure applied to the full sample, and report the velocity dispersions in Table~\ref{tab:datasets}.

Fig.~\ref{fig:velocity_dist} shows the observed LOSV and projected distances (top panels) and the binned velocity dispersion profiles (bottom panels) for all datasets.
For Draco, the MMT \citepalias{Walker23} and DESI \citepalias{Ding25} datasets have comparable radial coverage, spanning $[0.02, 12.5]\,\rhalf$ and $[0.11, 8.33]\,\rhalf$, respectively, though the DESI sample is predominantly distributed beyond the half-light radius with relatively few tracers at small projected radii.
The innermost DESI bin falls $\sim 2\sigma$ below the MMT bins at comparable radii.
We attribute this discrepancy to undersampling bias, as sparse sampling at small $R$ can systematically underestimate the true dispersion; we will explore this effect in more detail in Section~\ref{section:mock}).
More broadly, the low central completeness of the DESI sample reduces the kinematic information available in the inner region, which limits constraints on the inner density profile regardless of how well the selection function is modeled.

For \bootes, both \sfive and \sfivecomb \citepalias{Sandford26} have spectroscopic samples extending to large projected radii, up to $5.98 \, \rhalf$.
The tracer distributions of \sfivecomb are concentrated below \rhalf, primarily due to contributions from the archival VLT data \citep{Jenkins21}, while the \sfive samples are more uniform. 
Notably, the global velocity dispersion of the \sfive samples ($3.22^{+0.54}_{-0.47} \; \kms$) is systematically lower than those of the archival datasets, which range from approximately $\sigmalos =  3.6 - 4.3 \, \kms$.
The \sfivecomb samples thus have an intermediate global dispersion of $\sigmalos = 3.98^{+0.59}_{-0.39} \, \kms$, falling between \sfive and the remaining datasets.

Lastly, we note the effect of perspective rotation.
Perspective rotation is an observational effect in which the bulk transverse motion of a stellar system projects differently onto the LOSV at different positions across the system, thus inducing a spurious gradient that can mimic intrinsic rotation \citep{1961MNRAS.122..433F, 2002AJ....124.2639V, 2008ApJ...682L..93K, 2008ApJ...688L..75W, 2020MNRAS.495.3022P}.
This effect is most significant for nearby galaxies with large angular extents and proper motions, and has been primarily accounted for in analyses of classical dwarf spheroidals \citep[e.g.][]{2008ApJ...682L..93K, 2008ApJ...688L..75W, 2020MNRAS.495.3022P} and, more recently, ultra-diffuse galaxies \citep{2021ApJ...921...32J}.

Among the systems studied here, perspective rotation primarily affects \bootes due to its larger proper motion ($1.1\,\mathrm{mas\,yr^{-1}}$ versus $0.2\,\mathrm{mas\,yr^{-1}}$ for Draco; \citealt{2022ApJ...940..136P}), which produces a more substantial spurious gradient despite the smaller angular extent ($0.99^\circ$ versus $2.03^\circ$).
Previous analyses of \bootes by \citet{Hayashi23} and \citetalias{Sandford26}, with which we compare in Section~\ref{section:result_inner_density}, do not apply this correction.\footnote{\citetalias{Sandford26} note the effect but do not propagate it to their mass modeling; see discussion in Section~\ref{section:discussion}.}
To facilitate direct comparison with these analyses, the main results presented throughout this paper are computed \emph{without} correcting for perspective rotation, and we present a corresponding analysis with the correction applied in Appendix~\ref{app:rotation_corr}, showing that the inferred profiles are not substantially affected.

\subsection{LOSV Uncertainties}

\begin{figure}
    \centering
    \includegraphics[width=0.98\linewidth]{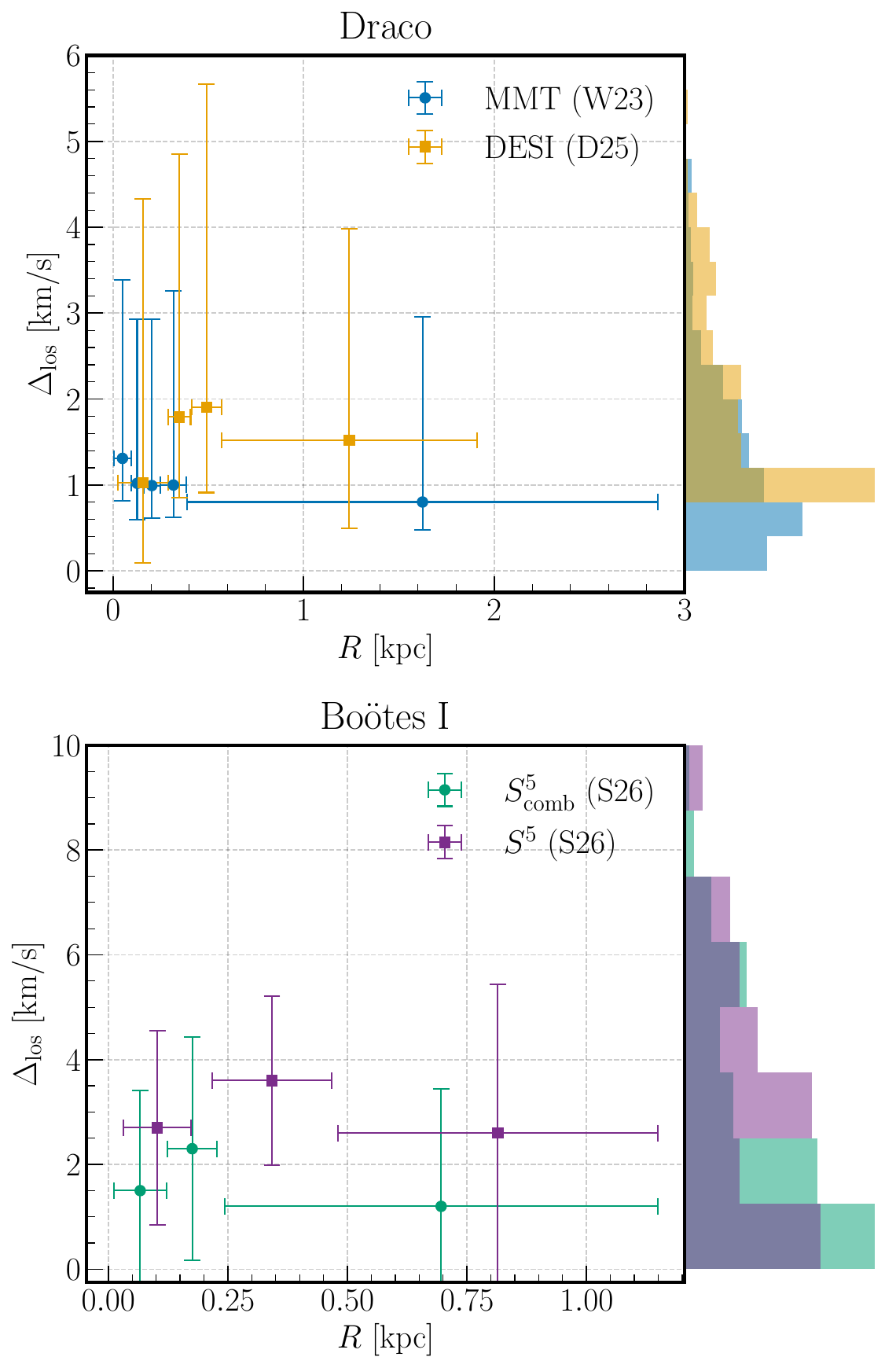}
    \caption{
    The LOSV measurement uncertainties $\Delta_\mathrm{los}$ as a function of projected radius for Draco (top) and \bootes (bottom), with normalized distributions shown on the right.
    The tracers of each dataset are divided into multiple bins, with the median and 68\% percentile of $\Delta_\mathrm{los}$ in each bin shown as data points with error bars. 
    For Draco, the MMT (\citetalias{Walker23}, blue circles) and DESI (\citetalias{Ding25}, orange squares) datasets are shown.
    For \bootes, the \sfivecomb (\citetalias{Sandford26}, teal circles) and \sfive (\citetalias{Sandford26}, purple squares) datasets are shown.
    }
    \label{fig:delta_vlos}
\end{figure}

Fig.~\ref{fig:delta_vlos} shows the median and 68\% percentile of the measurement uncertainties $\Delta_\mathrm{los}$ as a function of projected radius for all four datasets.
The average LOSV uncertainties $\langle\Delta_\mathrm{los}\rangle$ are also presented in Table~\ref{tab:datasets}.
We also note that the quoted LOSV measurements for DESI include a noise floor of approximately $0.9\,\kms$, added in quadrature to the formal measurement uncertainties.
Unless otherwise stated, every DESI value reported throughout this paper (including those in Table~\ref{tab:datasets}) already incorporates this noise floor.

From Fig.~\ref{fig:delta_vlos} and Table~\ref{tab:datasets}, the average LOSV uncertainties of Draco MMT and DESI are $\langle\Delta_\mathrm{los}\rangle = 1.25$ and $1.64 \, \kms$, respectively, and remain well below the intrinsic global LOSV dispersion of the system ($\sigmalos^\mathrm{global} \sim 9 \, \kms$).
On the other hand, for \bootes, $\langle\Delta_\mathrm{los}\rangle = 3.36$ and $2.56 \, \kms$ for the \sfive and \sfivecomb datasets, respectively. 
This is comparable to the intrinsic LOSV dispersion of \bootes ($\sigmalos^\mathrm{global} \sim 4-5 \, \kms$), so we expect the inference to be more sensitive to how well the uncertainties are characterized. 

 Lastly, we find that $\langle\Delta_\mathrm{los}\rangle$ is broadly flat as a function of projected radius $R$ for the Draco MMT and both \bootes datasets.
For the Draco DESI dataset, the uncertainty mildly increases with $R$, though this variation is small compared to the intrinsic dispersion, suggesting that the positional dependence of measurement uncertainties has a negligible impact on our inferences.

\section{Performance on Mock Data}
\label{section:mock}

\subsection{Generating Mock Galaxies}
\label{section:mock_generate}

We generate mock galaxies of \bootes and Draco using the same forward model described in Section~\ref{section:forward_model}.
For each galaxy, we consider two sets of parameters following OM anisotropy profiles, \textsc{CoreOM} and \textsc{CuspOM}, and two sets of parameters with isotropic velocity anisotropy profiles, \textsc{CoreIso} and \textsc{CuspIso}.
The naming convention and fiducial parameters are inspired by the \textsc{Gaia Challenge} catalog, which has been used extensively in the literature \citep[e.g.,][]{2011ApJ...742...20W, 2017MNRAS.471.4541R, 2021MNRAS.501..978R, 2026A&A...705A.212B}.
We note below the modifications to their fiducial parameters.

Instead of fixing the Plummer scale radius $a$ to $1 \, \mathrm{kpc}$ and $0.25 \, \mathrm{kpc}$ for cuspy and cored profiles, respectively, we choose it to match the observed half-light radius of each galaxy as reported in the LVDB~\citep{2024arXiv241107424P}.
These correspond to $\rhalf=192.24^{+8.97}_{-8.63}\, \mathrm{pc}$ and $\rhalf=229.24^{+4.78}_{-4.61} \, \mathrm{pc}$ for \bootes and Draco, respectively \citep{2018ApJ...860...66M}.

Additionally, we set the DM density normalization $\rho_s$ to satisfy the Wolf mass constraint \citep{2010MNRAS.406.1220W}.
The Wolf mass $M_{\mathrm{Wolf}}$ is defined as the enclosed mass at the radius $r_{\mathrm{Wolf}}$ where $d \ln \nu / d \ln r = -3$, which can be estimated from the observed velocity dispersion profile independent of the mass model or velocity anisotropy.
For a Plummer tracer density profile, this condition yields $r_{\mathrm{Wolf}} = \sqrt{3} \, r_\star \approx 1.73 \, r_\star$.
We adopt the Wolf masses from the LVDB, which are $\log_{10}(M_{\mathrm{Wolf}} / \modot) = 6.38^{+0.08}_{-0.08}$ for \bootes and $\log_{10} (M_{\mathrm{Wolf}} / \modot) = 7.17^{+0.12}_{-0.11}$ for Draco.
These are calculated using the relation $M_{\mathrm{Wolf}} = 930 \, \sigma^2 \rhalf$ \citep{2010MNRAS.406.1220W}, where the velocity dispersions are from \citetalias{Sandford26} for \bootes and \citet{2015MNRAS.448.2717W} for Draco, with the uncertainties propagated via Monte Carlo sampling.
We fix $r_s = 1 \, \mathrm{kpc}$ for all profiles rather than $\rho_s$, since $\rho_s$ scales linearly with the enclosed mass for fixed $r_s$ and $\gamma$.
We obtain $\rho_s = 2.94 \times 10^7 \, \modot \, \mathrm{kpc}^{-3}$ (cored) and $\rho_s = 5.04 \times 10^6 \, \modot \, \mathrm{kpc}^{-3}$ (cuspy) for \bootes, and $\rho_s = 1.19 \times 10^8 \, \modot \, \mathrm{kpc}^{-3}$ (cored) and $\rho_s = 2.33 \times 10^7 \, \modot \, \mathrm{kpc}^{-3}$ (cuspy) for Draco.

\begin{table}
\centering
\caption{Fiducial parameters for Draco and \bootes mocks. 
The parameters of all models are fixed to $\alpha=1$, $\beta=3$, $r_s = 1 \,\kpc$, and $\beta_0=0$. 
The Plummer scale radius $r_\star$ is set to the observed half-light radius of each galaxy.}
\renewcommand{\arraystretch}{1.3}
\begin{tabular}{lcccc}
\hline
Profile & $\gamma$ & $\rho_s$ (M$_\odot$ kpc$^{-3}$) & $r_\star$ (kpc) & $r_a$ (kpc)\\
\hline
\multicolumn{5}{c}{Draco} \\
\hline
\textsc{DraCoreOM} & 0 & $1.19 \times 10^8$ & 0.229 & 0.229 \\
\textsc{DraCuspOM} & 1 & $2.33 \times 10^7$ & 0.229 & 0.229 \\
\textsc{DraCoreIso} & 0 & $1.19 \times 10^8$ & 0.229 & $\infty$ \\
\textsc{DraCuspIso} & 1 & $2.33 \times 10^7$ & 0.229 & $\infty$  \\
\hline
\multicolumn{5}{c}{\bootes} \\
\hline
\textsc{BooICoreOM} & 0 & $2.94 \times 10^7$ & 0.192 & 0.192 \\
\textsc{BooICuspOM} & 1 & $5.04 \times 10^6$ & 0.192 & 0.192 \\
\textsc{BooICoreIso} & 0 & $2.94 \times 10^7$ & 0.192 & $\infty$ \\
\textsc{BooICuspIso} & 1 & $5.04 \times 10^6$ & 0.192 & $\infty$ \\
\hline
\end{tabular}
\label{tab:fiducial_models}
\end{table}
The complete set of fiducial parameters for all four models is listed in Table~\ref{tab:fiducial_models}.

We use \textsc{Agama} to sample the 6D phase-space coordinates of the tracers, which are then projected onto a randomly oriented sky plane to obtain $(x, y, v_{\mathrm{los}})$.
We then incorporate the selection function of the target galaxies by matching the radial distribution of the mock tracers to that of the observed data.
This is achieved by resampling the mock tracers with weights $w_i = N_{\mathrm{obs}}(R_i) / N_{\mathrm{sim}}(R_i)$, where $N_{\mathrm{obs}}(R_i)$ and $N_{\mathrm{sim}}(R_i)$ are the number of observed and simulated stars in radial bins centered on $R_i$, effectively applying an empirical selection function $\mathcal{S}(R) \propto w(R)$ to the mock data.
The total number of resampled tracers is set to match the observed sample size.

We incorporate measurement uncertainties by convolving each mock velocity with a Gaussian kernel $\mathcal{N}(v_{\mathrm{los},i}, \Delta_i^2)$, where $\Delta_i$ is drawn from the empirical uncertainty distribution of the observed spectroscopic sample.
This procedure matches the observed uncertainty distribution but does not capture correlations between uncertainties and tracer properties (e.g., magnitude, color, position).

Finally, we note that the mocks are generated from equilibrium DFs rather than $N$-body simulations and therefore do not account for tidal effects, substructure, or other dynamical processes that may affect the real systems.
Additionally, the fiducial parameters are chosen to approximate the observed properties of Draco and \bootes but are not intended to precisely reproduce their true density profiles or anisotropy.
These mocks serve primarily as a validation tool to assess the performance of our inference framework on realistic datasets with observational effects (selection functions, measurement uncertainties) matching the target galaxies, rather than as high-fidelity representations of the systems themselves.

\subsection{Results on Mock Galaxies}
\label{section:mock_result}

\begin{figure}
    \centering
    \includegraphics[width=\linewidth]{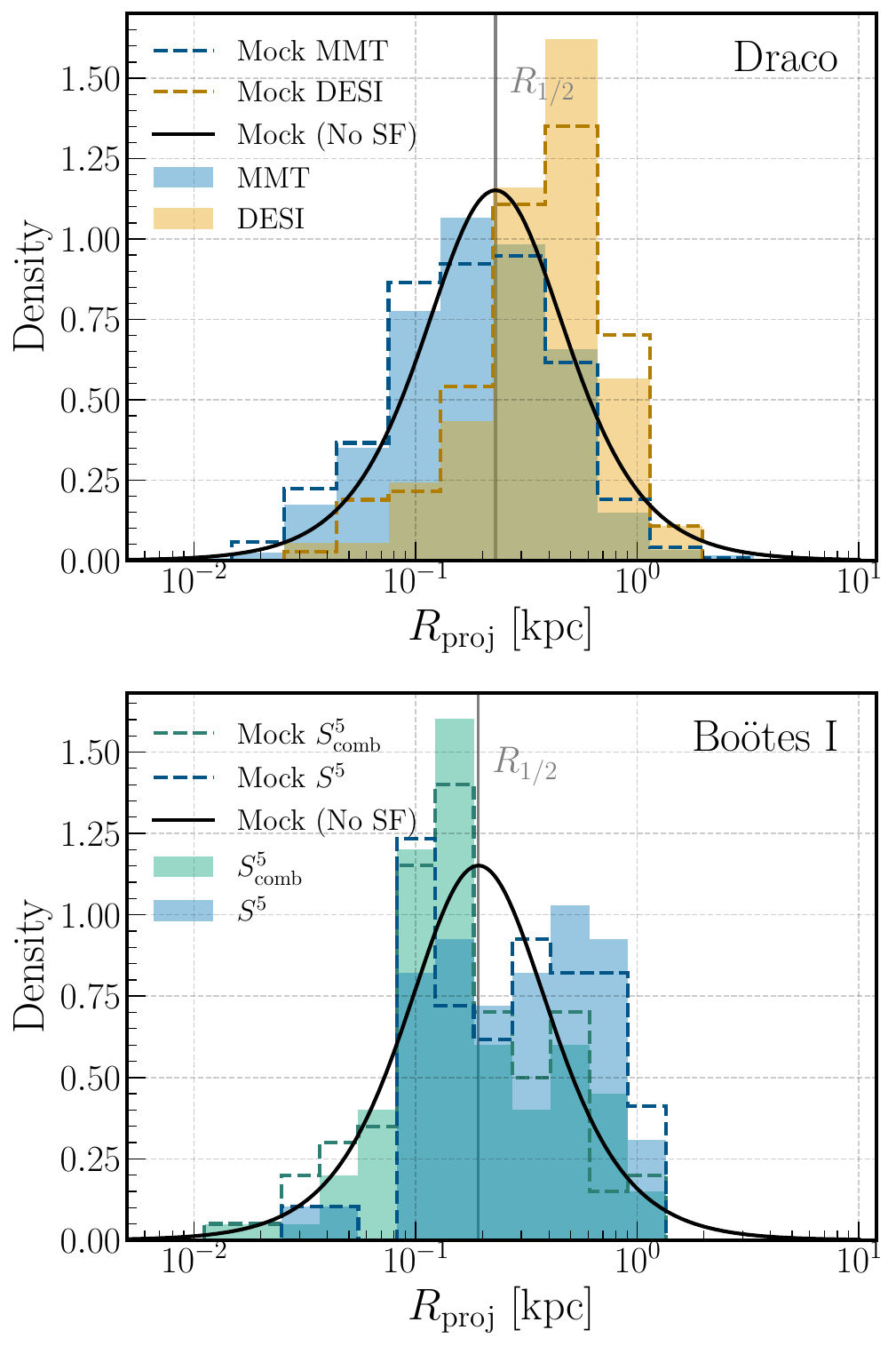}
    \caption{
    The projected radial distributions of spectroscopic tracers for Draco (top panel) and \bootes (bottom panel).
    Filled histograms show the observed spectroscopic samples: MMT \citepalias{Walker23} (blue) and DESI \citepalias{Ding25} (orange) for Draco, and \sfivecomb (teal) and \sfive (purple) \citetalias{Sandford26} for \bootes.
    Dashed histograms show the corresponding mock datasets after applying the empirical selection functions via resampling.
    The black solid curve shows the intrinsic radial distribution expected from the Plummer profile in the absence of spectroscopic selection effects.
    The vertical gray line marks the half-light radius $\rhalf$ from the LVDB~\citep{2024arXiv241107424P} for each galaxy.
    }
    \label{fig:selection_fn}
\end{figure}

\begin{figure*}
    \centering
    \includegraphics[width=\linewidth]{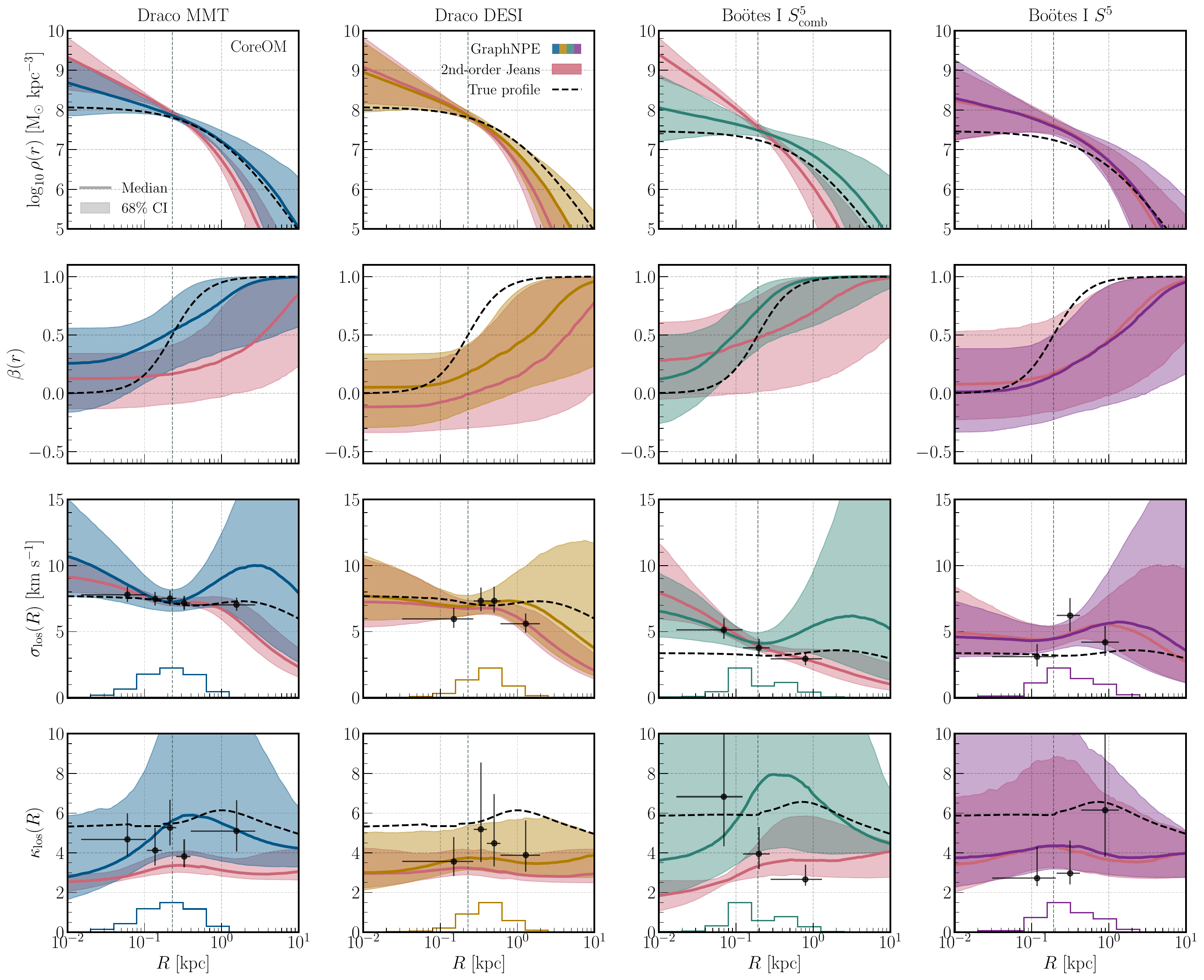}
    \caption{
    Comparison between the inferred profiles on mock galaxies with \textsc{CoreOM} parameters by \gnn and second-order Jeans modeling.
    From top to bottom, the rows show the DM density $\rho(r)$, velocity anisotropy $\beta(r)$, LOSV dispersion $\sigmalos(R)$, and LOSV kurtosis profiles $\kappalos(R)$.
    From left to right rows show results for Draco MMT, Draco DESI, \bootes \sfivecomb, and \bootes \sfive mocks, respectively.
    Shaded regions indicate 68\% credible intervals for each method, and solid lines show posterior medians.
    The dashed black line shows the true input profile.
    Data points show the binned LOSV dispersion and kurtosis profiles of the mock galaxies, shown for \textit{visualization only} as neither method uses them in the fit.
    The inset histograms in the LOSV dispersion panels show the radial distribution of tracers in the mocks.
    The vertical gray dashed line marks the observed half-light radius $\rhalf$ of each galaxy from the LVDB~\citep{2024arXiv241107424P}.
    }
    \label{fig:coreom_result}
\end{figure*}

\begin{figure*}
    \centering
    \includegraphics[width=0.95\linewidth]{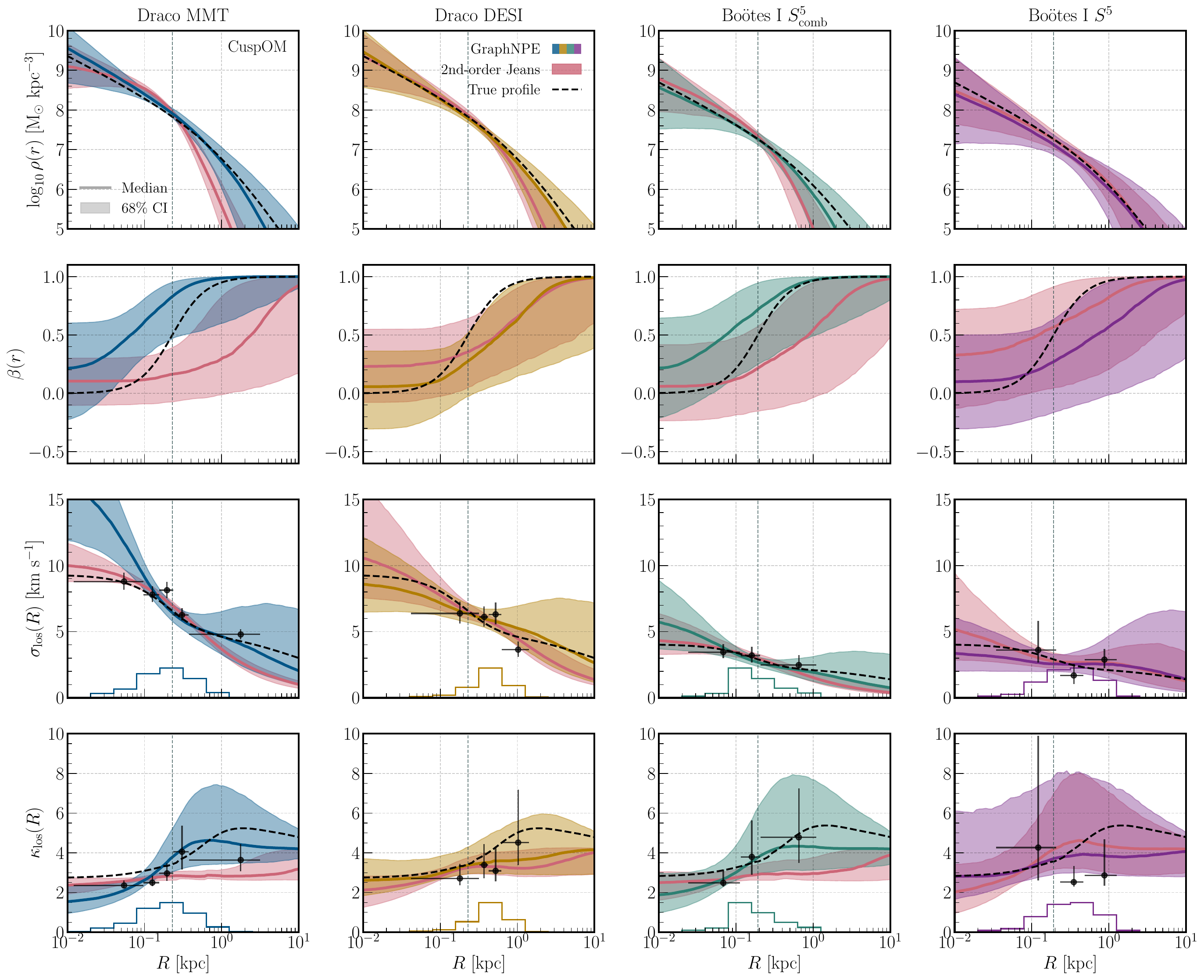}
    \caption{
    Same as Fig.~\ref{fig:coreom_result}, but for mock galaxies with \textsc{CuspOM} parameters.
    }
    \label{fig:cuspom_result}
\end{figure*}

\begin{table}
\centering
\caption{Comparison between the inferred inner slope $\gamma$ between \gnn and Jeans modeling on all mock datasets.}
\renewcommand{\arraystretch}{1.3}
\setlength{\tabcolsep}{2pt}
\begin{tabular}{lcccc}
\hline
& \multicolumn{2}{c}{\textsc{CoreOM} $(\gamma=0)$}  & \multicolumn{2}{c}{\textsc{CuspOM} $(\gamma=1)$}  \\
Mock & \gnn & Jeans & \gnn & Jeans \\
\hline
Draco MMT & $0.64^{+0.67}_{-0.47}$ & $0.95^{+0.75}_{-0.45}$ & $1.18^{+0.91}_{-0.47}$ & $0.37^{+0.66}_{-0.67}$ \\
Draco DESI  & $0.69^{+0.89}_{-0.64}$ & $0.82^{+0.73}_{-0.62}$ & $1.12^{+0.86}_{-0.49}$ & $0.93^{+0.76}_{-0.53}$ \\
\bootes \sfivecomb & $0.40^{+0.78}_{-0.70}$ & $1.27^{+0.66}_{-0.41}$ & $0.97^{+1.02}_{-0.63}$ & $0.77^{+0.72}_{-0.58}$ \\
\bootes \sfive & $0.49^{+0.73}_{-0.87}$ & $0.87^{+0.69}_{-1.03}$ & $0.31^{+0.61}_{-0.70}$ & $0.96^{+0.50}_{-0.71}$ \\
\hline
& \multicolumn{2}{c}{\textsc{CoreIso} $(\gamma=0)$}  & \multicolumn{2}{c}{\textsc{CuspIso} $(\gamma=1)$}  \\
Mock & \gnn & Jeans & \gnn & Jeans \\
\hline
Draco MMT & $0.59^{+0.63}_{-0.42}$ & $0.56^{+0.43}_{-0.29}$ & $1.08^{+0.81}_{-0.41}$ & $0.79^{+0.57}_{-0.42}$ \\
Draco DESI  & $0.10^{+0.57}_{-0.45}$ & $-0.14^{+0.35}_{-0.40}$ & $0.41^{+0.63}_{-0.47}$ & $0.80^{+0.46}_{-0.40}$ \\
\bootes \sfivecomb & $0.17^{+0.61}_{-0.53}$ & $0.15^{+0.45}_{-0.43}$ & $1.18^{+0.75}_{-0.41}$ & $1.01^{+0.66}_{-0.44}$ \\
\bootes \sfive & $0.05^{+0.62}_{-0.63}$ & $0.37^{+0.83}_{-0.88}$ & $0.28^{+0.38}_{-0.40}$ & $0.32^{+0.63}_{-0.53}$ \\
\hline
\end{tabular}
\label{tab:mock_gamma}
\end{table}

We generate mock galaxies and present results for MMT \citepalias{Walker23} and DESI \citepalias{Ding25} for Draco, and \sfivecomb and \sfive \citepalias{Sandford26} for \bootes.
Fig.~\ref{fig:selection_fn} compares the projected radial distributions of tracers of each mock dataset and their corresponding observed dataset.
The top and bottom panels show the distributions for Draco and \bootes, respectively.
The black curves show the expected intrinsic radial distributions for a Plummer profile with scale radius $r_\star = R_{1/2}$ from the LVDB~\citep{2024arXiv241107424P}, in the absence of spectroscopic selection effects, for their corresponding galaxies.

For each mock galaxy, we compare the inferred DM density $\rho(r)$, velocity anisotropy $\beta(r)$, and LOSV dispersion $\sigmalos(R)$ from \gnn\ and Jeans modeling.
To better demonstrate the methodological differences between \gnn and second-order Jeans, we additionally compute the fourth-order LOSV moment, parameterized by the LOSV kurtosis profile \citep{2002MNRAS.333..697L, 2013MNRAS.432.3361R}
\begin{equation}
    \kappalos(R) \equiv \frac{\braket{(\vlos - V_\mathrm{sys})^4}}{\sigmalos^4}.
\end{equation}
The kurtosis probes the tails of the velocity distribution, with $\kappalos > 3$ indicating a heavier-tailed (leptokurtic) distribution and $\kappalos < 3$ a lighter-tailed (platykurtic) one relative to a Gaussian.
The fourth-order Jeans equations for computing $\kappalos(R)$ and the procedure for fitting the binned $\kappalos(R)$ data points are summarized in Appendix~\ref{app:fourth_order_jeans}.

We emphasize that neither method uses $\kappalos(R)$ explicitly.
Jeans modeling fits only $\sigmalos(R)$ via Eq.~\ref{eq:loglike_jeans}, while \gnn operates on the full kinematic DF and thus implicitly captures higher-order moments including $\kappalos(R)$.
Comparing the inferred $\kappalos(R)$ against the data therefore tests how well \gnn\ captures the full velocity distribution beyond the second moment, and provides an independent diagnostic check on Jeans modeling.

Figs.~\ref{fig:coreom_result} and \ref{fig:cuspom_result} show the medians and 68\% percentiles of the inferred DM density, velocity anisotropy, LOSV dispersion, and LOSV kurtosis profiles for all \textsc{CoreOM} and \textsc{CuspOM} mock galaxies, respectively.
The true profiles with parameters listed in Table~\ref{tab:fiducial_models} are shown as black dashed lines.
The binned $\sigmalos(R)$ and $\kappalos(R)$ profiles are shown as black data points.
Additional results for the isotropic mocks (\textsc{CoreIso} and \textsc{CuspIso}) are shown in Appendix~\ref{app:isotropic_mock}.

\subsubsection{GraphNPE performance}

Figs.~\ref{fig:coreom_result} and \ref{fig:cuspom_result} show that \gnn recovers the true DM density and velocity anisotropy profiles across all mock datasets.
The 68\% credible intervals consistently cover the true profiles both within and beyond the radial range covered by the observed tracers, with only minor deviations in $\beta(r)$.
\gnn thus generalizes robustly to realistic observational conditions, despite the mock datasets employing selection functions and LOSV measurement uncertainty distributions that differ from the training distribution (Section~\ref{section:method_new}).

The two bottom rows of Figs.~\ref{fig:coreom_result} and \ref{fig:cuspom_result} show the inferred LOSV dispersion $\sigmalos(R)$ and kurtosis $\kappalos(R)$ profiles, alongside the binned data points computed using the procedure in Appendix~\ref{app:fourth_order_jeans}.
\gnn posteriors generally contain the binned data points within the 68\% credible intervals, with noticeably wider intervals on $\kappalos(R)$ than on $\sigmalos(R)$, reflecting the higher intrinsic variance of the fourth moment.
The recovery of both profiles, particularly $\kappalos(R)$, is notable, since \gnn does not fit either profile directly.
This demonstrates that \gnn, by operating on the full kinematic DF, accesses higher-order velocity moments beyond what the second-moment likelihood alone provides.
We caution, however, that good agreement with the binned data points does not by itself imply recovery of the true profiles, since the binned data points are subject to sampling noise and may themselves deviate from the truth.
We will revisit this point in more detail below.

Lastly, we note that the credible intervals on the \bootes mocks are also wider than those on the Draco DESI mocks, despite the two having comparable tracer counts.
This reflects the larger LOSV uncertainties and smaller intrinsic velocity dispersion of \bootes, which together reduce the effective signal-to-noise of the kinematic data.

\subsubsection{Comparison with second-order Jeans modeling}

We now turn to a direct comparison between \gnn and second-order Jeans modeling on the OM mock datasets.
Table~\ref{tab:mock_gamma} summarizes the inferred inner slope $\gamma$ for both methods across all mock datasets.

From Figs.~\ref{fig:coreom_result} and \ref{fig:cuspom_result}, \gnn recovers the inner density slope more accurately than second-order Jeans for cored profiles, while performance on cuspy profiles is broadly comparable between the two methods.
Specifically, for cored profiles, both methods are biased toward cuspy values, consistent with past studies using second-order Jeans methods \citep{2020MNRAS.498..144G, 2021MNRAS.507.4715C, 2021MNRAS.501..978R, 2026A&A...705A.212B}.
This bias likely reflects the fact that the LOSV dispersion profile is less sensitive to variations in $\gamma$ near zero than near unity, making cores intrinsically harder to detect than cusps (see Section~\ref{section:los_disp_v_gamma}).
\gnn is less biased than Jeans across all mock datasets, thanks to its implicit access to higher-order kinematic information through the DF, as demonstrated by its recovery of the binned $\kappalos(R)$ profile.
The most striking difference is seen in the \bootes \sfivecomb \textsc{CoreOM} mock, where Jeans infers $\gamma = 1.27^{+0.66}_{-0.41}$ while \gnn recovers $\gamma = 0.40^{+0.78}_{-0.70}$, consistent with the true value within the uncertainties.
We discuss this case in detail below.

Additionally, we find that \gnn consistently outperforms second-order Jeans in recovering the outer density profile.
We also attribute this to the higher-order moments accessed by \gnn, which provide information about the outer profile that $\sigmalos(R)$ alone does not encode.
\citet{2020MNRAS.498..144G} demonstrate that including higher-order moments such as the virial shape parameters in GravSphere \citep{2017MNRAS.471.4541R} significantly reduces scatter in the enclosed mass estimate at large radii.
We note, however, that for real satellite dwarf galaxies, the outer profile may be tidally truncated, thus limiting the direct interpretability of the extrapolated profile.
Nevertheless, \citet{gnn2} demonstrate that this extrapolation remains useful in practice, as \gnn can accurately recover pre-infall quantities such as $V_\mathrm{max}$ and the peak virial mass $M_\mathrm{vir}^\mathrm{peak}$ of FIRE-2 satellites.

\begin{figure}
    \centering
    \includegraphics[width=0.95\linewidth]{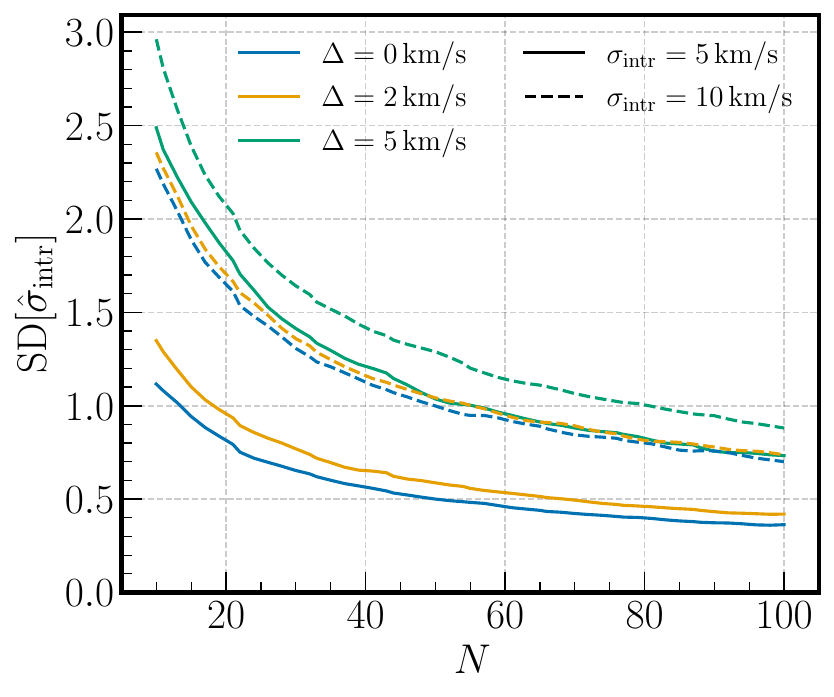}
    \caption{
    The expected scatter in the intrinsic velocity dispersion estimator $\mathrm{SD}[\hat{\sigma}_\mathrm{intr}]$, as a function of the number of tracer stars $N$ and measurement uncertainties $\Delta$, estimated empirically over $M=2000$ Monte Carlo realizations.
    Two levels of intrinsic dispersion are shown, $\sigma_\mathrm{intr} = (5, \, 10) \; \kms$, approximately representative of the observed velocity dispersions in \bootes and Draco, respectively.
    The estimator $\hat{\sigma}_\mathrm{intr}$ is the unbiased sample standard deviation with the known measurement uncertainty de-convolved in quadrature.
    }
    \label{fig:vdisp_mc_error}
\end{figure}

To better understand the performance differences between \gnn and second-order Jeans modeling, we now examine how well the recovered $\sigmalos(R)$ and $\kappalos(R)$ profiles match the binned data points and the true profiles.
Across all mock datasets, Jeans achieves a tighter fit to $\sigmalos(R)$ than \gnn.
This is by construction, as Jeans is explicitly optimized to fit the velocity dispersion (Eq.~\ref{eq:loglike_jeans}).
However, this tighter $\sigmalos(R)$ fit does not extend to $\kappalos(R)$, which Jeans systematically misses.\footnote{It is worth noting that $\sigmalos(R)$ and $\kappalos(R)$ are not fully independent: both are derived from the same $(\rho, \beta)$ posterior using the second- and fourth-order Jeans equations, respectively (Appendix~\ref{app:fourth_order_jeans}). As a result, the Jeans result can sometimes trace $\kappalos(R)$, even when the Jeans likelihood in Eq.~\ref{eq:loglike_jeans} only penalizes mismatch in $\sigmalos(R)$. A $(\rho, \beta)$ posterior that fits $\sigmalos(R)$ well can still predict a $\kappalos(R)$ that deviates from the data.}
\gnn, by contrast, matches both moments simultaneously and recovers the underlying density and velocity anisotropy profiles more accurately in most cases.
\textit{A tighter fit to $\sigmalos(R)$ alone is therefore not a reliable indicator that the underlying density and anisotropy profiles are accurately recovered; higher-order moments such as $\kappalos(R)$ are required to better assess recovery quality.}

We highlight two failure modes of second-order Jeans that contribute to the performance difference: (1) the sensitivity to sampling noise in $\sigmalos(R)$, and (2) the more well-known mass-anisotropy degeneracy.
We discuss each in turn below.

First, the observed $\sigmalos(R)$ can deviate from the true dispersion profile purely from sampling noise.
Consider the \bootes \sfivecomb \textsc{CoreOM} mock in Fig.~\ref{fig:coreom_result}, where the innermost $\sigmalos$ bin sits notably above the true profile due to statistical fluctuations.
This produces an artificially declining $\sigmalos(R)$ profile toward larger radii, which second-order Jeans incorrectly interprets as a cuspy density profile.
Fig.~\ref{fig:vdisp_mc_error} demonstrates this quantitatively.
For a system representative of \bootes with an intrinsic dispersion of $\sigma_\mathrm{intr} = 5 \; \kms$ and a mean LOSV uncertainty of $\langle \Delta \rangle = 2 \; \kms$, the scatter in the maximum likelihood estimator $\hat{\sigma}_\mathrm{intr}$ across 2000 Monte Carlo realizations is about $0.8-1.0 \; \kms$ ($16-20\%$ of $\sigma_\mathrm{intr}$) at $N \approx 40$ tracers (roughly the same number of tracers in each bin).
Even in the limit of negligible measurement uncertainty, i.e. $\langle \Delta \rangle = 0 \; \kms$, the scatter remains around $0.6 \; \kms$ due to the finite tracer counts.

From Fig.~\ref{fig:coreom_result}, we see that second-order Jeans infers a biased cuspy profile for the \bootes \sfivecomb \textsc{CoreOM} mock.
Perhaps most worrying, the Jeans posterior is both strongly biased and tightly constrained, yet provides a good fit to the binned $\sigmalos(R)$ data.
The $\sigmalos(R)$ fit alone thus provides no indication of the underlying bias.

On the other hand, \gnn is more robust as it operates on the full DF and thus can constrain $(\rho, \beta)$ using both $\sigmalos(R)$ and higher-order moments such as $\kappalos(R)$.
Since both moments depend on $(\rho, \beta)$, jointly constraining them yields more accurate posteriors and alleviates the bias from a noisy $\sigmalos(R)$ alone, even when neither moment is matched perfectly.
Similarly, we see that the \gnn posterior does not match the binned $\kappalos(R)$ exactly, but remains broadly consistent with both the data and the true profile, while Jeans stays flat and systematically underestimates them at intermediate radii.
We note, however, although the joint constraint mitigates the bias, it does not eliminate it: in the same mock, \gnn still overestimates the inner density normalization $\rho_s$, though substantially less than Jeans.

The second failure mode is the well-known mass-anisotropy degeneracy, where different combinations of the enclosed mass and velocity anisotropy profiles can produce identical $\sigmalos(R)$ profiles.
For example, the Draco MMT \textsc{CoreOM} mock result in Fig.~\ref{fig:coreom_result} illustrates that although the binned $\sigmalos$ aligns well with the true dispersion profile, Jeans still infers a biased inner slope ($\gamma = 0.95^{+0.75}_{-0.45}$) compared to \gnn ($\gamma = 0.64^{+0.67}_{-0.47}$).
The poorer recovery of $\beta(r)$ by Jeans suggests that the bias is driven by the mass-anisotropy degeneracy rather than noisy dispersion bins.
By contrast, \gnn breaks this degeneracy by recovering $\kappalos(R)$ through DF mapping.
This is consistent with the well-established role of higher-order moments in breaking the mass-anisotropy degeneracy \citep{2002MNRAS.333..697L, 2013MNRAS.432.3361R}.

Finally, we note an interesting pattern in the performance of second-order Jeans modeling across the different mock profiles.
Jeans modeling recovers the density profiles more accurately for the isotropic mocks (\textsc{CoreIso} and \textsc{CuspIso}) than for the OM mocks (\textsc{CoreOM} and \textsc{CuspOM}), as seen in Table~\ref{tab:mock_gamma} where the inferred inner slopes from Jeans are more comparable to \gnn for the isotropic cases.

\subsection{Sensitivity of velocity moments to the inner slope}
\label{section:los_disp_v_gamma}

\begin{figure}
    \centering
    \includegraphics[width=0.98\linewidth]{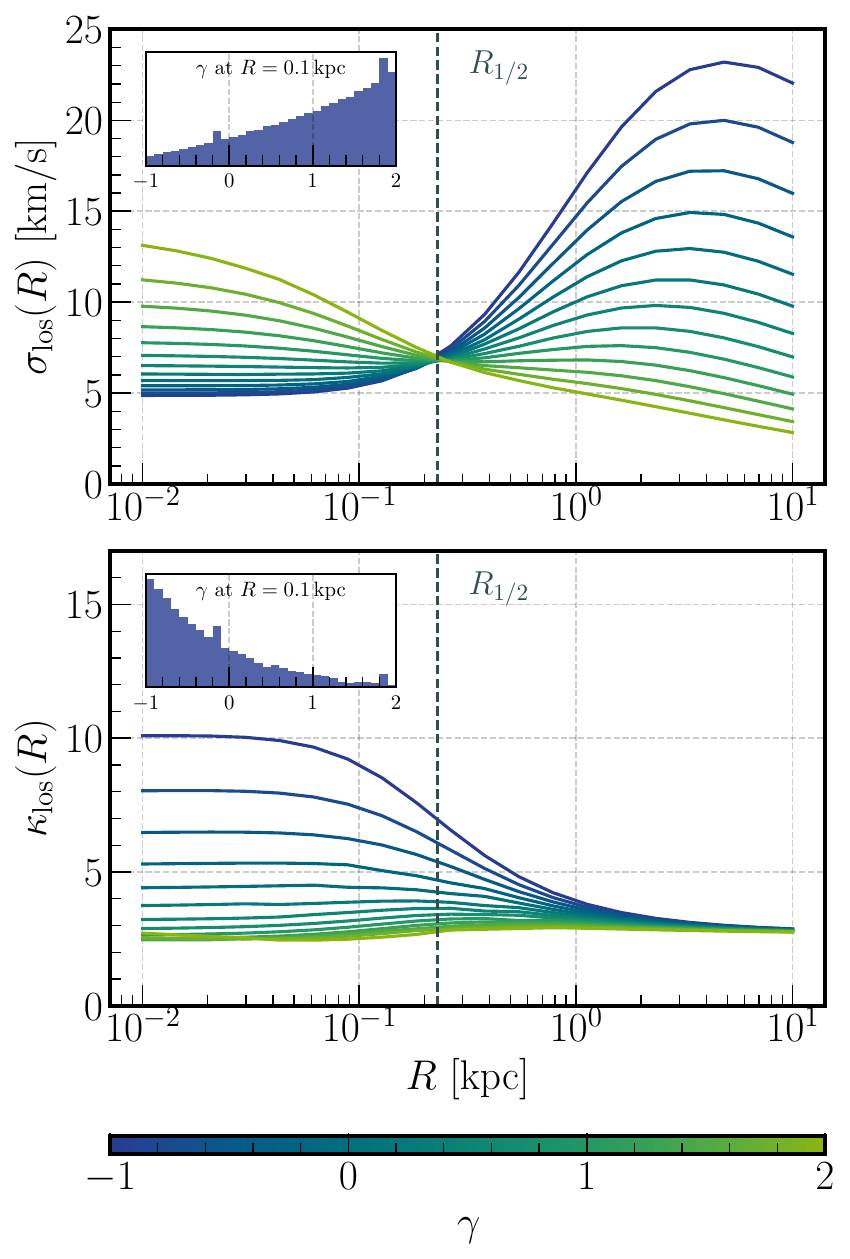}
    \caption{
    The LOSV dispersion $\sigmalos(R)$ (top) and kurtosis profiles $\kappalos(R)$ (bottom) as a function of projected radius for varying inner slopes $\gamma$, computed for an isotropic velocity profile ($\beta = 0$).
    The DM halo parameters are chosen to match the Wolf mass of Draco \citep{2010MNRAS.406.1220W}, with the dashed vertical line marking the half-light radius $\rhalf$.
    Each curve corresponds to a different value of $\gamma$, color-coded from cored ($\gamma = -1$) to cuspy ($\gamma = 2$), with a step size of $\Delta\gamma=0.25$.
    Inset histograms show the distribution of $\gamma$ recovered at $R = 0.1\,\kpc$ under uniform sampling of the corresponding moment ($\sigmalos$ for the top panels, $\kappalos$ for the bottom), illustrating the asymmetric sensitivity of each moment to the inner slope: \sigmalos preferentially constrains cuspy profiles, while \kappalos preferentially constrains cored profiles.
    }
    \label{fig:sigma_kappa_los}
\end{figure}

To gain further insight into the core-cusp sensitivity of second-order Jeans modeling, we calculate the LOSV dispersion $\sigmalos(R)$ and kurtosis $\kappalos(R)$ profiles as a function of the inner slope $\gamma$.
Fig.~\ref{fig:sigma_kappa_los} shows $\sigmalos(R)$ and $\kappalos(R)$ for 12 different values of $\gamma \in [-1, 2]$ (step size of $\Delta\gamma = 0.25$).
We fix $r_s = 1\,\kpc$ and choose $\rho_s$ such that the Wolf mass matches that of Draco, following the procedure described in Section~\ref{section:mock_generate}.
For simplicity, we assume an isotropic velocity profile ($\beta=0$).

For different values of $\gamma$, $\sigmalos$ converge near \rhalf but diverge significantly at both small and large radii.
Cuspy profiles ($\gamma \sim 1$) produce higher dispersions at small radii and lower dispersions at large radii compared to cored profiles ($\gamma \sim 0$).
Below \rhalf, the gradient of $\sigmalos$ with respect to $\gamma$ is much smaller near $\gamma \sim 0$ than near $\gamma \sim 1$, so cored profiles produce nearly identical dispersion profiles across a wider range of $\gamma$.
This asymmetric sensitivity has direct consequences for inference:
Jeans modeling, and any method relying solely on the velocity dispersion, is intrinsically less informative about cored profiles than about cuspy ones, as we now demonstrate quantitatively.

Consider the distribution of $\gamma$ recovered at a fixed projected radius of $R = 0.1\,\kpc$ under \textit{uniform sampling in $\sigmalos$}.
Specifically, we draw $\sigmalos$ values uniformly across the range spanned by the family of profiles and invert the $\sigmalos-\gamma$ relation at each step to obtain the corresponding $\gamma$.
The resulting distributions, shown as inset histograms in the top panels of Fig.~\ref{fig:sigma_kappa_los}, are strongly skewed toward cuspy values for both anisotropy assumptions.
This suggests that a uniform prior on the observable $\sigmalos$ maps onto a $\gamma$ posterior that disproportionately favors cusps.
Equivalently, a broad range of cored profiles is compressed into a narrow range of $\sigmalos$, so the data carry little information to distinguish among them.

Additional sources of uncertainty, such as large measurement errors and the mass-anisotropy degeneracy, further exacerbate this insensitivity in the cored regime.

The bottom panels of Fig.~\ref{fig:sigma_kappa_los} show $\kappalos(R)$ for the same range of $\gamma$.
At small radii, \kappalos varies strongly with $\gamma$, particularly in the cored regime, with cored profiles ($\gamma \sim 0$) producing notably higher kurtosis than cuspy ones ($\gamma \sim 1$).
At large radii, the curves converge.
Repeating the uniform-sampling exercise on \kappalos at $R = 0.1\,\kpc$ yields distributions skewed toward cored values.
This reveals the exact opposite asymmetry from $\sigmalos(R)$, i.e. below \rhalf, \sigmalos best discriminates cuspy profiles, while \kappalos best discriminates cored ones.

The two moments therefore carry complementary information on $\gamma$.
Constraining the inner slope near $\gamma \sim 0$ from $\sigmalos(R)$ alone requires measurements extending to larger radii, where the curves diverge more significantly.
An alternative is to incorporate higher-order moments such as $\kappalos(R)$, which carry complementary information on $\gamma$ in the cored regime.
\gnn naturally exploits both moments through DF mapping, allowing it to constrain the inner slope across the full range of $\gamma$.

Lastly, this insensitivity to core density profiles can also be interpreted as a realization of the prior-volume effect identified by \citet{2026A&A...705A.212B}.
Notably, \citet{2026A&A...705A.212B} shows that second-order Jeans models can appear biased toward cuspy profiles not because cored profiles fit the data more poorly, but because a much larger volume of cuspy configurations achieves comparable likelihood.
This arises because $\sigmalos$ alone provides no direct constraint on the anisotropy parameter $\beta$, so the effective prior volume is dominated by the full range of $(\gamma, \beta)$ combinations consistent with the data.
Since this volume extends preferentially toward low-$\beta$, cuspy solutions, marginalizing over $\beta$ biases inference toward cusps, even when cored models fit equally well.
Incorporating $\kappalos$ breaks this degeneracy by directly constraining $\beta$, thus disfavoring the low-$\beta$ portion of the prior volume responsible for the cuspy bias.
We refer readers to Section 4.3 of \citet{2026A&A...705A.212B} for a more detailed discussion.

\medskip

In summary, the mock results validate the updated \gnn framework under realistic measurement uncertainties and selection effects, and give insights into the key performance differences between \gnn and second-order Jeans modeling.
\gnn recovers cored density profiles with less bias than Jeans modeling, while both methods perform comparably well on cuspy inner profiles.
Additionally, \gnn more accurately extrapolates the outer density profile across all mock datasets, regardless of inner slope.
Importantly, these results highlight that the LOSV dispersion profile $\sigmalos(R)$ is an incomplete summary of the underlying DM density structure: (1) sampling noise and measurement uncertainties can bias Jeans inferences even when the binned dispersion appears well-fit, and (2) even in the absence of such noise, the mass--anisotropy degeneracy allows fundamentally different density and anisotropy profiles to produce nearly identical dispersion profiles.
Higher-order moments such as $\kappalos(R)$ carry complementary information that mitigates both failure modes and serve as an important diagnostic of the inferred posterior.
By operating on the full DF, \gnn naturally exploits this higher-order information, which is the key driver of its improved performance over the second-order Jeans model.

\section{Results on Observational Data}
\label{section:result}

We now present our results on the observational data of Draco and \bootes.
We show the inferred density and anisotropy profiles (Section~\ref{section:result_profiles}), inner densities (Section~\ref{section:result_inner_density}), and $J$- and $D$-factors (Section~\ref{section:result_DJ}).
For Draco, we analyze the MMT \citepalias{Walker23} and DESI \citepalias{Ding25} datasets; for \bootes, we analyze the \sfivecomb and \sfive datasets from \citetalias{Sandford26}.

Table~\ref{tab:posterior} summarizes the \gnn posteriors for all four datasets.
For the inner DM density slope $\gamma$, we adopt a uniform prior $\gamma \in [-1, 2]$ throughout the main text, and report results with the restricted prior $\gamma \in [0, 2]$ in Appendix~\ref{app:results_gm0} for direct comparison with literature analyses that exclude cored profiles by construction.

\begin{table*}
\centering
\caption{Posterior summaries (median with $1\sigma$ uncertainties) for the \gnn inference on Draco and \bootes.
The top and middle sections show results without perspective rotation correction, under the full prior $\gamma \in [-1, 2]$ and the restricted prior $\gamma \in [0, 2]$ used for comparison with literature analyses (see Appendix~\ref{app:results_gm0}), respectively.
The bottom section shows results with the perspective rotation correction applied, under the full $\gamma$ prior; see Appendix~\ref{app:rotation_corr} for the corresponding analysis.
}
\setlength{\tabcolsep}{2pt}
\renewcommand{\arraystretch}{1.5}
\small
\begin{tabular}{llcccccccccc}
\hline
Galaxy & Dataset & $\beta_0$ & $\log r_a$ & $\log \rho_s$ & $\log r_s$ & $\alpha$ & $\beta$ & $\gamma$ & $\rho_{150}$ & $\log J$ & $\log D$ \\
& & & $\kpc$ & $\modot\,\kpc^{-3}$ & $\kpc$ & & & & $10^8 \, \modot\, \kpc^{-3}$ & $\mathrm{GeV}^2\,\mathrm{cm}^{-5}$ & $\mathrm{GeV}\,\mathrm{cm}^{-2}$\\
\hline
\multicolumn{12}{c}{\textit{Prior:} $\gamma \in [-1, 2]$, \textit{Without Perspective Rotation Correction}} \\
\hline
Draco   & MMT        & $0.00^{+0.35}_{-0.33}$  & $0.94^{+1.00}_{-0.96}$  & $7.56^{+0.92}_{-1.28}$  & $0.41^{+1.00}_{-0.61}$  & $1.76^{+0.84}_{-0.79}$  & $5.71^{+2.83}_{-2.55}$  & $0.56^{+0.51}_{-0.74}$  & $1.62^{+0.45}_{-0.38}$  & $18.73^{+0.27}_{-0.20}$  & $18.45^{+0.24}_{-0.23}$ \\
        & DESI       & $-0.25^{+0.28}_{-0.19}$ & $1.12^{+0.90}_{-0.95}$  & $7.60^{+0.99}_{-1.40}$  & $0.42^{+1.01}_{-0.63}$  & $1.78^{+0.83}_{-0.79}$  & $5.77^{+2.83}_{-2.58}$  & $0.57^{+0.53}_{-0.86}$  & $1.88^{+0.65}_{-0.50}$  & $18.86^{+0.27}_{-0.20}$  & $18.50^{+0.23}_{-0.20}$ \\
\bootes & \sfive AAT & $0.17^{+0.38}_{-0.44}$  & $0.08^{+1.39}_{-0.83}$  & $6.62^{+1.27}_{-1.85}$  & $0.33^{+1.04}_{-0.72}$  & $1.69^{+0.89}_{-0.83}$  & $6.02^{+2.69}_{-2.65}$  & $0.89^{+0.67}_{-1.03}$  & $0.26^{+0.17}_{-0.11}$  & $17.27^{+0.80}_{-0.49}$  & $17.39^{+0.28}_{-0.26}$ \\
        & Combined   & $0.24^{+0.33}_{-0.44}$  & $-0.03^{+1.55}_{-0.82}$ & $7.16^{+0.88}_{-1.40}$  & $0.39^{+1.05}_{-0.66}$  & $1.81^{+0.80}_{-0.84}$  & $5.79^{+2.77}_{-2.57}$  & $0.41^{+0.64}_{-0.79}$  & $0.36^{+0.15}_{-0.11}$  & $17.58^{+0.37}_{-0.26}$  & $17.76^{+0.24}_{-0.23}$ \\
\hline
\multicolumn{12}{c}{\textit{Prior:} $\gamma \in [0, 2]$, \textit{Without Perspective Rotation Correction}} \\
\hline
Draco   & MMT        & $-0.03^{+0.36}_{-0.31}$ & $0.96^{+0.99}_{-0.96}$  & $7.19^{+0.82}_{-1.12}$  & $0.64^{+0.88}_{-0.68}$  & $1.82^{+0.81}_{-0.82}$  & $5.67^{+2.85}_{-2.58}$  & $0.73^{+0.40}_{-0.44}$  & $1.64^{+0.42}_{-0.35}$  & $18.73^{+0.25}_{-0.20}$  & $18.46^{+0.23}_{-0.22}$ \\
        & DESI       & $-0.26^{+0.28}_{-0.18}$ & $1.12^{+0.90}_{-0.94}$  & $7.16^{+0.85}_{-1.17}$  & $0.69^{+0.86}_{-0.69}$  & $1.82^{+0.81}_{-0.80}$  & $5.74^{+2.84}_{-2.64}$  & $0.77^{+0.40}_{-0.47}$  & $1.90^{+0.59}_{-0.46}$  & $18.86^{+0.27}_{-0.19}$  & $18.52^{+0.22}_{-0.20}$ \\
\bootes & \sfive AAT & $0.18^{+0.38}_{-0.44}$  & $0.12^{+1.37}_{-0.84}$  & $6.16^{+1.22}_{-1.61}$  & $0.52^{+0.94}_{-0.80}$  & $1.72^{+0.87}_{-0.87}$  & $5.93^{+2.75}_{-2.64}$  & $1.09^{+0.53}_{-0.66}$  & $0.26^{+0.16}_{-0.10}$  & $17.31^{+0.94}_{-0.52}$  & $17.37^{+0.27}_{-0.25}$ \\
        & Combined   & $0.22^{+0.33}_{-0.43}$  & $0.05^{+1.54}_{-0.82}$  & $6.63^{+0.80}_{-1.22}$  & $0.66^{+0.90}_{-0.75}$  & $1.88^{+0.76}_{-0.88}$  & $5.74^{+2.82}_{-2.59}$  & $0.69^{+0.49}_{-0.44}$  & $0.36^{+0.13}_{-0.10}$  & $17.54^{+0.32}_{-0.24}$  & $17.73^{+0.20}_{-0.22}$ \\
\hline
\multicolumn{12}{c}{\textit{Prior:} $\gamma \in [-1, 2]$, \textit{With Perspective Rotation Correction}} \\
\hline
Draco & MMT & $0.02^{+0.34}_{-0.33}$ & $0.91^{+1.00}_{-0.96}$ & $7.56^{+0.91}_{-1.30}$ & $0.44^{+1.00}_{-0.63}$ & $1.77^{+0.83}_{-0.79}$ & $5.74^{+2.85}_{-2.58}$ & $0.55^{+0.50}_{-0.72}$ & $1.60^{+0.45}_{-0.38}$ & $18.70^{+0.23}_{-0.20}$ & $18.34^{+0.19}_{-0.18}$ \\
      & DESI & $-0.24^{+0.29}_{-0.19}$ & $1.08^{+0.92}_{-0.90}$ & $7.60^{+0.98}_{-1.35}$ & $0.42^{+1.00}_{-0.62}$ & $1.79^{+0.83}_{-0.80}$ & $5.84^{+2.82}_{-2.64}$ & $0.56^{+0.53}_{-0.86}$ & $1.89^{+0.66}_{-0.51}$ & $18.87^{+0.28}_{-0.20}$ & $18.50^{+0.24}_{-0.21}$ \\
\bootes & \sfive AAT & $0.13^{+0.37}_{-0.41}$ & $0.38^{+1.25}_{-0.96}$ & $6.59^{+1.20}_{-1.70}$ & $0.59^{+0.94}_{-0.79}$ & $1.72^{+0.88}_{-0.83}$ & $5.89^{+2.75}_{-2.63}$ & $0.68^{+0.74}_{-0.94}$ & $0.23^{+0.16}_{-0.10}$ & $17.28^{+0.67}_{-0.47}$ & $17.50^{+0.33}_{-0.28}$ \\
        & Combined & $0.25^{+0.31}_{-0.41}$ & $0.22^{+1.39}_{-0.94}$ & $7.35^{+0.75}_{-1.05}$ & $0.69^{+0.87}_{-0.72}$ & $1.82^{+0.80}_{-0.85}$ & $5.78^{+2.80}_{-2.59}$ & $0.11^{+0.59}_{-0.61}$ & $0.30^{+0.13}_{-0.10}$ & $17.68^{+0.37}_{-0.28}$ & $17.82^{+0.21}_{-0.18}$ \\
\hline
\end{tabular}
\label{tab:posterior}
\end{table*}

\subsection{Density profiles}
\label{section:result_profiles}

\begin{figure*}
    \centering
    \includegraphics[width=0.98\linewidth]{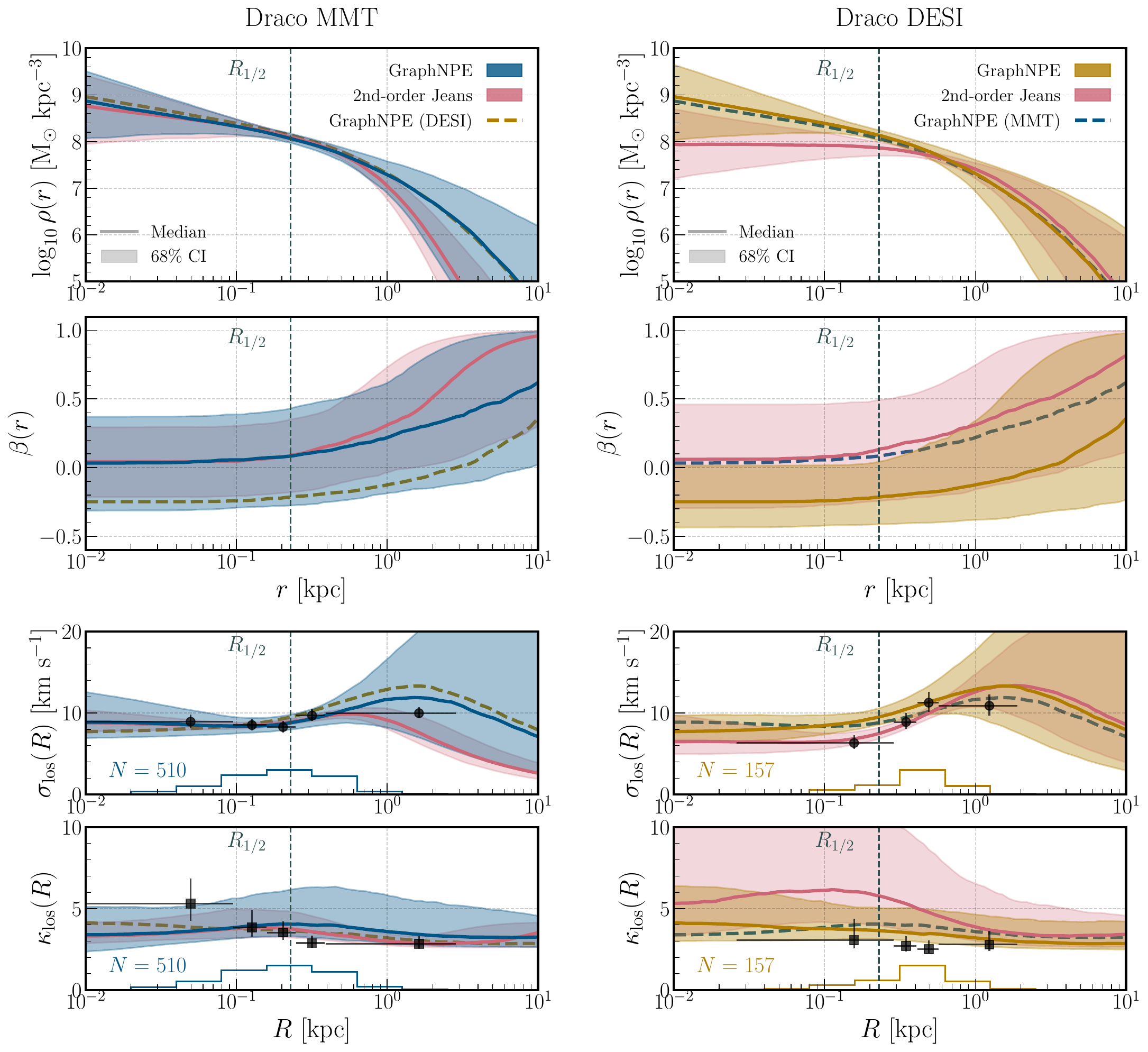}
    \caption{
    From top to bottom: the inferred DM density $\rho(r)$, velocity anisotropy $\beta(r)$, LOSV dispersion $\sigmalos(R)$, and LOSV kurtosis $\kappalos(R)$ profiles for Draco from \gnn and second-order Jeans modeling.
    The left and right columns show results for MMT \citepalias{Walker23} and DESI \citepalias{Ding25}, respectively.
    The solid lines and shaded regions indicate the posterior median and 68\% credible intervals.
    For cross-comparison between dataset, in each dataset panel, the posterior median of \gnn from the other dataset is also shown.    
    The vertical gray dashed line marks the observed half-light radius $\rhalf$ of each galaxy from the LVDB~\citep{2024arXiv241107424P}.
    In the bottom two panels, data points show the binned LOSV dispersion and kurtosis profiles, shown for \textit{visualization only} as neither method uses them in the fit.
    The inset histograms in the LOSV dispersion panels show the observed radial distribution of tracers.
    }
    \label{fig:draco_profiles}
\end{figure*}

We first present the inferred density $\rho(r)$ and velocity anisotropy $\beta(r)$ profiles for Draco and \bootes.
We additionally compare the inferred LOSV dispersion $\sigmalos(R)$ and kurtosis $\kappalos(R)$ profiles with the binned data.
As discussed in Section~\ref{section:mock}, jointly examining both $\sigmalos(R)$ and $\kappalos(R)$ is essential for assessing inference quality, since a model that fits $\sigmalos$ well can still bias in $\rho(r)$ and $\beta(r)$, with the mismatch only revealed through $\kappalos$.

\subsubsection{Draco}

The top two rows of Fig.~\ref{fig:draco_profiles} show the density and anisotropy profiles for MMT (left column) and DESI (right column).
For MMT, the density profiles from \gnn and second-order Jeans are in good agreement, with the posterior medians and 68\% credible intervals consistent across the full radial range probed by the data.
Both methods marginally prefer a cuspy inner profile, with the asymptotic inner slope $\gamma = 0.56^{+0.51}_{-0.74}$ and  $0.41^{+0.54}_{-0.60}$ for \gnn and Jeans, though a cored profile cannot be ruled out at the 68\% level.
For DESI, by contrast, $\rho(r)$ differs significantly between the two methods.
\gnn continues to favor a mildly cuspy inner profile ($\gamma=0.57^{+0.53}_{-0.86}$), consistent with the MMT result, whereas Jeans now favors a cored profile ($\gamma = 0.00^{+0.54}_{-0.48}$) with a low central density.
We defer the discussion of this disagreement to the joint examination of the LOSV dispersion and kurtosis below.

For the velocity anisotropy, \gnn infers a near-isotropic profile ($\beta_0 \sim 0.00$) for MMT and a mildly tangentially anisotropic profile ($\beta_0 \sim -0.25$) for DESI, up to the outermost tracers.
Beyond this radius, the rise toward radial anisotropy is likely driven by the COM anisotropy model (Eq.~\ref{eq:velani_COM}) rather than the data.
By contrast, Jeans infers more strongly radially anisotropic profiles for both datasets.

To better understand the discrepancy between the \gnn and second-order Jeans results, we now examine the $\sigmalos(R)$ and $\kappalos(R)$ profiles in the bottom two rows of Fig.~\ref{fig:draco_profiles}.
For MMT, we find that both models fit the binned \sigmalos data well, while, for DESI, Jeans follows the two innermost bins closely, while \gnn sits above them.
As discussed in Section~\ref{section:vdisp_profile}, these bins also fall $\sim 2\sigma$ below the MMT measurements at comparable radii, which we attribute to undersampling bias from low tracer counts at small projected radii (see Section~\ref{section:mock}).
By fitting these biased bins, Jeans is pulled toward a lower density normalization $\rho_s$ and a cored inner profile, which drives the disagreement with MMT.
This interpretation is further supported by the $\kappalos(R)$ panels.
\gnn matches the binned $\kappalos$ for both, while Jeans matches MMT but overestimates it significantly on DESI.
This mismatch reveals a bias in the Jeans posterior that is not apparent from the $\sigmalos$ fit alone.

Taken together, the consistency of the \gnn density profiles between MMT and DESI, combined with the $\kappalos$ mismatch in the Jeans posterior on DESI, implies that \gnn correctly recovers the underlying DM distribution of Draco, while second-order Jeans is biased by the inner $\sigmalos$ bins of DESI.

However, we note that these noisy bins still affect the \gnn inference of the velocity anisotropy.
The more tangential anisotropy inferred from DESI likely reflects partial compensation for the suppressed inner $\sigmalos$.
Additionally, although \gnn generally recovers $\kappalos(R)$ better than second-order Jeans, the median prediction does not always trace the observed radial variation.
For Draco MMT, the median $\kappalos(R)$ underestimates the innermost bin and fails to capture the rising trend toward small $R$.
For Draco DESI, the inferred profile slightly overestimates all bins, with the observed $\kappalos(R)$ being flatter than the median prediction.
This could reflect a mismatch between the observed DF of Draco and those in the training set, as we discuss in more detail in Section~\ref{section:systematics}.

\subsubsection{Bo\"{o}tes I}

\begin{figure*}
    \centering
    \includegraphics[width=0.98\linewidth]{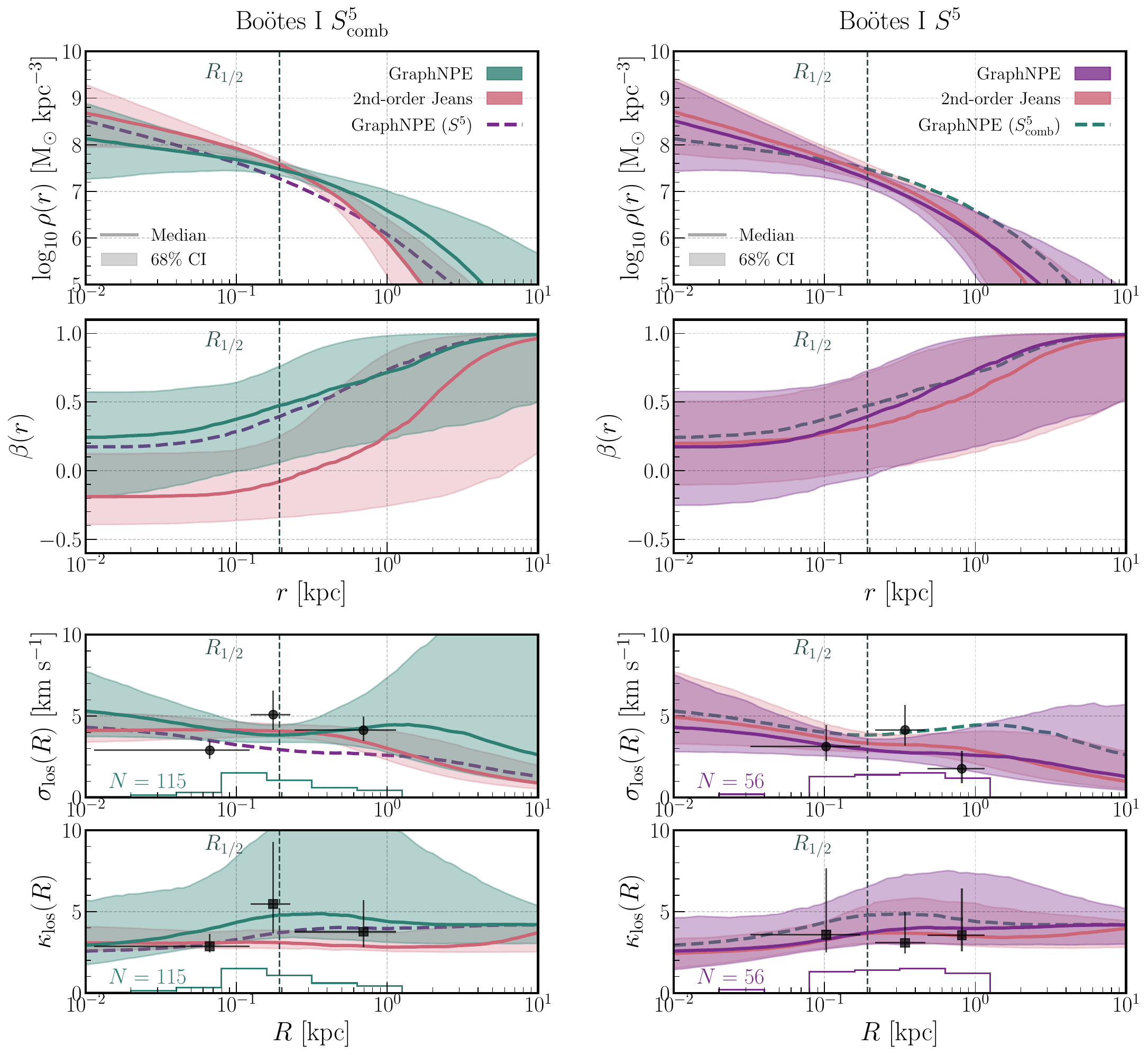}
    \caption{
    Same as Fig.~\ref{fig:draco_profiles}, but for \bootes.
    The left and right columns show results for \sfivecomb and \sfive \citepalias{Sandford26}, respectively.
    }
    \label{fig:bootes_profiles}
\end{figure*}

In Fig.~\ref{fig:bootes_profiles}, the top two rows similarly show the inferred density and velocity anisotropy profiles of \bootes for \sfivecomb (left column) and \sfive (right column).
For \sfive, the density profiles from \gnn and second-order Jeans are nearly indistinguishable and slightly favor a cusp, with $\gamma = 0.89^{+0.67}_{-1.03}$ and $0.92^{+0.58}_{-0.76}$, respectively.
However, due to the small sample size ($N=56$) and the large velocity uncertainties relative to the intrinsic dispersion ($\sigma_\mathrm{los}^\mathrm{global}\sim 3.22 \, \kms$ and $\braket{\Delta_\mathrm{los}}\sim 3.36 \, \kms$), the 68\% credible intervals are broad, and thus a core cannot be fully excluded. 
Both methods recover mildly radial anisotropy near $R_{1/2}$ ($\beta \sim 0.1-0.3$), with credible intervals broad enough to admit isotropic and tangentially biased solutions.

For \sfivecomb, \gnn and Jeans prefer qualitatively different inner density profiles.
\gnn favors a shallower inner slope with $\gamma = 0.41^{+0.64}_{-0.79}$, while Jeans infers a steeper slope with $\gamma = 0.61^{+0.65}_{-0.69}$.
The 68\% credible intervals overlap, so the disagreement reflects a difference in central preference rather than a statistically significant tension.
This disagreement is plausibly driven by the inclusion of archival VLT data in \sfivecomb (see Section~\ref{section:data}), which provides additional kinematic coverage below the half-light radius $R_{1/2}$.
The two methods also yield different anisotropy profiles, with \gnn recovering a stronger radial bias ($\beta \sim 0.3$ at $R_{1/2}$) than Jeans ($\beta \sim 0.05$).

We now examine the LOSV dispersion and kurtosis profiles, shown in the two bottom rows of Fig.~\ref{fig:bootes_profiles}.
For \sfive, both methods are consistent with the binned $\sigmalos$ and $\kappalos$ within their credible intervals.
For \sfivecomb, the two methods fit the binned $\sigmalos$ comparably well despite disagreeing on the density profile.
Both posterior predictions are also consistent with the binned $\kappalos$ within their credible intervals, though \gnn tracks the second radial bin more closely.
However, the $\kappalos$ uncertainty in this bin of \sfivecomb is substantially larger than in the corresponding bin of \sfive, likely reflecting a few stars in the velocity tails introduced by the archival data that inflate the higher moments of the binned distribution.
Therefore, unlike in the Draco DESI case, the $\kappalos$ uncertainties on \bootes are too large to favor either \gnn or Jeans modeling.

Overall, the limited statistical power of current \bootes data prevents a definitive determination of its inner density profile.
Resolving the disagreement between \gnn and second-order Jeans will require larger spectroscopic samples that better constrain both the inner kinematics and the higher-order moments of the velocity distribution.
We return to the interpretation of these results in Section~\ref{section:bootes_core}.

\subsection{Inner DM densities}
\label{section:result_inner_density}

\begin{figure*}
    \centering
    \includegraphics[width=0.95\linewidth]{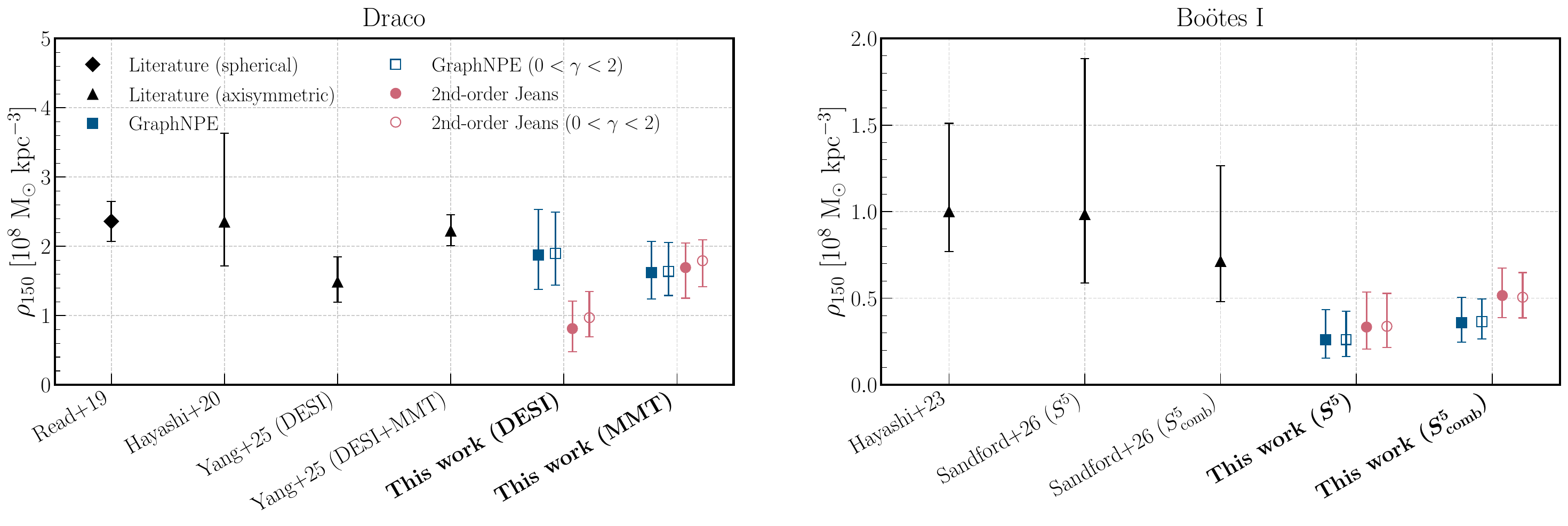}
    \caption{
    The DM density at $150 \, \pc$, \rhoinner, for Draco (left) and \bootes (right).
    Colored data points show measurements from this work, with filled and open symbols corresponding to the $\gamma \in [-1, 2]$ and $\gamma \in [0, 2]$ priors, respectively.
    Blue and pink points show results from \gnn and second-order Jeans modeling, respectively.
    Black data points show literature values, with diamonds denoting spherical Jeans modeling and triangles denoting non-spherical Jeans modeling.
    For Draco, literature values are from \citet{2018MNRAS.481..860R}, \citet{Hayashi2020}, and \citet{Yang25}.
    For \bootes, literature values are from \citet{Hayashi23} and \citetalias{Sandford26}.
    }
    \label{fig:rho_inner}
\end{figure*}

We now present measurements of the inner DM density, quantified as the DM density at $150 \, \pc$, denoted as $\rhoinner \equiv \rho(150 \, \pc)$.
\cite{2019MNRAS.484.1401R} demonstrate that \rhoinner is less model- and prior-dependent than the inner density slope $\gamma$ and provides a more robust distinction between cored and cuspy profiles for luminous dwarfs.
Additionally, \cite{2019MNRAS.490..231K} find an anti-correlation between \rhoinner and the pericentric distances $r_\mathrm{peri}$ for bright Milky Way dwarf spheroidals, and suggest that this anti-correlation can be used as test of self-interacting DM. 

Fig.~\ref{fig:rho_inner} shows our measurements of \rhoinner, including both \gnn and second-order Jeans modeling, compared to past literature values. 
For each method, we show the median and 68\% credible intervals as colored data points, with filled and open symbols corresponding to the $\gamma \in [-1, 2]$ and $\gamma \in [0, 2]$ priors, respectively.\footnote{Since the prior on $\gamma$ is uniform, posteriors for $\gamma \in [0, 2]$ is obtained by simply excluding posterior samples with $\gamma < 0$.}
We show the $\gamma \in [0, 2]$ results for ease of comparison with the literature, most of which adopt a $\gamma \geq 0$ prior.
However, we do not expect the results to differ significantly, since \rhoinner is by design insensitive to the choice of the DM parameters and their priors \citep{2019MNRAS.484.1401R}.
Literature values are shown as black data points. 

In Fig.~\ref{fig:rho_inner}, literature measurements based on spherical Jeans modeling are shown as black diamonds \citep{2018MNRAS.481..860R}, and those based on axisymmetric Jeans modeling as black triangles \citep{Hayashi2020, Hayashi23, Yang25, Sandford26}.
The axisymmetric studies also assume a constant velocity anisotropy, defined as
\begin{equation}
    \beta_z = 1 - \frac{\sigma_z^2}{\sigma_R^2},
\end{equation}
where $\sigma_z$ and $\sigma_R$ are the velocity dispersions in the cylindrical $z$ and $R$ coordinates \citep{2008MNRAS.390...71C}.
We note that in the limit of spherical symmetry, a constant $\beta_z$ does \textit{not} reduce to a constant $\beta$.
The anisotropy assumption between these works and ours are fundamentally different, and thus direct comparison of the inferred density profiles across modeling frameworks should be made with care.
Lastly, \citet{Hayashi2020}, \citet{Hayashi23}, and \citetalias{Sandford26} assume a triaxial halo with an additional axis ratio parameter $Q$, which is typically under-constrained, resulting in substantially larger uncertainties in the inferred density profiles.

\subsubsection{Draco}

We compare our \rhoinner measurements of Draco to \cite{2018MNRAS.481..860R}, \cite{Hayashi2020}, and \cite{Yang25} (hereafter \citetalias{Yang25}).
The first two use MMT/Hectochelle samples from \citet{2015MNRAS.448.2717W}, while \citetalias{Yang25} reports results for the DESI dataset \citepalias{Ding25} and a combined DESI+MMT sample using the re-analysis of MMT/Hectochelle samples from \citetalias{Walker23}.

Overall, we find that the \gnn \rhoinner measurements are consistent between the two datasets, with the MMT and DESI datasets yielding $1.62^{+0.45}_{-0.38}$ and $1.88^{+0.65}_{-0.50} \times 10^8 \, \modot \, \mathrm{kpc}^{-3}$, respectively.
The MMT $\rho_{150}$ measurement has smaller uncertainties than the DESI value, as expected from the higher tracer count and lower LOSV uncertainties of MMT, but also from the better central completeness of the MMT sample, which provides stronger kinematic constraints in the inner region where $\rho_{150}$ is most sensitive.
By contrast, second-order Jeans modeling yields $\rhoinner = 1.69^{+0.36}_{-0.44} \times 10^8 \, \modot \, \mathrm{kpc}^{-3}$ for MMT, consistent with \gnn, but a notably lower value of $0.81^{+0.40}_{-0.34} \times 10^8 \, \modot \, \mathrm{kpc}^{-3}$ for DESI.
We attribute this difference to the lower inferred density normalization $\rho_s$ from Jeans modeling on DESI.
As discussed in Section~\ref{section:result_profiles}, the failure of Jeans to recover the binned kurtosis profile $\kappalos(R)$ on DESI provides independent evidence that this lower normalization reflects an inference bias rather than a genuine difference between the datasets.

Both \gnn and second-order Jeans modeling yield median \rhoinner values lower than those reported by \citet{2018MNRAS.481..860R}, \citet{Hayashi2020}, and the \citetalias{Yang25} DESI+MMT measurement, which span $2.2-2.4 \times 10^8 \, \modot \, \mathrm{kpc}^{-3}$, although all measurements remain consistent within $1\sigma$.
Our \gnn measurement agrees more closely with the DESI result of \citetalias{Yang25} ($1.48^{+0.36}_{-0.29} \times 10^8 \, \modot \, \mathrm{kpc}^{-3}$).
This agreement is likely coincidental, however, since their DESI value is itself lower than their DESI+MMT value ($2.22^{+0.24}_{-0.21} \times 10^8 \, \modot \, \mathrm{kpc}^{-3}$).
Because the Jeans model in \citetalias{Yang25} also relies solely on the velocity dispersion, this discrepancy suggests that their DESI inference may be affected by the same undersampling bias as with our DESI Jeans analysis.

As a separate note, we find our Jeans measurements of \rhoinner for both MMT and DESI data to be lower than those from \citetalias{Yang25}, although the values are still consistent within $1\sigma$.
We attribute this offset to differences in modeling choice and sample selection.
\citetalias{Yang25} solves the axisymmetric Jeans equations assuming a free but constant velocity anisotropy $\beta_z$, which does not reduce to a constant $\beta$ in the limit of spherical symmetry, and also a radial cut at $2.5 \, \rhalf$ to their kinematic sample.
Additional differences in the stellar population model (e.g., \citetalias{Yang25} adopts two populations with independent kinematics) and membership selection may also contribute, though we do not expect these effects to be significant.

\subsubsection{Bo\"{o}tes I}

For \bootes, we compare our results with measurements from \citetalias{Sandford26} and from \citet{Hayashi23}, the latter using the VLT data of \citet{Jenkins21}.
Since \citetalias{Sandford26} do not report values of \rhoinner directly, we compute them from the posterior chains provided by the authors.

The \rhoinner measurements from \gnn and Jeans modeling agree within their uncertainties on each individual dataset, in contrast to the offset seen in Draco.
For \sfive, \gnn and Jeans report values of $0.26^{+0.17}_{-0.11}$ and $0.33^{+0.20}_{-0.13}\,\times\,10^8\,\modot\,\kpc^{-3}$, respectively, while for \sfivecomb the corresponding values are $0.36^{+0.15}_{-0.11}$ and $0.52^{+0.16}_{-0.14}\,\times\,10^8\,\modot\,\kpc^{-3}$.
The lower \rhoinner from \gnn is consistent with the more cored inner profile inferred in Section~\ref{section:result_profiles}.
Both methods also recover a smaller \rhoinner from \sfive than from \sfivecomb, which may reflect the lower velocity dispersion of \sfive (see Section~\ref{section:data}).

Compared to literature values, our analysis yields substantially lower \rhoinner than \citet{Hayashi23} and \citetalias{Sandford26}, whose values lie in the range $0.75-1.0\,\times\,10^8\,\modot\,\kpc^{-3}$.
We further note that \citetalias{Sandford26} report a higher median \rhoinner for \sfive than for \sfivecomb, with their inferred $\rho(r)$ appearing more cuspy for \sfive than for \sfivecomb (their Figs.~10 and C3).
This is opposite to the trend in our analysis, where \sfivecomb yields the more cored profile, and is in tension with the lower velocity dispersion of \sfive relative to \sfivecomb.

Several methodological differences may account for these discrepancies.
\citet{Hayashi23} and \citetalias{Sandford26} both employ axisymmetric Jeans modeling with a constant velocity anisotropy $\beta_z$, whereas we adopt a spherical model with a radially varying anisotropy.
The constant-anisotropy assumption is known to introduce systematic differences in the inferred inner density, and we identify an analogous offset in our Draco analysis (Section~\ref{section:result_inner_density}).
Isolating the source of this offset would require reanalyzing the literature data under identical modeling assumptions, which we defer to future work.
Here, we treat our \bootes results as a complementary constraint informed by higher-order velocity moments through the \gnn framework.

\subsection{DM annihilation and decay}
\label{section:result_DJ}

\begin{figure*}
    \centering
    \includegraphics[width=0.95\linewidth]{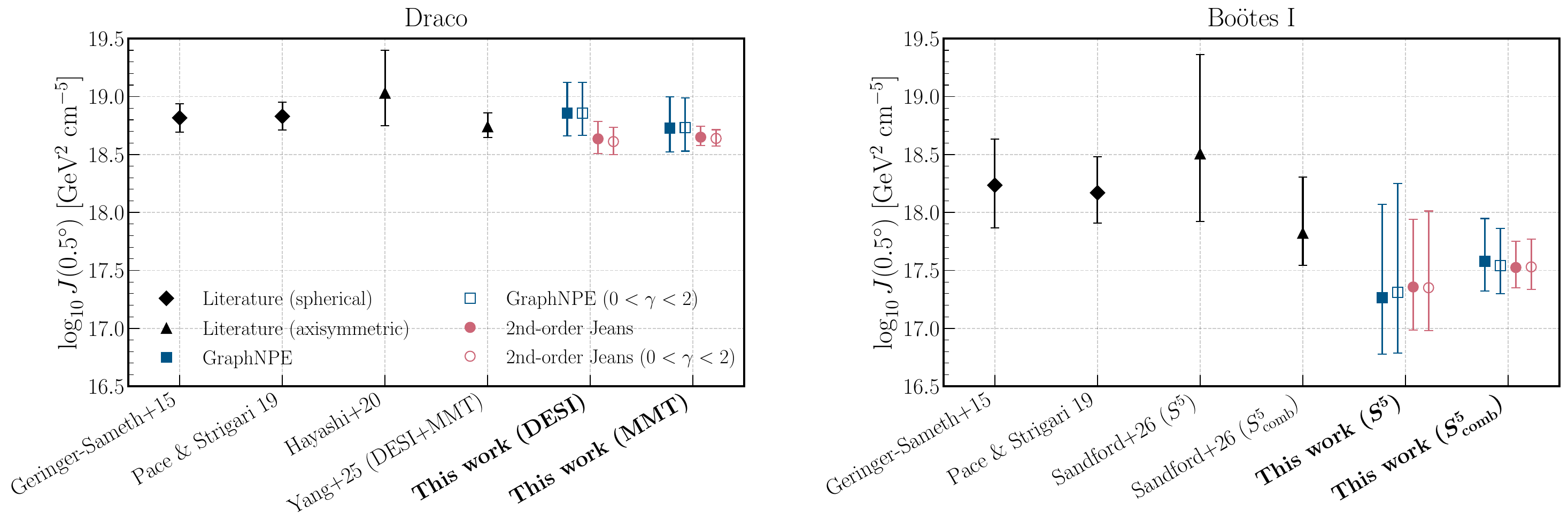}
    \includegraphics[width=0.95\linewidth]{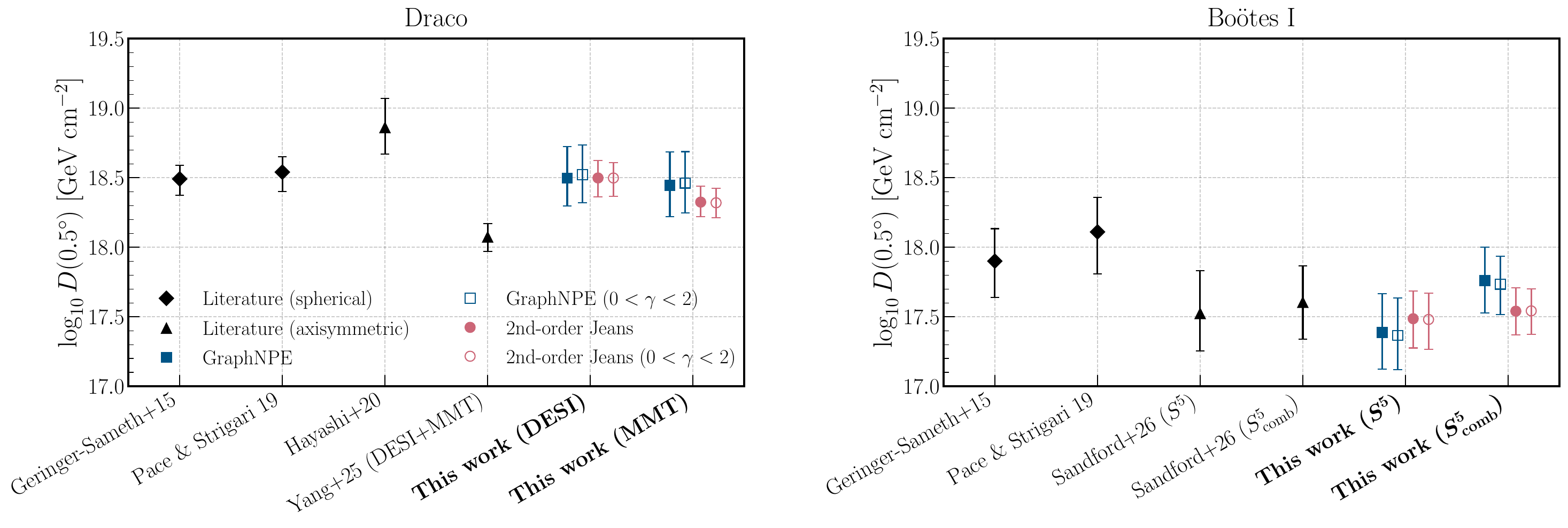}
    \caption{
    The $J$-factors (top row) and $D$-factors (bottom row) for Draco (left column) and \bootes (right column).
    Colored data points show measurements from this work, with filled and open symbols corresponding to the $\gamma \in [-1, 2]$ and $\gamma \in [0, 2]$ priors, respectively.
    Blue and pink points show results from \gnn and second-order Jeans modeling, respectively.
    Black data points show literature values, with diamonds denoting spherical Jeans modeling and triangles denoting non-spherical Jeans modeling.
    For Draco, literature values are from \citet{GS2015}, \citet{Hayashi2020}, and \citetalias{Yang25}.
    For \bootes, literature values are from \citet{GS2015} and \citetalias{Sandford26}.
    }
    \label{fig:DJ_factor}
\end{figure*}

Fig.~\ref{fig:DJ_factor} presents the inferred $J$- and $D$-factors and compares them against literature values.
For Draco, we compare with \citet{GS2015}, \citet{Pace2019}, \citet{Hayashi2020}, and \citetalias{Yang25}; for \bootes, with \citet{GS2015}, \citet{Pace2019}, and \citetalias{Sandford26}.
As in the \rhoinner comparison, we denote measurements from spherical and axisymmetric Jeans models with black diamonds and black triangles, respectively, and note that non-spherical analyses \citep{Hayashi2020, Hayashi23, Sandford26} report larger uncertainties.

To compute the $J$ and $D$-factors, we integrate the density profiles $\rho(r)$ within a fixed solid angle of $0.5^\circ$ from the halo center, as is standard in the literature.
Additionally, to exclude contributions from the outer halo regions, which are likely tidally stripped for satellites, we truncate the halo at the observed projected radius of the outermost member star, $R_\mathrm{max}$.
Since datasets of the same galaxy can have different radial extents, we adopt the largest $R_\mathrm{max}$ available for each system, which are $2.86\,\kpc\,(12.5\,\rhalf)$ and $1.15\,\kpc\,(5.98\,\rhalf)$ for Draco and \bootes, respectively. 
Since \citetalias{Sandford26} do not report $J$- or $D$-factors directly, we obtain their posterior samples through private communication and compute the factors following the procedure of \citet{2016MNRAS.461.2914H} for non-spherical halos.
All $J$- and $D$-factors are quoted in units of $\mathrm{GeV}^2\,\mathrm{cm}^{-5}$ and $\mathrm{GeV}\,\mathrm{cm}^{-2}$, respectively.

\subsubsection{Draco}

For Draco, both \gnn and second-order Jeans modeling yield consistent $J$- and $D$-factors across the MMT and DESI datasets.
\gnn infers $\log_{10} J = 18.73^{+0.27}_{-0.20}$ and $18.86^{+0.27}_{-0.20}$, and $\log_{10} D = 18.45^{+0.24}_{-0.23}$ and $18.50^{+0.23}_{-0.20}$, for MMT and DESI respectively.
Excluding $\gamma < 0$ samples has little effect on the inferred factors.
From Fig.~\ref{fig:DJ_factor}, the \gnn uncertainties on both factors are larger than those from Jeans, reflecting the more extended outer halo profiles discussed in Sections~\ref{section:mock_result} and~\ref{section:result_profiles}.

Compared to \citetalias{Yang25}, our $D$-factor estimates from both methods are systematically higher.
We attribute this primarily to differences in radial coverage. \citetalias{Yang25} restrict their kinematic sample to stars within $2.5\,\rhalf$, truncating their halo at a considerably smaller radius and reducing the integrated mass along the line of sight to which the $D$-factor is most sensitive.
The $J$-factors are broadly consistent within $1\sigma$, as they are less sensitive to the outer truncation radius.

The \gnn $J$- and $D$-factors are in closer agreement with \citet{GS2015, Pace2019, Hayashi2020}, all of which use MMT data from \citet{2015MNRAS.448.2717W} and have comparable spatial coverage.
Our values are marginally smaller than those of \citet{Hayashi2020}, consistent with their higher inferred \rhoinner (Fig.~\ref{fig:rho_inner}).
\citet{GS2015} adopt a slightly smaller truncation radius, obtained by deprojecting the 2D position of the outermost member star to $1.87^{+0.72}_{-0.32}\,\kpc$, compared to our projected $R_\mathrm{max} = 2.86\,\kpc$.
Despite this difference, our $J$- and $D$-factors remain in good agreement with theirs.

\subsubsection{Bo\"{o}tes I}

For \bootes, the choice of $\gamma$ prior again has little effect on the inferred $J$- and $D$-factors from either method, as seen in Fig.~\ref{fig:DJ_factor}.
\gnn recovers $\log_{10} J = 17.27^{+0.80}_{-0.49}$ and $17.58^{+0.37}_{-0.26}$, and $\log_{10} D = 17.39^{+0.28}_{-0.26}$ and $17.76^{+0.24}_{-0.23}$, on \sfive and \sfivecomb, respectively, in agreement with the corresponding Jeans values.

Compared to \citetalias{Sandford26}, our $D$-factors are in good agreement, while our $J$-factors are systematically lower by $\sim 0.5\,\mathrm{dex}$ for both \sfive and \sfivecomb.
Since $J \propto \rho^2$ is more sensitive to the inner density than $D$, this offset is consistent with the lower \rhoinner we infer relative to \citetalias{Sandford26} (Section~\ref{section:result_inner_density}), and reflects the methodological differences discussed there (i.e. axisymmetric versus spherical Jeans modeling, constant versus radially varying anisotropy).

\citet{GS2015} and \citet{Pace2019} report substantially higher values than both our analysis and \citetalias{Sandford26}, by $\sim 0.5-1.0\,\mathrm{dex}$ in $J$ and $\sim 0.3-0.4\,\mathrm{dex}$ in $D$.
We attribute this offset to a combination of differences in data, profile parametrization, and prior choices.
Both studies use the VLT/FLAMES samples of \citet{Koposov11}, which contain few tracers beyond the half-light radius (Table~\ref{tab:datasets}), so the inferred density profile is primarily constrained by the innermost tracers and the outer profile is model-driven.

The two studies differ from each other and from our work in their prior and profile choices in ways that are difficult to disentangle.
\citet{GS2015} adopts a generalized Zhao profile similar to ours (Eq.~\ref{eq:zhao}), but do not impose $r_s > a$, i.e. they do not require the DM scale radius to exceed the stellar Plummer radius.
This allows small-$r_s$ solutions that tend to produce substantially larger $J$-factors (see Fig.~2 of \citealt{Pace2019}).
They also apply a cosmological cut to their posterior that rejects halo profiles requiring extremely rare peaks in the primordial density field (their Section~6.4), effectively excluding $\rho_s > 10^9\,\modot\,\kpc^{-3}$ and thus lowering their inferred factors.
On the other hand, \citet{Pace2019} imposes $r_s > a$ similarly to our work, but assume a pure NFW profile, which forces a cuspy inner slope.
Furthermore, they truncate the NFW profile at the tidal radius, which is substantially larger than the truncation radius used in either our analysis or \citet{GS2015}, thus increasing their inferred $J$-factors.
That the two studies nonetheless report broadly consistent values suggests these effects partially cancel, and points to the shared shallow kinematic coverage of \citet{Koposov11} as the dominant driver of the offset relative to our analysis.
Given the substantial scatter in literature values for \bootes and the methodological differences across studies, we caution against drawing strong conclusions from any direct numerical comparison.

\section{Hints of a core in Bo\"{o}tes I}
\label{section:bootes_core}

\subsection{Inner DM Densities and Profile Shapes in Context}

In this section, we interpret our inferred inner densities and slopes in the context of galaxy formation and cosmology.
Section~\ref{sec:rho_inner} places the \gnn inferences in the $\rhoinner-\mvir$ relation of \cite{2019MNRAS.484.1401R}, and Section~\ref{sec:inner_profile} compares the inferred inner slopes against the simulation predictions of \cite{2016MNRAS.456.3542T, 2020MNRAS.497.2393L}.

\subsubsection{The \rhoinner--\mvir relation} 
\label{sec:rho_inner}

\begin{figure*}
    \centering
    \includegraphics[width=0.90\linewidth]{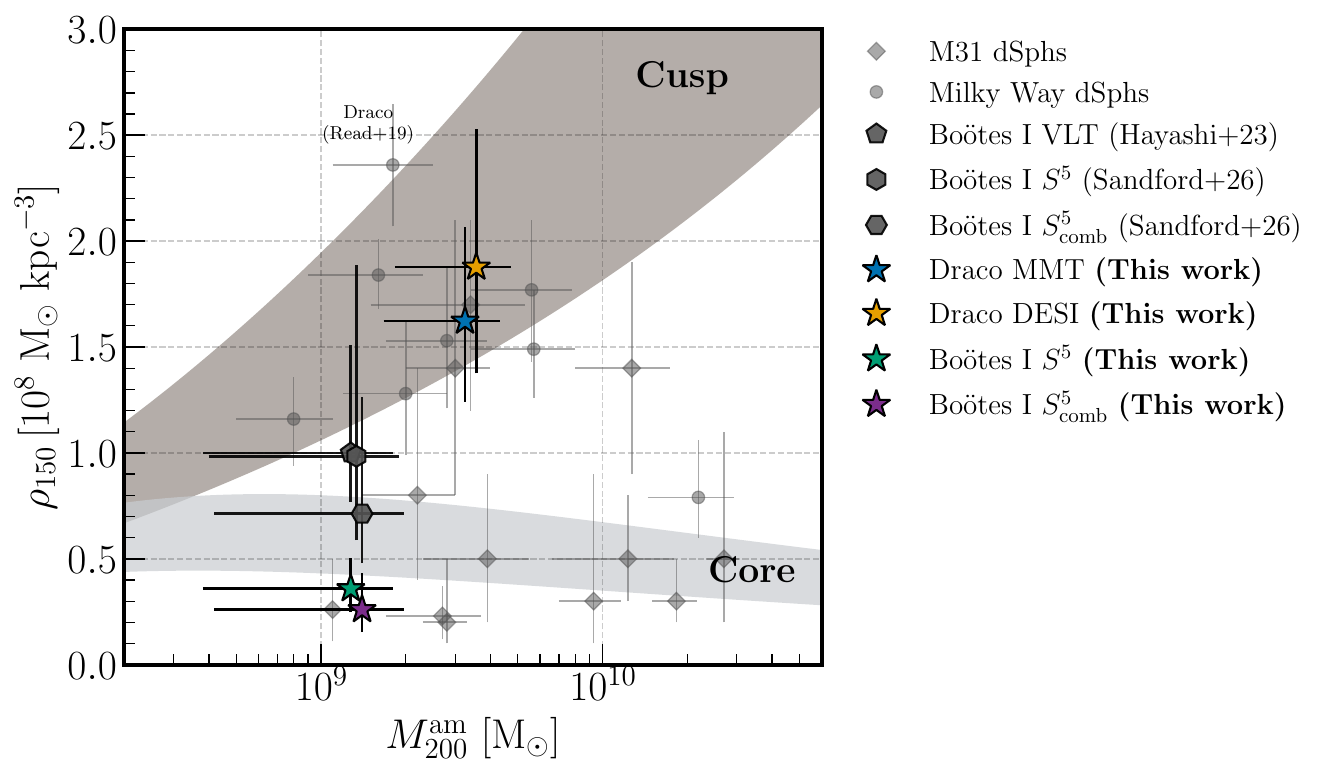}
    \caption{
    The inner DM density at $150\,\pc$ as a function of pre-infall halo mass \mvir, estimated by abundance matching (AM). 
    For better visualization, we apply a small offset to \mviram of the same galaxy.  
    Literature values are compiled from the following sources: Milky Way dwarf spheroidals (gray circles) are from \citet{2019MNRAS.484.1401R}, with satellite \mvir from the \citet{2019MNRAS.487.5799R} AM; M31 dwarf spheroidals (gray diamonds) are from \citet{2021MNRAS.505.5686C, 2023MNRAS.521.3527C, 2025MNRAS.540.1701P, 2026arXiv260600221P}.
    For reference, the Draco measurement of \citet{2019MNRAS.487.5799R} is indicated by a text label.
    The cusp and core bands show the predicted $\rho(150\,\pc)$ as a function of \mvir for NFW and cored profiles, respectively, adopting the \citet{2014MNRAS.441.3359D} mass-concentration relation; the band widths correspond to its $1\sigma$ scatter in concentration.
    }
    \label{fig:rho150_m200}
\end{figure*}

\citet{2019MNRAS.484.1401R} show analytically that cuspy and cored halos occupy distinct regions in the $\rhoinner-\mvir$ plane, where \mvir denotes the \textit{pre-infall} virial mass of the halo.\footnote{\citet{2019MNRAS.484.1401R} adopt the pre-infall mass because the present-day \mvir is not well-defined for satellites that have likely undergone significant tidal stripping.}
Assuming the mass-concentration relation of \citet{2014MNRAS.441.3359D}, they find that cuspy NFW halos above $\mvir \gtrsim 10^8\,\modot$ exhibit a \rhoinner that rises with \mvir, whereas cored halos remain nearly flat with mass.

In order to compare our results with the theoretical predictions of \cite{2019MNRAS.484.1401R}, we first need to define the masses for Draco and \bootes.
We estimate the pre-infall \mvir for Draco and \bootes via abundance matching (AM).
Starting from the $V$-band luminosities $M_V$ of \citet{2018ApJ...860...66M} and adopting a stellar mass-to-light ratio of $\Upsilon_\star = 1.6 \, \modot/L_\odot$ \citep{2008MNRAS.390.1453W}, we obtain stellar masses of $\log_{10}(M_\star/\modot) = 5.67^{+0.02}_{-0.02}$ for Draco and $4.54^{+0.10}_{-0.10}$ for \bootes.
Applying the stellar-to-halo mass relation of \citet{2021ApJ...923...35M} then yields $\log_{10}(\mviram/\modot) = 9.53^{+0.17}_{-0.17}$ for Draco and $9.13^{+0.23}_{-0.23}$ for \bootes.\footnote{The superscript am in \mviram refer to \mvir obtained by abundance matching.}
The reported uncertainties propagate the $M_V$ errors together with the best-fit intrinsic scatter of the stellar-to-halo mass relation in \citet{2021ApJ...923...35M}.

Near the assumed stellar mass of \bootes ($\sim 10^{4.5}\,\modot$), the stellar-to-halo mass relation of \citet{2021ApJ...923...35M} exhibits substantial scatter that is not captured by their best-fit parametrization (their Fig.~2).
Nonetheless, we adopt this relation for consistency with Draco and with previous studies \citep{Hayashi2020, Hayashi23}, while cautioning that the lower bound on \mviram likely extends well below the formal error budget quoted above.

Fig.~\ref{fig:rho150_m200} shows \rhoinner as a function of \mvir for Draco and \bootes.
The shaded bands indicate the expected $\rhoinner-\mvir$ relation and $1\sigma$ scatter for cuspy and cored profiles from \citet{2019MNRAS.484.1401R}.
For context, we overplot literature measurements for Milky Way and M31 dwarf spheroidal galaxies.
We also plot the \bootes measurements from \cite{Hayashi23} and \cite{Sandford26} (using the same value of \mviram) for direct comparison with our \bootes result. 
Following \citet{2019MNRAS.487.5799R}, \citet{2019MNRAS.484.1401R} performs AM using the $\braket{\mathrm{SFR}}-\mvir$ relation, where $\braket{\mathrm{SFR}}$ is the star formation rate averaged over the active star-forming period, since it has lower scatter for satellites compared to $M_\star-\mvir$.
Their inferred \mviram therefore differs systematically from ours.
We indicate the \citet{2019MNRAS.484.1401R} value for Draco with a text label for direct comparison.

Fig.~\ref{fig:rho150_m200} places Draco firmly within the cusp band in both panels, consistent with the cuspy inner profile inferred in Section~\ref{section:result} and in broad agreement with previous measurements in the literature.

On the other hand, \bootes presents a more ambiguous picture.
Previous measurements from \citet{Hayashi23} and \citetalias{Sandford26} yield $\rhoinner \sim 0.7-1.0 \times 10^8\,\modot\,\kpc^{-3}$. 
The \sfivecomb measurement of \citetalias{Sandford26} falls within the core band, while \citet{Hayashi23} and the \sfive measurement of \citetalias{Sandford26} sit just above it, in the intermediate region between the cusp and core bands.
Our measurements fall well below this range and below the core band, identifying \bootes as one of the lowest density galaxies reported for dwarfs of comparable stellar mass.

As discussed in Section~\ref{section:data}, to facilitate comparison with literature results from \citet{Hayashi23, Sandford26}, we do not correct for perspective rotation in our analysis.
Here, we note that our \rhoinner measurements are robust to this choice.
With the perspective rotation correction applied, \gnn recovers $\rhoinner = 0.23^{+0.16}_{-0.10}$ and $0.30^{+0.13}_{-0.10} \times 10^8\,\modot\,\kpc^{-3}$ for \sfive and \sfivecomb respectively (see Table~\ref{tab:posterior} and Appendix~\ref{app:rotation_corr}), shifted slightly downward but well within $1\sigma$ of our fiducial values.
Indeed, applying the correction would further strengthen the cored interpretation discussed above.

Lastly, we note that the shallow inner profiles for \bootes are similar to the low \rhoinner values reported for several M31 satellites \citep{2021MNRAS.505.5686C, 2023MNRAS.521.3527C, 2025MNRAS.540.1701P, 2026arXiv260600221P}, several of which sit within or below the core band in Fig.~\ref{fig:rho150_m200}.

\begin{figure*}
    \centering
    \includegraphics[width=0.90\linewidth]{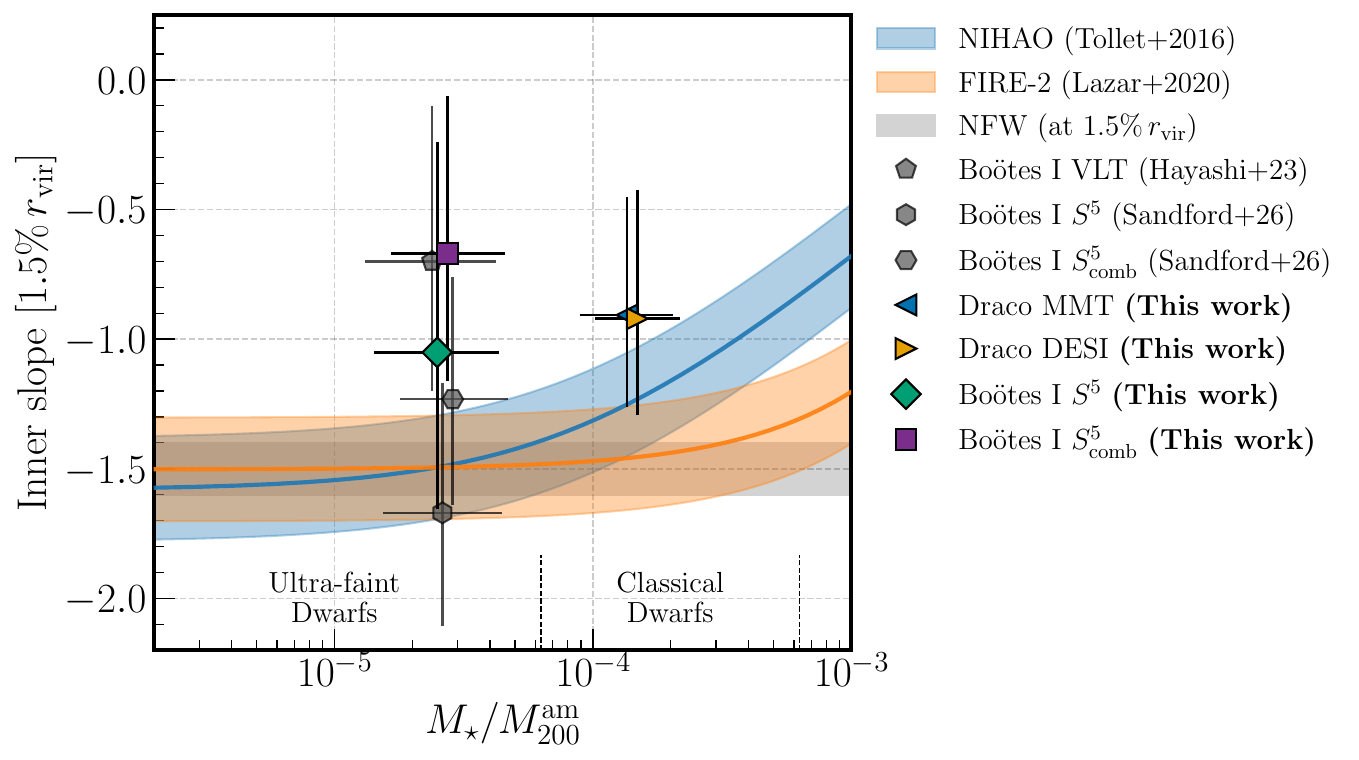}
    \caption{
    Inner slope at $1.5\%$ of the virial radius $r_\mathrm{vir}$ as a function of the stellar mass fraction $M_\star/\mvir$, with the pre-infall halo mass estimated by AM.
    For better visualization, we apply a small offset to $M_\star/\mvir$  of the same galaxy.      
    The blue and orange bands show theoretical predictions from the NIHAO \citep{2016MNRAS.456.3542T} and FIRE-2 \citep{2020MNRAS.497.2393L} hydrodynamic simulations, respectively, while the gray band shows the NFW expectation from DM-only $N$-body simulations \citep{2016MNRAS.456.3542T}.
    }
    \label{fig:alpha_shmr}
\end{figure*}

\subsubsection{The inner slope}
\label{sec:inner_profile}

We next examine the inner slope of the density profile,
\begin{equation} \label{eq:slope}
    \gammain \equiv \left.\frac{d\log\rho}{d\log r}\right|_{0.015\, r_\mathrm{vir}},
\end{equation}
where $r_\mathrm{vir}$ is the virial radius of the halo, defined as the radius at which the average density is equal to 200 times the critical density of the Universe.
We obtain the virial radius by extrapolating the density profile $\rho(r)$.

It is important to differentiate the inner slope in Eq.~\ref{eq:slope} from the \textit{asymptotic} inner slope $\gamma$ of Eq.~\ref{eq:zhao}, where we also note the opposite sign convention as adopted by \citet{2020MNRAS.497.2393L}. 
The $\gammain$ definition from Eq.~\ref{eq:slope} is more accessible in simulations and therefore more straightforward as a comparison tool. 
Indeed, evaluating the slope at a finite radius rather than in the asymptotic limit avoids extrapolating beyond the radial range probed by the stellar tracers and is therefore less sensitive to the choice of parametric profile family, although knowledge of the virial radius already involves some extrapolation and error. 

Fig.~\ref{fig:alpha_shmr} shows \gammain as a function of $M_\star/\mvir$ for Draco and \bootes, with the halo masses estimated via AM as before.
For context, we overlay literature values for \bootes from \citet{Hayashi23} and \citetalias{Sandford26}, with the \citetalias{Sandford26} values calculated from their posterior samples.  
Additionally, predictions from the NIHAO \citep{2016MNRAS.456.3542T} and FIRE-2 \citep{2020MNRAS.497.2393L} cosmological hydrodynamic simulations are shown as bands.
Both predict that stellar feedback can transform a primordial NFW cusp into a shallower inner profile, with the strength of the effect set primarily by $M_\star/\mvir$.
Core formation peaks at $M_\star/\mvir \sim 10^{-3}-10^{-2}$, while at lower stellar mass fractions feedback energy is insufficient to redistribute the inner DM and therefore the inner slope is expected to match NFW.

From Fig.~\ref{fig:alpha_shmr}, we see that Draco sits near or slightly above the NIHAO band depending on \mvir, broadly consistent with the predicted level of feedback-driven core formation at the classical dwarf regimes.

Our \gammain measurements for \bootes are consistent with each other within $1\sigma$ and lie above both simulation bands, with $\gammain = -0.43^{+0.49}_{-0.42}$ and $-0.85^{+0.71}_{-0.65}$ for \sfivecomb and \sfive, respectively.
The \sfivecomb measurement is consistent with a cored profile, while \sfive is only mildly shallower than the NFW expectation.
Notably, our \sfivecomb result agrees with the independent measurement of \citet{Hayashi23} on VLT data \citep{Jenkins21}, despite their higher inferred \rhoinner.
The \citetalias{Sandford26} measurements sit closer to the NFW expectation in this diagnostic.

\subsection{Implications}
\label{section:implication_bootes}

\bootes is an interesting case study given that it is an ultra-faint galaxy, and therefore we expect it to be dominated by DM. 
Additionally, its proximity, given that it is only $d=66.4 \, \kpc$ away \citep{2006ApJ...653L.109D}, makes it a great target for spectroscopic studies as well as indirect detection measurements of DM. 
In this section, we seek to understand its properties as found within our \gnn framework.  

As shown in Fig.~\ref{fig:rho150_m200}, the density at $150\,\pc$ of \bootes is quite low, consistent with (and even below) the theoretical predictions of \citet{2019MNRAS.484.1401R} for a cored profile.
However, the AM estimate of \mvir is uncertain enough that if the true \bootes mass is much lower than the current estimate of $\sim 10^9\,\modot$, the core-cusp distinction based on \rhoinner alone becomes ambiguous, given that the cusp and core bands converge at low \mvir.
As noted above, \mviram likely admits substantially lower values than the formal error budget of the \citet{2021ApJ...923...35M} relation suggests.
This mass uncertainty affects our measurement and those of \citet{Hayashi23} and \citetalias{Sandford26} asymmetrically, because each sits at a different \rhoinner relative to the core and cusp bands.
Our \rhoinner is low enough to remain at or below the core band across the plausible range of \mvir.
Only near $\mvir \sim 3 \times 10^7\,\modot$, where the core and cusp bands merge, does \rhoinner cease to distinguish a core from a cusp.
The two literature values are higher, and toward the low-mass end of the range they can cross into the cusp band.
Our inference of a core is therefore robust to the \mvir uncertainty, while the core or cusp classification of the literature measurements is not.

As shown in Fig.~\ref{fig:alpha_shmr}, the \bootes measurements from the two datasets predict a median larger than the expected NFW $(\gammain \sim -1.5)$ for $M_\star/\mviram \sim 3\times 10^{-5}$. 
The \sfivecomb median lies $\sim 1\sigma$ above the upper edge of the theoretical bands.
We note, however, that theoretical predictions in the ultra-faint regime are highly uncertain.
Although the inner profiles of ultra-faint galaxies are expected to converge towards NFW, the minimum ratio of stellar mass to halo mass above which baryonic feedback starts coring the inner profile might depend on the subgrid physics implemented, and accurate representation of that requires more detailed studies of ultra faints at sufficiently high resolution in simulations.

The two diagnostics presented above paint a consistent picture.
\bootes exhibits both a low inner density and a shallow inner slope, placing it away from the NFW expectation across the plausible \mvir range.
Several physical mechanisms could in principle produce this signature, which we discuss in turn.

First, in principle, supernova-driven core formation \citep{2012MNRAS.421.3464P, 2014Natur.506..171P} is the canonical $\Lambda$CDM mechanism invoked for cored profiles in classical dwarfs. 
Different sub-grid models implement it differently, which leads to the variation in the theoretical curves by~\cite{2016MNRAS.456.3542T,2020MNRAS.497.2393L} in Fig.~\ref{fig:alpha_shmr}. 
However, this is unlikely to apply to \bootes given that this mechanism requires $M_\star/\mvir \gtrsim 5\times 10^{-4}$ and sustained star formation for efficient coring \citep{2016MNRAS.459.2573R}.
From Fig.~\ref{fig:alpha_shmr}, \bootes lies well below the $M_\star/\mvir$ threshold under \mvir estimates from both dynamics and abundance matching.
Furthermore, \citet{2014ApJ...796...91B} and, more recently, \citet{2025ApJ...992..106D} show that \bootes formed essentially all of its stars in a single short burst $\sim 13.3-13.4\,\mathrm{Gyr}$ ago, far too brief to sustain feedback-driven coring \citep{2016MNRAS.459.2573R}.

Second, another possible explanation for the low values of \rhoinner is past tidal stripping and shocking by the Milky Way, which can progressively lower the inner DM density of a satellite once the bound mass has been significantly reduced \citep{2006MNRAS.366..429R, 2010MNRAS.406.1290P}. 
Indeed, the orbital history of \bootes is consistent with disruption, with a small pericenter of $r_\mathrm{peri} = 37.9^{+7.5}_{-6.8}\,\kpc$ \citep{2022ApJ...940..136P}, thus placing it within the strong tidal regime of the Milky Way (though see the discussion in Section~\ref{section:systematics_tides}).
Additional kinematic evidence in \bootes comes from \citetalias{Sandford26}, which reports an intrinsic $4\sigma$ LOSV gradient of $1.2^{+0.4}_{-0.3}\, \mathrm{km\,s^{-1}\,\rhalf^{-1}}$, aligned with the orbital motion after accounting for perspective rotation (Section~\ref{section:vdisp_profile}).
However, recent papers have shown using idealized $N$-body, high-resolution simulations (thus bypassing issues of artificial disruption) that tides alone cannot turn a primordial cusp into a core~\citep{2020MNRAS.491.4591E,2021MNRAS.505...18E}. 
Specifically, tidal stripping removes mass from the halo outskirts and lowers the characteristic density and radius along well-defined tidal tracks \citep{2018MNRAS.474.3043V, 2020MNRAS.491.4591E, 2021MNRAS.505...18E}.
In these scenarios, the inner slope is retained, and the halo approaches an exponentially truncated cusp rather than a core.
For tidal effects to reduce the density such that the tidal radius penetrates the cusp region, the stellar component would itself be heavily disrupted \citep{2010MNRAS.406.1290P}.

Third, \bootes could have formed in an unusually diffuse halo, drawn from the low-concentration end of the \citet{2014MNRAS.441.3359D} distribution.
Such an outcome is statistically unlikely but cannot be ruled out for a single system. 

Fourth, alternative DM models can also produce cores.
For example, self-interacting DM generates cores through gravothermal heating from DM self-scattering, which redistributes energy in the inner halo \citep[e.g.,][]{2000PhRvL..84.3760S,2000ApJ...534L.143B, 2013MNRAS.430...81R, 2013MNRAS.430..105P, 2017MNRAS.472.2945R, 2018PhR...730....1T}.
Within the gravothermal collapse framework, galaxies can occupy different stages of core expansion and core collapse, accommodating a broad population of dwarfs, though collapse can be accelerated in tidal fields, which complicates model predictions \citep{2020PhRvD.101f3009N, Ando2025}.
Fuzzy DM provides a second way to form cores, since quantum pressure supports a ground-state soliton that forms a constant-density core at the center of each halo \citep{2000PhRvL..85.1158H, 2014PhRvL.113z1302S, 2014NatPh..10..496S}.
However, existing constraints on the fuzzy DM particle mass make this an unlikely explanation for \bootes.
The survival of the central star cluster in Eridanus~II implies $m_\mathrm{DM} \gtrsim 10^{-19}\,\mathrm{eV}$ through gravitational heating by soliton oscillations and granule fluctuations \citep[e.g.,][]{2019PhRvL.123e1103M}, while the sizes and stellar kinematics of Segue~1 and Segue~2 tighten this to $m_\mathrm{DM} \gtrsim 3\times10^{-19}\,\mathrm{eV}$ \citep{2022PhRvD.106f3517D}.
At these masses, the de Broglie scale $\lambda \sim \hbar / (m_\mathrm{DM}\,\sigma_\mathrm{DM}) \sim \mathcal{O}(1 \, \mathrm{pc})$\footnote{Assuming that the velocity dispersion of DM particles is similar to that of the observed tracers.} for a \bootes-mass halo, far below the radii we constrain.
In addition, \citet{2023A&A...676A..63B} find that the observed core sizes and masses in Local Group dwarf irregulars are in tension with fuzzy DM subhalo abundances and the predicted core scaling relations.

We also note that the density profile beyond $\rhalf$ may provide additional insights into the interpretation of \bootes's density profiles.
Both baryonic feedback and self-interacting DM gravothermal heating produce cores by redistributing energy within the inner halo, and the resulting density suppression is therefore largely confined to $r \lesssim \rhalf$ \citep{2005MNRAS.356..107R, 2012MNRAS.421.3464P, 2014Natur.506..171P, 2016MNRAS.459.2573R}.
A system shaped by either mechanism alone would therefore be expected to exhibit a low $\rho(150\,\pc)$ alongside an outer density consistent with an NFW cusp.
In this work, the density at $300\,\pc$, which lies beyond $\rhalf \sim 200\,\pc$ for \bootes, is also found to be low (see Appendix~\ref{app:rho300}), which may instead favor processes that suppress the density at all radii, such as a low-concentration halo or tidal stripping.
Nonetheless, the shallow $\gammain$ inferred across the plausible \mvir range (Fig.~\ref{fig:alpha_shmr}) provides independent evidence that the inner profile of \bootes is shallower than expectations from NFW, irrespective of its physical origin.

Taken together, the low $\rhoinner$ and the shallow $\gammain$ paint a consistent picture of an unusually low-density halo, whose inner profile is shallower than NFW, but whose physical origin remains ambiguous.
The degeneracy among these scenarios, whether alternative DM, tidal stripping, or a low-concentration halo, possibly in combination, cannot be resolved from the kinematics of \bootes alone.
Distinguishing them will require both deeper spectroscopic data for \bootes and a broader statistical sample of dwarf galaxies spanning a range of orbital histories, star formation durations, and halo masses, so that the relative contributions of these mechanisms can be disentangled at the population level.

\section{Discussion}
\label{section:discussion}

\subsection{Draco DESI: a case for higher-order moments}
\label{section:discussion_draco}

A key result of Section~\ref{section:result} is the disagreement between \gnn and second-order Jeans modeling on the Draco DESI dataset.
While the two methods fit the binned $\sigmalos(R)$ profile comparably well, the Jeans model infers a substantially lower \rhoinner on DESI than on MMT.
\gnn, by contrast, recovers consistent density profiles across the two datasets.
The two methods nonetheless fit $\sigma_{\rm los}(R)$ comparably well, so the dispersion alone provides no indication that one inference is biased.

The kurtosis profile $\kappalos(R)$ breaks this apparent tie.
Fig.~\ref{fig:draco_profiles} shows that \gnn reproduces $\kappalos(R)$ on both datasets, whereas the second-order Jeans model overestimates it on DESI.
Since neither method fits $\kappalos(R)$ directly, its recovery serves as an independent diagnostic of the inferred posterior.
The Jeans mismatch on DESI thus provides evidence that the low \rhoinner reflects an inference bias rather than a real difference between the two Draco samples.
This is consistent with our mock results of Section~\ref{section:mock}, where sampling noise in $\sigmalos(R)$ can drive second-order Jeans toward biased posteriors that nonetheless fit the dispersion well.

A similar pattern can be seen in \citetalias{Yang25}, which analyzes the same DESI dataset using the second-order Jeans model from \citet{2013MNRAS.436.2598W, 2016MNRAS.462.4001Z, 2016MNRAS.463.1117Z}.
Their DESI inference yields a significantly lower \rhoinner than their DESI+MMT value, despite the two samples probing the same galaxy.
We interpret this offset as a manifestation of the same DESI-specific bias seen in our Jeans analysis, since both methods rely on $\sigmalos(R)$ alone.

More broadly, the Draco DESI case highlights a limitation of dispersion-based inference that is not specific to any one implementation.
A model that fits $\sigmalos(R)$ well cannot, on its own, be taken as evidence that the underlying density and anisotropy profiles are correctly recovered.
Multiple combinations of mass and velocity anisotropy can produce nearly identical dispersion profiles \citep{2002MNRAS.333..697L, 2013MNRAS.432.3361R, 2017MNRAS.471.4541R}.
Additionally, as shown in Section~\ref{section:los_disp_v_gamma}, $\sigmalos(R)$ is intrinsically less sensitive to variations in the inner slope $\gamma$ in the cored regime than in the cuspy regime, while $\kappalos(R)$ exhibits the opposite asymmetry.
The two moments therefore carry complementary information on both the inner slope and the velocity anisotropy, and jointly break degeneracies that $\sigmalos$ alone cannot.

As current and upcoming surveys \citep[e.g.][]{2014PASJ...66R...1T, 2019Msngr.175....3D} expand the spectroscopic samples available for Milky Way dwarfs, we argue that incorporating $\kappalos(R)$, either explicitly through higher-order Jeans equations \citep[e.g.][]{2013MNRAS.432.3361R, 2017MNRAS.471.4541R, 2026A&A...705A.212B} or implicitly through DF-based methods \citep{2002MNRAS.330..778W, 2018MNRAS.480..927P, 2021MNRAS.501..978R, 2025A&A...700A..77P, 2026arXiv260424855P}, including \gnn, will be essential to fully exploit the statistical power of these datasets.
At the very least, $\kappalos(R)$ should be reported alongside $\sigmalos(R)$ as an independent posterior diagnostic.

\subsection{Systematics}
\label{section:systematics}

\subsubsection{Tidal disruption}
\label{section:systematics_tides}

We now discuss the modeling systematics that can affect our inference results.
As with Jeans modeling, \gnn assumes dynamical equilibrium, which can break down for systems undergoing significant tidal disruption.
For example, \citet{2022ApJ...941..108W} shows that systems experiencing strong tidal effects can have contraction motions, and that dynamical modeling may underestimate the inner density profile.
However, tidal disruption does not necessarily bias dynamical inferences, provided that contamination from unbound stars remains negligible \citep{2018MNRAS.481..860R, 2020MNRAS.498..144G, 2024MNRAS.535.1015D, gnn2}.

Assuming a flat Milky Way rotation curve and circular orbits for both Draco and \bootes, the tidal radius is
\begin{equation}
    r_t = r_\mathrm{peri} \left(\frac{M_\mathrm{sat}}{2 M_\mathrm{MW} (< r_\mathrm{peri})}\right)^{1/3},
\end{equation}
where $r_\mathrm{peri}$, $M_\mathrm{sat}$, and $M_\mathrm{MW}$ are the pericentric distance, satellite mass, and enclosed Milky Way mass, respectively.
We adopt pericenters of $r_\mathrm{peri} = 58\,\kpc$ for Draco and $r_\mathrm{peri} = 38\,\kpc$ for \bootes from \cite{2022ApJ...940..136P}, and estimate the enclosed Milky Way mass using the \texttt{MWPotential2014} potential from \cite{Bovy2015}.
Following \citetalias{Sandford26}, we estimate $M_\mathrm{sat}$ by integrating the inferred mass profiles by \gnn up to the scale radius of the DM halo.

For Draco, the median tidal radius is $r_t \sim 5.0\,\kpc$, with a conservative $16$th-percentile lower bound of $2.5\,\kpc$, consistent across the MMT and DESI datasets.
For \bootes, the median tidal radii are $\sim 3.1$ and $1.9\,\kpc$ for \sfivecomb and \sfive respectively, consistent with the median value $2.4 \, \kpc$ reported by  \citetalias{Sandford26}, with $16$th-percentile lower bounds of $1.4$ and $1.0\,\kpc$.
At these lower bounds, only the two outermost stars in Draco MMT fall outside the inferred tidal radius, while both \bootes samples remain within the \sfivecomb lower bound of $1.4\,\kpc$.
Adopting the smaller \sfive lower bound of $1.0\,\kpc$ as a limiting case, only the outermost star in both \bootes samples would also fall outside.
In all cases, the bulk of each sample sits well within the median tidal radius, thus suggesting that contamination from unbound stars is unlikely to bias our inference.

Although the tidal radii estimated above suggest that severe contamination is unlikely, we note that \citetalias{Sandford26} detect an intrinsic velocity gradient of $1.2^{+0.4}_{-0.3}\, \mathrm{km \, s^{-1} \, \rhalf^{-1}}$ in \bootes along its orbital motion, after correcting for perspective rotation.
While this gradient may be indicative of tidal disruption, \citetalias{Sandford26} conclude that its origin remains ambiguous: dynamical simulations of a \bootes-like system predict a tidally induced velocity gradient only beyond $\sim 7\,\rhalf$, outside the extent of current spectroscopic samples, and the gradient direction is consistent with both prolate rotation and a past dwarf-dwarf merger \citep[e.g.,][]{Frebel2016}.
We therefore treat this as a caveat on the dynamical equilibrium assumption for \bootes and note that any bias introduced by disequilibrium would affect both \gnn and Jeans modeling, leaving their relative comparison valid.

Independently of the equilibrium assumption, we note that neither \gnn nor Jeans modeling explicitly truncates the DM halo at the tidal radius.
As a result, inferred density profiles beyond this radius reflect an extrapolation of the untruncated Zhao profile, rather than the true tidally stripped distribution, and should be interpreted with caution.
While recent studies have explored explicit modeling of tidal truncation \citep{2021MNRAS.505.5686C}, \citet{gnn2} demonstrates that the un-truncated generalized NFW profile can nonetheless be extrapolated to give an unbiased estimate of global halo parameters, including $V_\mathrm{max}$ and the peak virial mass $M_\mathrm{200}^\mathrm{peak}$, albeit with large uncertainties.

Finally, we note that the equilibrium assumption in \gnn is encoded in the training simulations rather than imposed analytically as in Jeans modeling, and can in principle be relaxed by incorporating non-equilibrium systems into the training set.
Similar SBI approaches have been applied to non-equilibrium systems such as stellar streams \citep[e.g.][]{2025ApJ...987...96M, 2025arXiv251207960N}, though these works are currently limited to mock data, in part due to the significantly larger parameter space required to describe non-equilibrium configurations.
Extending \gnn to non-equilibrium dwarf galaxies is therefore a natural but non-trivial direction for future work.

\subsubsection{Tracer density profile}

To account for the spectroscopic selection function, we estimate the true tracer mass density profile from the photometric light profile.
We model the light profile as a Plummer sphere and condition \gnn on the Plummer scale radius $r_\star$.
This requires two key assumptions.
First, as discussed in Section~\ref{section:method}, we assume that the light profile traces the mass profile, which requires an approximately constant mass-to-light ratio with projected radius.
This is a reasonable approximation for old, metal-poor stellar populations, which holds for both Draco \citep{Kirby2011, Ding25} and \bootes \citep{Longeard22, Sandford26}.

Second, and more critically, we assume the stellar tracer density follows a Plummer profile, with structural parameters fixed to the values reported in the LVDB~\citep{2024arXiv241107424P}.
While the Plummer profile is a standard and widely adopted choice for dwarf galaxies, deviations from this form, particularly in the inner and outer slopes, can introduce systematic biases in mass estimators \citep{2026arXiv260211273S}.
Observationally, \citet{2020ApJ...892...27M} find evidence for steepened outer stellar density profiles in several MW dwarfs (Fornax, Leo~I, Leo~II, and Reticulum~II), though no significant central cusps in the stellar distribution are detected.

We note that most of the literature results we compare against \citep{GS2015, Hayashi2020, Hayashi23, Sandford26} also adopt a Plummer profile, with the exception of \citet{2018MNRAS.481..860R}, which models the stellar distribution with a three-component Plummer profile, and \citet{Yang25}, which models the light profile non-parametrically. 
As such, systematic biases arising from the Plummer assumption are expected to affect our results and the literature comparisons in a broadly similar manner.
While it is straightforward to condition \gnn\ on the parameters of a different parametric profile (e.g.\ S\'ersic, King), this would require training separate models for each assumed form.
A more flexible but considerably more difficult alternative is to feed a non-parametric light profile directly to \gnn.
We leave both extensions for future work.

\subsubsection{Halo and anisotropy profile assumptions}
\label{section:systematics_halo}

In Sections~\ref{section:result_inner_density} and \ref{section:result_DJ}, we show that the inferred \rhoinner, $J$-, and $D$-factors of both Draco and \bootes can vary significantly across analyses that differ in their assumed halo geometry and treatment of the velocity anisotropy.
For \bootes, our spherical inferences yield \rhoinner systematically lower than those of \citet{Hayashi23} and \citetalias{Sandford26}, both of which use axisymmetric Jeans modeling with a non-spherical halo and a constant velocity anisotropy $\beta_z$.
As noted in Section~\ref{section:result_inner_density}, $\beta_z$ is defined in cylindrical coordinates and does \textit{not} reduce to the spherical anisotropy $\beta$ even in the spherical halo limit \citep{2008MNRAS.390...71C}.
The Draco analysis shows a similar pattern: our results sit below \citet{Hayashi2020}, which adopts the same framework as \citet{Hayashi23, Sandford26, Yang25}, which uses axisymmetric Jeans with a spherical halo but still constant $\beta_z$.
These offsets cannot be unambiguously attributed to any single modeling choice without re-running each analysis under matched assumptions, but they illustrate the magnitude of the systematic uncertainty introduced by these choices.

A natural way to disentangle these effects is to validate such assumptions against realistic mock data, similar to the approach of \citet{2021MNRAS.501..978R}.
\citet{2020MNRAS.498..144G} applied \textsc{GravSphere} \citep{2017MNRAS.471.4541R} to APOSTLE satellites, and \citet{2026MNRAS.547ag279T} extended similar tests to tailored simulations of several Milky Way dwarfs, both identifying tidal effects and outer-profile parametrization as the dominant sources of bias in spherical inferences.
However, these tests do not systematically vary the underlying halo geometry, nor do they explore the impact of different anisotropy parametrizations.
A controlled comparison applying spherical, axisymmetric, and non-spherical Jeans inferences, each under different anisotropy assumptions, to a common set of mocks spanning a range of true geometries would clarify how these methodological choices interact with the underlying physics and help interpret the offsets in the present \bootes literature.

For \gnn, an analogous stress test has been performed in \citet{gnn2}, which applies the framework to FIRE-2 satellites \citep{2018MNRAS.480..800H} including tidally disrupting systems, and demonstrates that the inferred density profiles remain unbiased even when the underlying physics departs from the equilibrium and spherical-symmetry assumptions of the training distribution.
These tests are most directly applicable to galaxies in the classical-dwarf regime represented by the FIRE-2 sample, such as Draco.
\bootes lies below the mass and size range probed by current FIRE-2 dwarfs, so the effect of resolution-dependent physics on \gnn inferences in the ultra-faint regime remains untested.
Extending these stress tests to higher-resolution cosmological simulations of ultra-faint dwarfs, alongside the methodological comparisons proposed above, is a natural direction for future work.

\subsubsection{Other modeling assumptions and caveats}

We discuss other modeling assumptions and limitations of \gnn.
As described in Section~\ref{section:forward_model}, the training dataset is generated using the \texttt{df::QuasiSphericalCOM} class in \textsc{Agama}, which assumes a specific functional form for the DF.
This restricts the class of DFs that \gnn is trained on, and real dwarf galaxies may have DFs that fall outside this family.

A related limitation arises from the COM DF itself.
As detailed in Appendix~\ref{app:kurtosis}, the COM family imposes a specific relationship between the fourth- and second-order anisotropy structure of the DF, which need not hold for general DFs.
Since \gnn is trained exclusively on COM realizations, it implicitly assumes this relationship for all inferences, including the kurtosis profiles computed in this work.
Systems whose fourth-order velocity anisotropy deviates from the COM family may therefore not be accurately captured by \gnn.
We regard both of these as caveats on the current implementation and defer a systematic exploration of more general DFs to future work.

It is also worth noting that higher-order moments such as $\kappalos$ are more sensitive to outliers than $\sigmalos$.
Since the kurtosis is determined by the tails of the LOSV distribution, a small fraction of foreground or background contaminants could potentially bias the inferred $\kappalos(R)$, thus affecting the \gnn inference (and DF-based methods) more severely than $\sigmalos(R)$ alone.
In practice, differences in membership determination, selection cuts, and data quality across datasets may introduce additional variation in the inferred $\kappalos(R)$ that is difficult to disentangle from physical signal.

As discussed in Section~\ref{section:systematics_halo}, \citet{gnn2} tests \gnn on FIRE-2 dwarf galaxies and finds the inferred density profiles to be robust even under departures from the training assumptions.
Additionally, \citet{gnn2} selects member stars for these mock dwarfs in a way that introduces potential contamination from unbound stars, and finds the model to be largely robust to this effect.
However, these tests are performed on galaxies with realistic DFs but without the uncertainties and selection functions characteristic of real spectroscopic datasets.
Conversely, the mock validation in this work incorporates realistic observational uncertainties and selection functions, but uses in-distribution COM realizations rather than fully realistic DFs.
Thus, a stress test combining both realistic non-equilibrium DFs and realistic observational effects may be worth revisiting in future work.

\section{Summary and Outlook}
\label{section:summary}

The mass density profiles of dwarf spheroidal galaxies are among the most powerful probes of DM physics.
In this work, we extend the \gnn framework of \citet{gnn1, gnn2} to account for measurement uncertainties and spectroscopic selection effects.
Compared to second-order Jeans modeling, which fits the LOSV dispersion, \gnn performs inference using the full DF and thus implicitly exploits higher-order moments of the velocity distribution.

We apply \gnn to the classical dwarf Draco and the ultra-faint dwarf \bootes.
For Draco, we use two independent spectroscopic samples from MMT/Hectochelle \citep{Walker23} and DESI \citep{Ding25}.
For \bootes, we use recent observations from \sfive \citep{Sandford26}, as well as a combined sample that supplements the \sfive observations with archival data from VLT/FLAMES \citep{Jenkins21} and MMT/Hectochelle \citep{Walker23}.

Using controlled mock datasets matched to the observational properties of each galaxy, we benchmark \gnn against second-order Jeans modeling in Section~\ref{section:mock}.
We summarize our findings below:

\begin{itemize}

\item Any method relying solely on the LOSV dispersion profile $\sigmalos(R)$, including second-order Jeans modeling, is biased toward cuspy density profiles, \textit{even in the absence of the mass-anisotropy degeneracy}.
This bias is driven by the intrinsic insensitivity of \sigmalos to the inner slope $\gamma$ below the half-light radius in the cored regime, and by sampling noise in the innermost dispersion bins that can mimic the signature of a cuspy profile (Section~\ref{section:los_disp_v_gamma}).
The LOSV kurtosis profile $\kappalos(R)$ exhibits the opposite asymmetry, preferentially constraining cored profiles, so \sigmalos and \kappalos together resolve the degeneracy.
Thus, any inference framework that exploits \kappalos in addition to \sigmalos will mitigate the same bias, whether through higher-order LOSV moments \citep[e.g.][]{2013MNRAS.432.3361R, 2017MNRAS.471.4541R, 2025ApJ...982..167W, 2026A&A...705A.212B} or through distribution-function-based methods such as \gnn.
More broadly, this cuspy bias can be interpreted a prior-volume effect, whereby the space of $(\gamma,\beta)$ combinations consistent with $\sigmalos(R)$ alone is dominated by cuspy solutions even when cored models fit equally well (see \citealt{2026A&A...705A.212B}).

\item More broadly, our results suggest that $\kappalos(R)$ is a useful posterior diagnostic alongside $\sigmalos(R)$, even for analyses that fit only the dispersion.
A model that matches $\sigmalos(R)$ but fails to reproduce \kappalos may not reliably recover the underlying density profile, since the mass-anisotropy degeneracy allows distinct $(\rho, \beta)$ combinations to produce nearly identical \sigmalos while predicting distinct \kappalos.
This diagnostic can be applied retroactively to existing analyses without rerunning the inference, and we recommend it as a standard sanity check for future dwarf galaxy dynamical studies.

\end{itemize}

In Section~\ref{section:result}, we apply both methods to observational data and summarize our key results below:

\begin{itemize}

\item For Draco, \gnn yields consistent density profiles across the MMT and DESI samples, preferring a cuspy inner profile with $\rhoinner = 1.62^{+0.45}_{-0.38}$ and $1.88^{+0.65}_{-0.50} \times 10^8\,\modot\,\kpc^{-3}$ for MMT and DESI, respectively, in good agreement with past literature from \citet{2018MNRAS.481..860R, Hayashi2020, Yang25}.
The MMT inference is better constrained, reflecting both the higher tracer count and lower LOSV uncertainties of that sample and the better central completeness, which provides stronger kinematic information in the inner region where $\rhoinner$ is most sensitive.
On the other hand, the second-order Jeans model infers a substantially lower $\rhoinner$ on DESI than on MMT.
Both methods fit \sigmalos comparably well, but only \gnn reproduces the observed \kappalos, providing direct evidence that the Jeans inference on DESI is biased rather than reflecting a real difference between the two spectroscopic samples.
A similar offset between dispersion-only DESI and DESI+MMT inferences is seen in \citet{Yang25}, plausibly tracing to the same sampling-noise sensitivity and central incompleteness of the DESI sample.
The Draco DESI case thus provides a real-data illustration of the failure mode predicted by our mock analysis.

\item For \bootes, the limited statistical power of current spectroscopic samples prevents a definitive determination of the asymptotic inner slope $\gamma$.
On the \sfive sample alone, \gnn and second-order Jeans are nearly indistinguishable and both marginally favor a cusp, with credible intervals broad enough to admit a core.
On the larger \sfivecomb sample, \gnn favors a shallower inner slope while Jeans prefers a steeper slope, although the $68\%$ credible intervals overlap.
The binned $\kappalos(R)$ uncertainties on \bootes are too large to discriminate between the two methods as cleanly as in Draco, so the asymptotic slope itself remains an open question.
More robustly, \gnn recovers $\rhoinner = 0.26^{+0.17}_{-0.11}$ and $0.36^{+0.15}_{-0.11} \times 10^8\,\modot\,\kpc^{-3}$ on the \sfive and \sfivecomb samples, respectively, significantly lower than past non-spherical Jeans analyses \citep{Hayashi23, Sandford26}.
This places \bootes among the most diffuse inner halos reported for dwarfs of comparable mass, and provides the first hint for a core in an ultra-faint dwarf.

\end{itemize}

In Section~\ref{section:bootes_core}, we explored the implications of the low $\rhoinner$ of \bootes by placing it in the $\rhoinner-\mvir$ relation of \citet{2019MNRAS.484.1401R} and the $\gammain-M_\star/\mvir$ relation predicted by the NIHAO \citep{2016MNRAS.456.3542T} and FIRE-2 \citep{2020MNRAS.497.2393L} cosmological simulations.
Across both diagnostics, \bootes sits below the cuspy NFW expectation and below the predicted inner slopes from feedback-driven core formation, identifying it as one of the most diffuse inner halos reported for dwarfs of comparable mass.
We discussed four physical interpretations:

\begin{itemize}
    \item First, we disfavor feedback-driven core formation.
    \bootes lies below the $M_\star/\mvir$ threshold required for efficient coring (Fig.~\ref{fig:alpha_shmr}), and past studies have shown that it formed essentially all of its stars in a single short burst at early times \citep{2014ApJ...796...91B, 2025ApJ...992..106D}.
    We note, however, that the \mvir of \bootes is poorly constrained, and at sufficiently low halo masses the $M_\star/\mvir$ threshold for feedback-driven coring may be reached even with a short star formation history.
    Nonetheless, the single burst constraint from \citet{2014ApJ...796...91B} argues against sustained feedback, which is required for efficient coring regardless of halo mass \citep{2016MNRAS.459.2573R}.

    \item Past tidal processing by the Milky Way is plausible but might not be sufficient on its own.
    The small pericenter of \bootes \citep{2022ApJ...940..136P} places it well within the strong tidal regime, and the intrinsic LOSV gradient detected by \citet{Sandford26} is consistent with tidal heating from previous pericentric passages.
    The parallel with the population of tidally processed M31 satellites further supports this picture \citep{2021MNRAS.505.5686C, 2023MNRAS.521.3527C, 2025MNRAS.540.1701P}.
    However, tidal stripping primarily removes mass from the outskirts of a halo, and several studies have shown that it cannot, on its own, transform a primordial cusp into a core \citep{2010MNRAS.406.1290P, 2018MNRAS.474.3043V, 2020MNRAS.491.4591E, 2021MNRAS.505...18E}.
    Tidal effects can lower $\rhoinner$ once the bound mass is substantially reduced, but this requires the inner profile to be already shallow before mass loss begins.

    \item It is possible, though statistically unlikely, that \bootes was formed in an unusually diffuse halo at the low-concentration end of the mass-concentration relation from \citet{2014MNRAS.441.3359D}. 

    \item Alternative DM models predict shallower inner profiles at low halo masses.
    Self-interacting DM generates cores through gravothermal heating from DM self-scattering, with the inner profile evolving over time as the system transitions from core expansion to gravothermal collapse \citep{2000PhRvL..84.3760S, 2013MNRAS.430...81R, 2017MNRAS.472.2945R, 2018PhR...730....1T}.
    Within this framework, the ultra-faint population is expected to span a range of inner densities at fixed halo mass, reflecting different stages of gravothermal evolution.
    Fuzzy DM produces core-like solitons from wave interference on galactic scales \citep{2014NatPh..10..496S, 2014PhRvL.113z1302S}, though prior works \citep[e.g.][]{2019PhRvL.123e1103M, 2022PhRvD.106f3517D, 2023A&A...676A..63B} have placed strong constraints on fuzzy DM and disfavors this interpretation.
\end{itemize}

Distinguishing between these scenarios, and between any of them and tidal processing, will require dynamical mass modeling of a larger population of ultra-faints spanning a range of masses and orbital histories.
The case for population-level ultra-faint dynamics is thus ever more compelling.
\bootes is the brightest and best-sampled ultra-faint dwarf galaxy, yet here we show that its inner slope cannot be unambiguously determined from current data.
Other ultra-faint dwarfs have smaller spectroscopic samples and will provide weaker constraints, which both motivates the upcoming spectroscopic samples, and demands inference frameworks that extract maximum information from sparse kinematics.

Our results demonstrate the power of DF-based inference for extracting DM properties from sparse stellar kinematics in dwarf galaxies.
By operating on the full kinematic DF rather than the LOSV dispersion alone, \gnn naturally exploits higher-order velocity moments, mitigating both sampling-noise and mass-anisotropy biases that affect second-order Jeans modeling.
In a follow-up work, we will apply \gnn to larger samples of dwarf galaxies, both classical and ultra-faint, to systematically test for cores in low-mass halos and place population-level constraints on DM physics.
Further extensions include incorporating proper motions to decisively break the mass-anisotropy degeneracy, and relaxing the spherical-symmetry and equilibrium assumptions to model triaxial and tidally disrupting systems.
Together, these directions position SBI as a versatile and robust framework for the next generation of dwarf galaxy dynamical analyses.

\section*{Acknowledgements}
We thank Eugene Vasiliev, Alex Ji, Wenting Wang, and Tjitske Starkenburg for discussions on mass modeling, and Josh Speagle for discussions on uncertainty modeling in simulation-based inference. 

TN is supported by the CIERA Postdoctoral Fellowship.
LN is supported by the DOE grant DE-SC0024112, and the Sloan Fellowship. This work is also supported by the National Science Foundation under Cooperative Agreement PHY-2019786 (The NSF AI Institute for Artificial Intelligence and Fundamental Interactions, \url{http://iaifi.org/}).
TSL acknowledges financial support from Natural Sciences and Engineering Research Council of Canada (NSERC) through grant RGPIN-2022-04794.
JIR acknowledges support from STFC grants ST/Y002857/1 and ST/Y002865/1.
ABH acknowledges support from the funding received from the European Union through the grant ``UNDARK'' of the Widening participation and spreading excellence programme (project number 101159929) and MICINN through the grant ``DarkMaps'' PID2022-142142NB-I00.
HY is supported by NSFC (12573022, 12595312, 12273021), the National Key R\&D Program of China (2023YFA1605600, 2023YFA1605601), and the Office of Science and Technology, Shanghai Municipal Government (grant Nos. 24DX1400100, ZJ2023-ZD-001).
CAFG was supported by NSF through grants AST-2108230 and AST-2307327; by NASA through grants 80NSSC22K0809, 80NSSC22K1124 and 80NSSC24K1224; by STScI through grant JWST-AR-03252.001-A; and by BSF through grant \#2024262.
TN and CAFG also gratefully acknowledge the support of the NSF-Simons AI-Institute for the Sky (SkAI) via grants NSF AST-2421845 and Simons Foundation MPS-AI-00010513. 
NRS acknowledges financial support from an Arts \& Science Postdoctoral Fellowship at the University of Toronto and from the Natural Sciences and Engineering Research Council of Canada (NSERC) through grants RGPIN-2020-04712 and RGPIN-2022-04794.
KH was supported by JSPS KAKENHI Grant Nos. 25H01553, 24K00669, 26H02044, and 26K07153.

\section*{Data Availability}

The spectroscopic datasets used in this work are publicly available from the following sources: MMT Draco data \citep{Walker23}; MMT DESI data \citep{Ding25}; and \sfive and archival data for Bootes \citep{Sandford26}. 
The \gnn repository is available via \url{https://doi.org/10.5281/zenodo.18854552} \citep{jgnn_zenodo} and \url{https://github.com/trivnguyen/jgnn}.
The trained model weights and an example code for running inference and plotting density profiles is available at \url{https://zenodo.org/records/20558847}.

\section*{Software}
This research makes use of the following packages:
\textsc{Agama}~\citep{2019MNRAS.482.1525V},
\textsc{Matplotlib}~\citep{2007CSE.....9...90H},
\textsc{NumPy}~\citep{harris2020array},
\textsc{PyTorch}~\citep{2019arXiv191201703P}, 
\textsc{PyTorch Geometric}~\citep{2019arXiv190302428F}, 
\textsc{PyTorch Lightning}~\citep{william_falcon_2020_3828935},
\textsc{SciPy}~\citep{2020SciPy-NMeth},
\textsc{zuko}~\citep{2023zndo...7625672R}.

\bibliographystyle{mnras}
\bibliography{bib}

\appendix

\section{Model architecture and Training}
\label{app:training}

We provide additional details on the model architecture and training process.
As described in Section~\ref{section:npe}, \gnn consists of two primary components: (1) a GNN embedding network that reads in the kinematic data and reduces into a low-dimensional summary vector, and (2) a conditional density estimator that models that posterior distribution using this summary vector.

The GNN embedding network consists of five Chebyshev convolutional layers, each with a hidden size of 128.
Each layer performs message passing over the graph, where neighborhood aggregation is carried out via a spectral filter expressed as a truncated expansion in Chebyshev polynomials of order $K=8$ over the graph Laplacian \citep{2016arXiv160609375D}.
This ensures that each node aggregates information strictly from its $K$-hop neighborhood, preserving spatial locality in the graph.
The GNN layers are followed by a global mean pooling layer, which averages node embeddings across the graph to produce a fixed-dimensional, permutation-invariant summary vector.
This summary vector is then passed through a two-layer multi-layer perceptron (MLP) block with a hidden size of 128.
The ReLU activations \citep{Nair2010RectifiedLU} are applied after each GNN and MLP layer except the final one.

As noted in Section~\ref{section:method_new}, to account for the spectroscopic selection effect, the Plummer scale radius $r_\star$ is passed through a two-layer MLP with a hidden size of 128 and a ReLU activation between the layers, producing a 128-dimensional embedding.
This embedding is added element-wise to the output of the MLP block described above, forming the final summary vector passed to the density estimator.

The summary vector is then passed to the density estimator, which consists of a neural spline flow \citep{2019arXiv190604032D} with six transformation layers.
Each transformation layer is conditioned on a two-layer MLP block with a hidden size of 128, 8 spline knots, and Tanh activations.
The flow models the posterior distribution over the seven DM and anisotropy parameters $\bm{\theta}=(\alpha, \beta, \gamma, \rho_s, r_s, \beta_0, r_a)$.

During training, the 6D kinematic data is first passed through a stochastic forward model that projects the data to the 3D observed space, perturbs each observable within its measurement uncertainty, and applies the selection function, before being passed into the embedding network (see Section~\ref{section:method_new}).
This procedure significantly increases sample efficiency, as each training step produces a distinct data realization for the same parameter vector $\boldsymbol{\theta}$, effectively providing an infinite stream of augmented training samples.

We train the embedding network and the density estimator simultaneously.
We experimented with the sequential training procedure of \citet{2025A&A...697A.162L}, where each component is trained separately, but find that joint training yields better performance.
Since we expand the parameter space compared to \citet{gnn1, gnn2}, we increase our training set to $2 \times 10^6$ simulations, split into a $90-10$ training-validation ratio.
The loss is optimized using the AdamW optimizer \citep{adamw2019, kingma2014adam} with a peak learning rate of $5 \times 10^{-4}$ and a weight decay of $0.01$.
We use a cosine annealing learning rate scheduler \citep{2016arXiv160803983L} with a total of $\sim 2.8 \times 10^6$ steps, of which the first $5\%$ constitute a linear warmup, with the remaining steps following cosine decay.
We use a batch size of 64, which over $2.8 \times 10^6$ steps corresponds to roughly $100$ epochs over the training set.
The training takes about 52 hours on a NVIDIA Tesla A100 GPU.

\section{Additional Results}

\subsection{Posterior Calibration Test}
\label{app:coverage}

\begin{figure}
    \centering
    \includegraphics[width=0.85\linewidth]{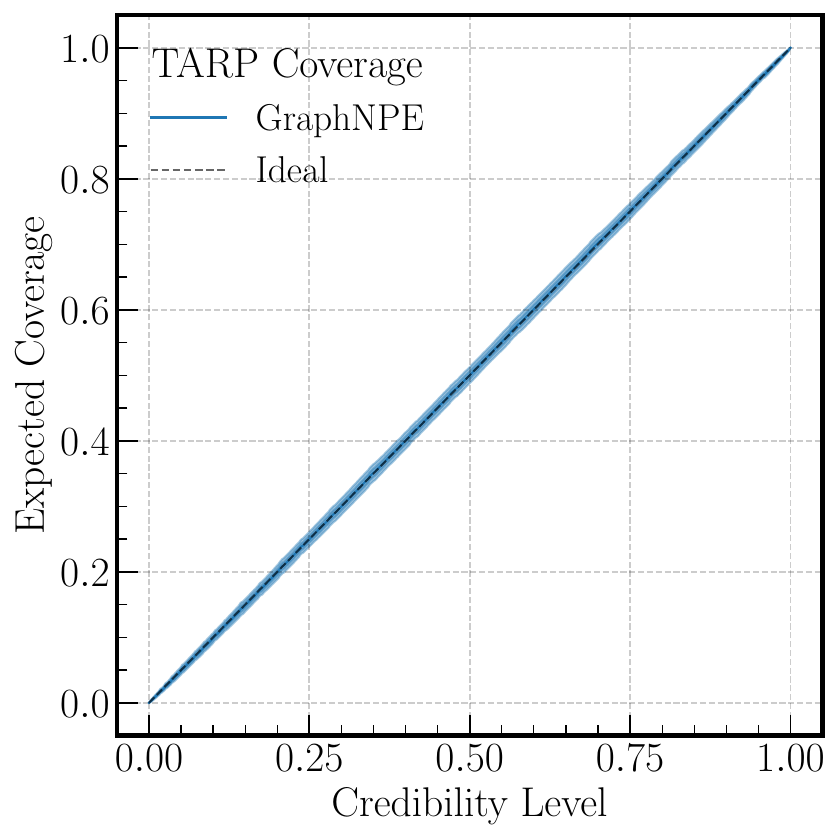}
    \caption{
    The expected coverage probability versus the credibility level of \gnn.
    The mean, $68\%$ and $95\%$ confidence intervals are shown as blue solid line and shaded bands, respectively.
    For comparison, the perfectly calibrated case is shown as black lines.
    }
    \label{fig:tarp}
\end{figure}

Posterior calibration is critical to ensure that the credible intervals returned by the posterior reflect the true frequentist coverage probabilities.
An uncalibrated posterior can lead to systematically over- or underestimated uncertainties, which in turn compromises the scientific conclusions drawn from the inferred DM density profiles.
We assess the calibration of \gnn posteriors using TARP \citep{2023PMLR..20219256L}, which computes the expected coverage probability as a function of the credibility level by drawing random reference points in parameter space and checking whether the true parameters fall within the corresponding highest-posterior-density regions.
Crucially, TARP is both necessary and sufficient for posterior calibration, and remains valid in high-dimensional parameter spaces \citep{2023PMLR..20219256L}.
We apply TARP to $5000$ independent test galaxies, each with $1000$ posterior samples drawn from \gnn.
To estimate the uncertainty on the TARP curve, we perform 100 bootstrap resamplings of the test set and compute the $68\%$ and $95\%$ confidence intervals across the bootstrap samples.
Fig.~\ref{fig:tarp} shows the resulting mean TARP curve and its confidence intervals.
A perfectly calibrated posterior follows the diagonal, while an overconfident or underconfident posterior produces an S-shaped deviation above or below the diagonal, respectively.
We find that \gnn\ provides a near-perfectly calibrated posterior, with the mean TARP curve closely following the diagonal within the $68\%$ confidence interval across all credibility levels.

\subsection{Perspective rotation correction}
\label{app:rotation_corr}

\begin{figure*}
    \centering
    \includegraphics[width=0.98\linewidth]{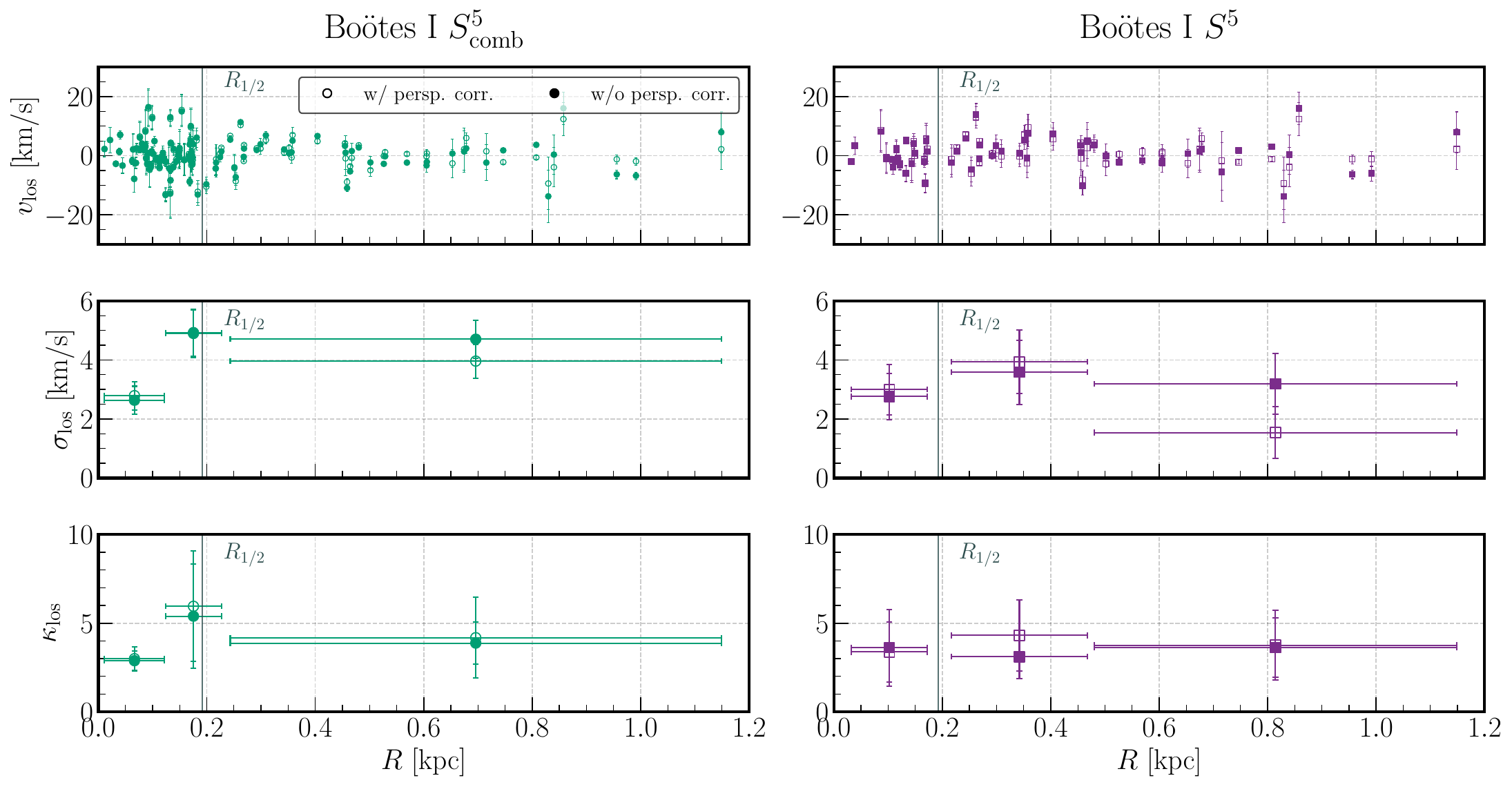}
    \caption{
    The effect of perspective rotation on the LOSV of tracer stars in the \sfivecomb  (left column) and \sfive (right column) samples of \bootes. 
    The top rows shows the LOSV as a function of the projected radii, while the middle and bottom rows show the binned LOSV dispersion and kurtosis profiles, respectively.  
    Data without and with perspective rotation correction are denoted with filled and unfilled markers, respectively.
    Gray vertical lines denote the observed half-light radius $\rhalf$ of \bootes.
    }
    \label{fig:bootes_pcorr}
\end{figure*}

We present analysis results on \bootes without correcting for perspective rotation, an observational effect in which a galaxy's bulk space motion projects onto the LOSV differently across its face, inducing a spurious gradient that can mimic intrinsic rotation \citep{1961MNRAS.122..433F, 2002AJ....124.2639V, 2008ApJ...682L..93K, 2008ApJ...688L..75W, 2020MNRAS.495.3022P}.
As discussed in Section~\ref{section:vdisp_profile}, this effect mainly affects \bootes among our datasets, owing to its larger systemic proper motion ($\mu_{\mathrm{tot}} \sim 1.1\,\mathrm{mas\,yr^{-1}}$).
Because the induced gradient adds a position-dependent component to the LOSV, it inflates the inferred velocity dispersion, with the largest effect at large projected radii.

Fig.~\ref{fig:bootes_pcorr} compares the \sfive and \sfivecomb samples with and without the correction.
From top to bottom, the panels show the LOSV \vlos as a function of projected radius $R$, and the binned \sigmalos and \kappalos profiles.
As expected, stars at larger $R$ are more strongly affected than those near the center.
The dispersion in the outermost bin, however, increases after correction, contrary to the naive expectation.
This is consistent with the aligned LOSV gradient reported by \citetalias{Sandford26}, which they argue is at least partly intrinsic to \bootes.

\begin{figure*}
    \centering
    \includegraphics[width=0.98\linewidth]{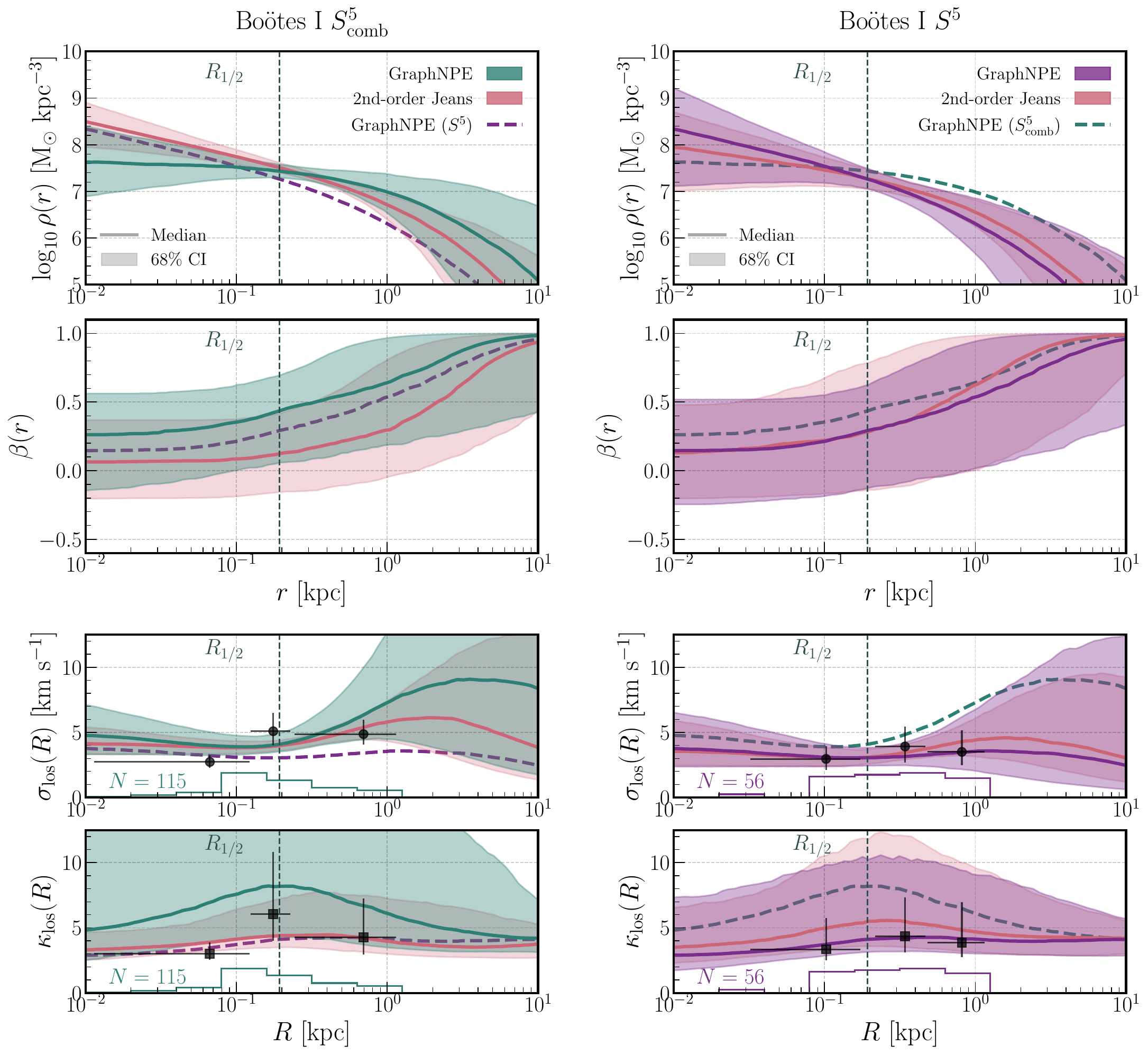}
    \caption{
    From top to bottom: the inferred DM density $\rho(r)$, velocity anisotropy $\beta(r)$, LOSV dispersion $\sigmalos(R)$, and LOSV kurtosis $\kappalos(R)$ profiles for \bootes from \gnn and Jeans modeling.
    The left and right columns show results for the \sfivecomb and \sfive samples \citepalias{Sandford26}, respectively.
    Panels are the same as Fig.~\ref{fig:bootes_profiles}, but with the perspective rotation correction applied.
    }
    \label{fig:bootes_profiles_pcorr}
\end{figure*}

Fig.~\ref{fig:bootes_profiles_pcorr} presents the inferred density, velocity anisotropy, $\sigmalos(R)$, and $\kappalos(R)$ profiles of \bootes after correcting for perspective rotation.
Compared to the fiducial results in Section~\ref{section:result_profiles}, both \gnn and Jeans modeling shift slightly toward cored profiles, while remaining consistent with the uncorrected inferences within their credible intervals.
The \gnn fit to $\kappalos(R)$ on \sfivecomb is also somewhat noisier than in the uncorrected case, which could be due to the residual intrinsic LOSV gradient as discussed above, though the large uncertainty on the middle $\kappalos$ bin makes a definitive interpretation difficult.
The corresponding inner densities are $\rhoinner = 0.23^{+0.16}_{-0.10}$ and $0.30^{+0.13}_{-0.10} \times 10^8\,\modot\,\kpc^{-3}$ for \sfive and \sfivecomb respectively, shifted marginally lower than the uncorrected values reported in Section~\ref{section:result_inner_density}.
Overall, our \bootes conclusions are robust to the perspective rotation correction, with the corrected values reinforcing the cored interpretation discussed in Section~\ref{section:bootes_core}.

\subsection{Isotropic mock datasets}
\label{app:isotropic_mock}

\begin{figure*}
    \centering
    \includegraphics[width=\linewidth]{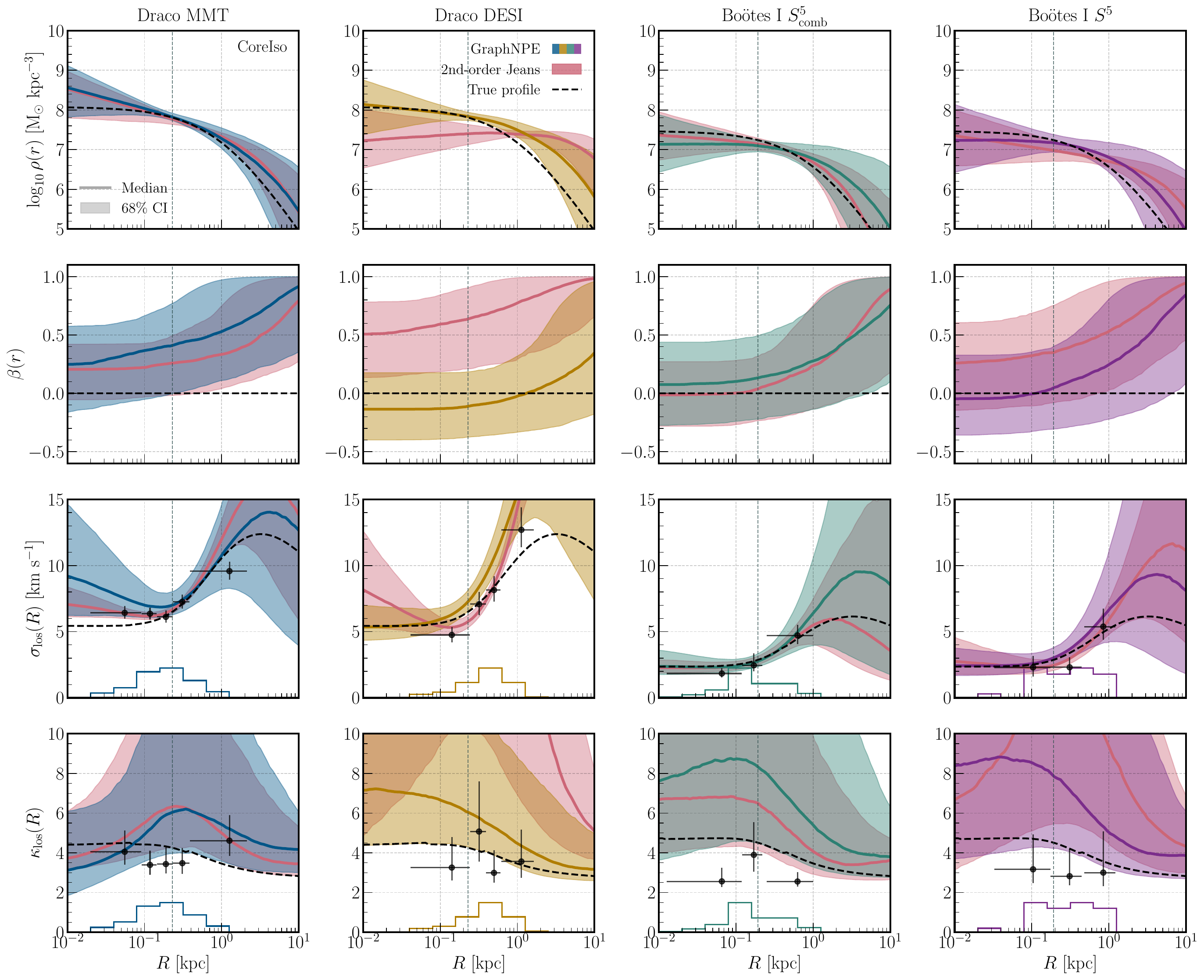}
    \caption{
    Same as Fig.~\ref{fig:coreom_result}, but for mock galaxies with \textsc{CoreIso} parameters.
    }
    \label{fig:coreiso_result}
\end{figure*}
\begin{figure*}
    \centering
    \includegraphics[width=\linewidth]{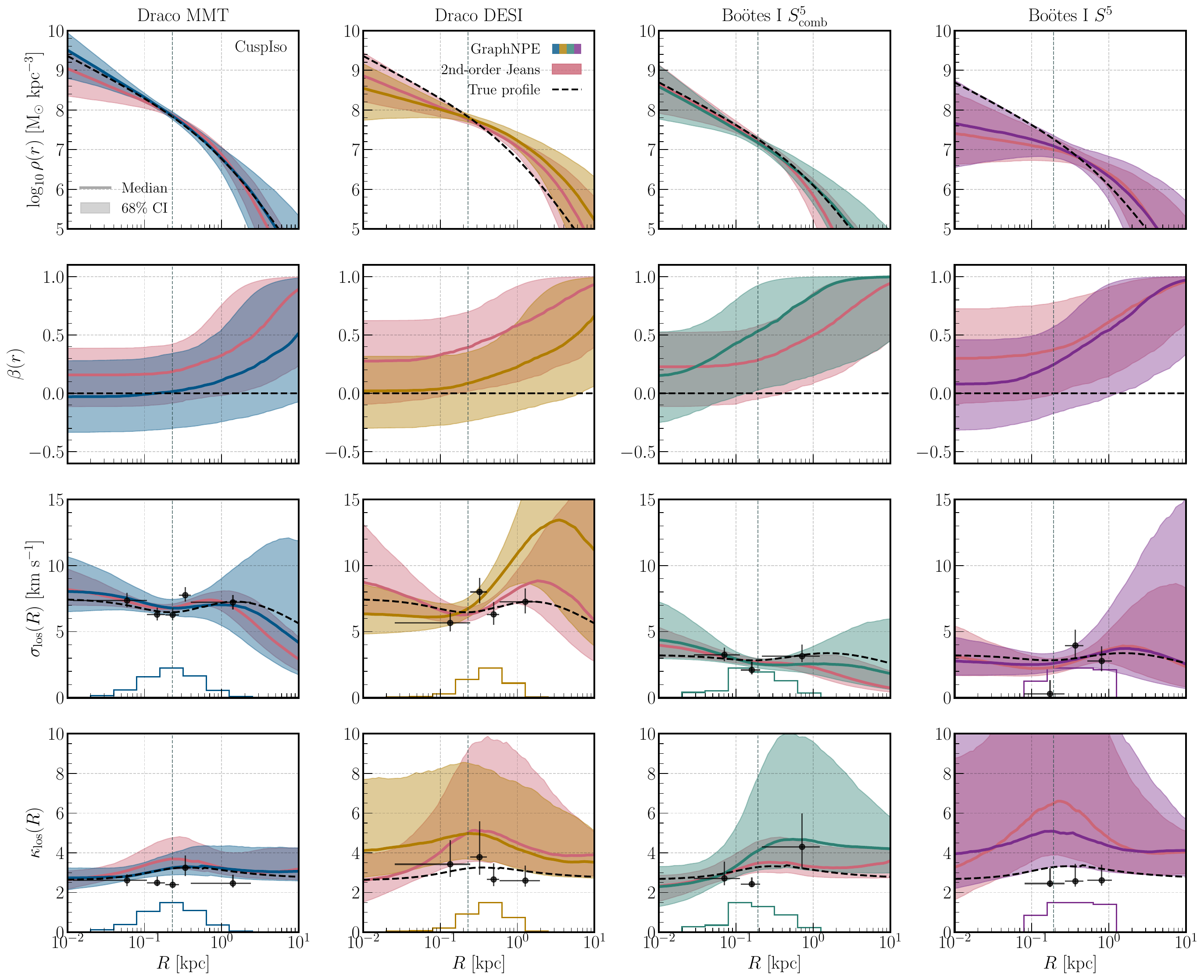}
    \caption{
    Same as Fig.~\ref{fig:coreiso_result}, but for mock galaxies with \textsc{CuspIso} parameters.
    }
    \label{fig:cuspiso_result}
\end{figure*}

We present results on \textsc{CoreIso} and \textsc{CuspIso} mock datasets in Figures~\ref{fig:coreiso_result} and \ref{fig:cuspiso_result}, respectively.
Consistent with the OM results in Section~\ref{section:mock_result}, \gnn successfully recovers the density and anisotropy profiles within the radial range covered by the observed tracers within the 68\% credible intervals.
In particular, \gnn correctly identifies the constant anisotropy despite this configuration not being explicitly represented in the training distribution, as the COM profile requires $r_a \to \infty$ to reproduce the $\beta = 0$ limit.
Beyond the outermost tracer, the inferred $\beta(r)$ naturally rises toward radially biased values as the model reverts to the COM profile in the absence of data constraints.

Compared to the OM cases, the performance of \gnn and Jeans is more comparable for most isotropic mocks.
This may be attributed to the greater sensitivity of $\sigma_\mathrm{los}(R)$ to $\gamma$ in the isotropic case: in the OM case, the radially varying anisotropy profile suppresses this sensitivity, making different inner slopes harder to distinguish from the velocity dispersion alone (see Section~\ref{section:los_disp_v_gamma}).
Despite this, Jeans tends to infer a rising $\beta(r)$ even within the tracer range, suggesting that the mass-anisotropy degeneracy remains unresolved.
A notable example is the Draco DESI \textsc{CoreIso} mock, where Jeans overfits the binned velocity dispersion and infers strongly biased density and anisotropy profiles, while \gnn remains more robust.

\subsection{Observational data with $\gamma > 0$}
\label{app:results_gm0}

\begin{figure*}
    \centering
    \includegraphics[width=0.92\linewidth]{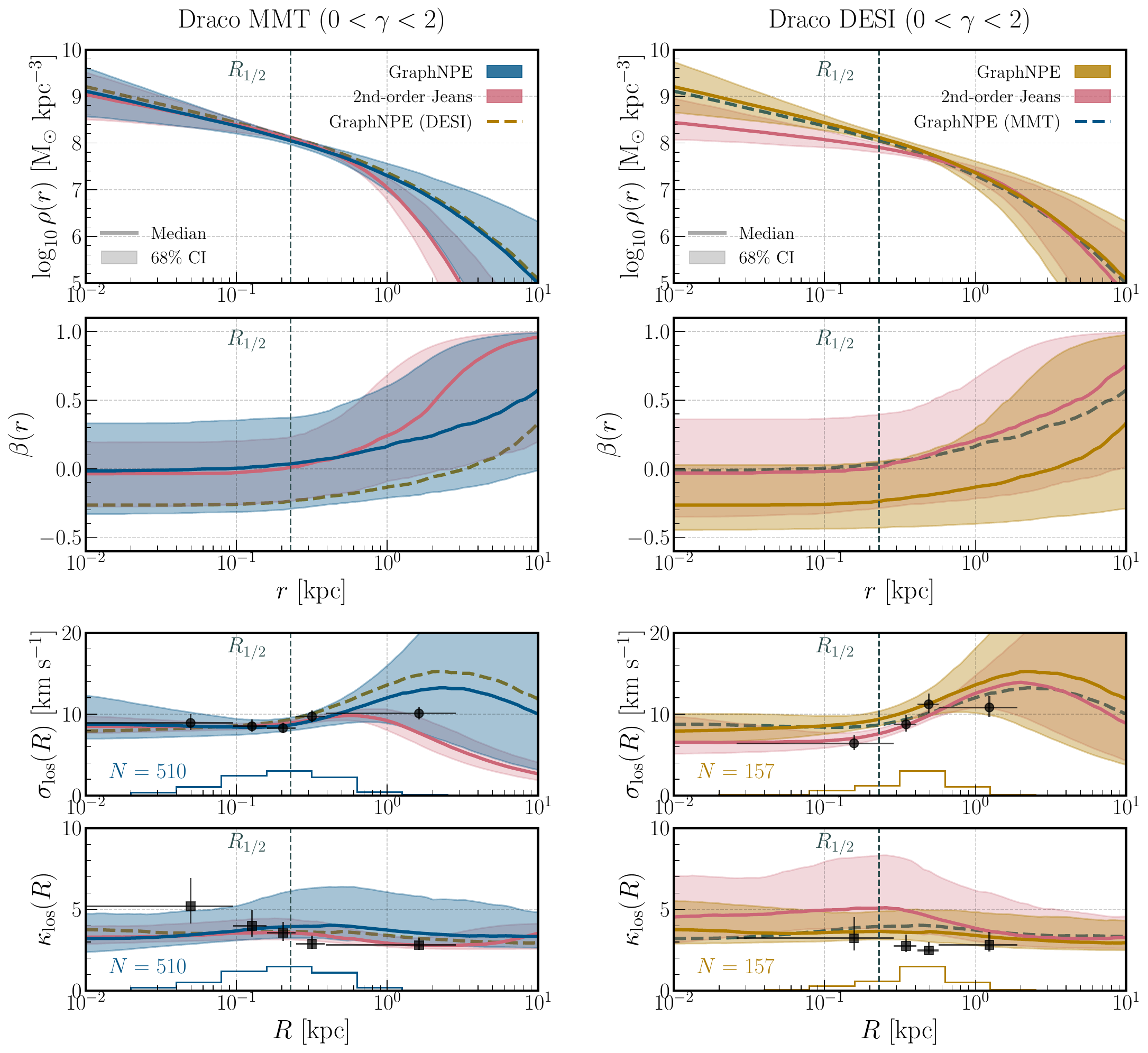}
    \caption{
    From top to bottom: the inferred DM density $\rho(r)$, velocity anisotropy $\beta(r)$, LOSV dispersion $\sigmalos(R)$, and LOSV kurtosis $\kappalos(R)$ profiles for Draco from \gnn and Jeans modeling.
    The left and right columns show results for MMT \citepalias{Walker23} and DESI \citepalias{Ding25}, respectively.
    Panels are the same as Fig.~\ref{fig:draco_profiles}, but only including the posterior samples with $\gamma \in [0, 2]$.
    }
    \label{fig:draco_profiles_gm0}
\end{figure*}
\begin{figure*}
    \centering
    \includegraphics[width=0.92\linewidth]{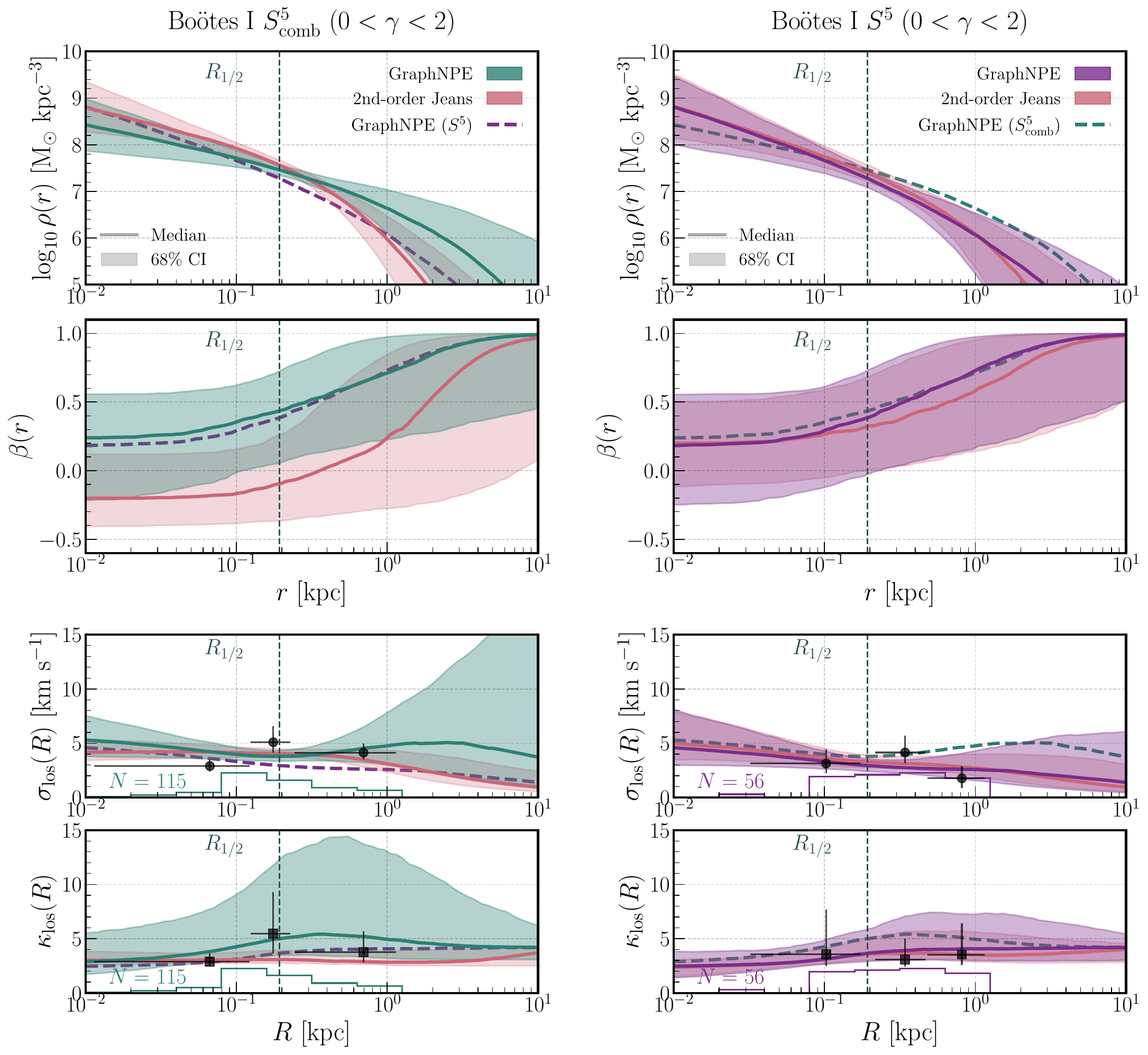}
    \caption{
    From top to bottom: the inferred DM density $\rho(r)$, velocity anisotropy $\beta(r)$, LOSV dispersion $\sigmalos(R)$, and LOSV kurtosis $\kappalos(R)$ profiles for \bootes from \gnn and Jeans modeling.
    The left and right columns show results for the \sfivecomb and \sfive \citepalias{Sandford26}, respectively.
    Panels are the same as Fig.~\ref{fig:bootes_profiles}, but only including the posterior samples with $\gamma \in [0, 2]$.
    }
    \label{fig:bootes_profiles_gm0}
\end{figure*}

We present the inference results of \gnn and Jeans modeling using the inner slope prior of $\gamma \in [0, 2]$.
The medians and $68\%$ percentiles of the \gnn posteriors for each model parameters, as well as \rhoinner, $J$- and $D$- factors, are shown in Table~\ref{tab:posterior} in the main text.
Figures~\ref{fig:draco_profiles_gm0} and \ref{fig:bootes_profiles_gm0} show the inferred DM density, velocity anisotropy, and LOSV  dispersion profiles of \gnn and Jeans modeling for Draco and \bootes, respectively.

\subsection{DM density at $300\,\pc$} 
\label{app:rho300}

\begin{figure*}
    \centering
    \includegraphics[width=0.90\linewidth]{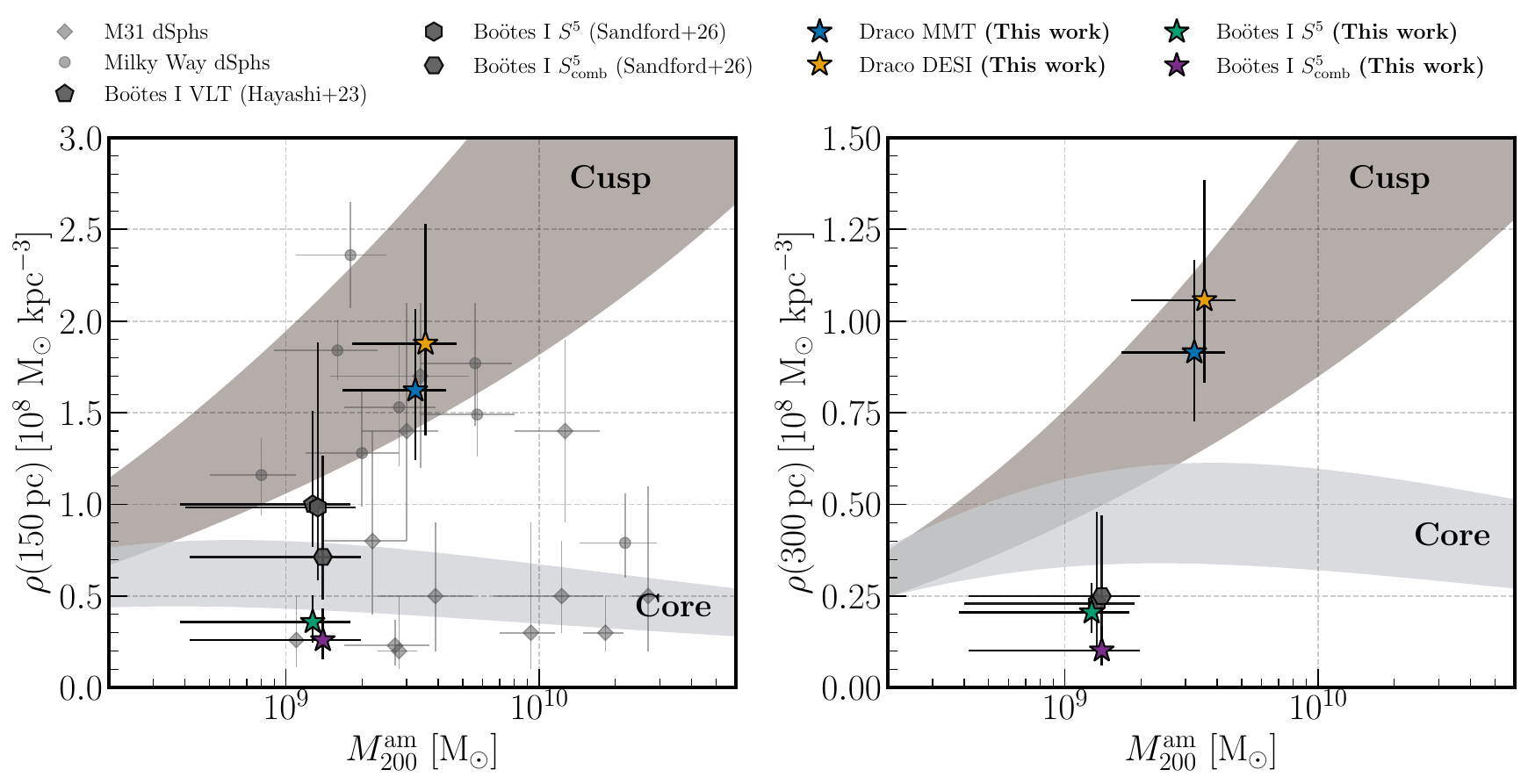}
    \caption{
    The DM density at $150\,\pc$ (left) and $300\,\pc$ (right) as a function of pre-infall halo mass \mvir estimated by AM.
    For better visualization, we apply a small offset to \mviram of the same galaxy.
    The left panel reproduces Fig.~\ref{fig:rho150_m200} for reference.
    For the right panel, literature values at $300\,\pc$ are only available for \bootes from \citet{Sandford26}, computed directly from their posterior samples; Milky Way and M31 dwarf measurements are not shown as their density profiles at this radius are not reported.
    The cusp and core bands show the predicted $\rho$ as a function of \mvir for NFW and cored profiles, respectively, adopting the \citet{2014MNRAS.441.3359D} mass-concentration relation; the band widths correspond to its $1\sigma$ scatter in concentration.
    }
    \label{fig:rho150_rho300_m200}
\end{figure*}

Fig.~\ref{fig:rho150_rho300_m200} shows the DM density at $150\,\pc$ and $300\,\pc$ as a function of \mviram for Draco and \bootes, with the left panel reproducing Fig.~\ref{fig:rho150_m200} for reference.
For the right panel, we note that literature density profiles evaluated at $300\,\pc$ are not broadly available; we therefore show only the \citet{Sandford26} \bootes measurements, computed from their posterior samples.
Draco sits within the cusp band in both panels, consistent with the cuspy profile inferred in Section~\ref{section:result}.
For \bootes, both \rhoinner and $\rho(300\,\pc)$ fall below the core band.
Interestingly, while the \citetalias{Sandford26} measurements lie above ours at $150\,\pc$, their $\rho(300\,\pc)$ values are consistent with our own, suggesting that the discrepancy between the two analyses is largely confined to the inner profile.
Since $300\,\pc$ lies well beyond the half-light radius of \bootes ($\rhalf \sim 200\,\pc$), this low outer density is unlikely to be attributed to core-forming processes that operate primarily within the stellar distribution.
We refer the reader to Section~\ref{section:implication_bootes} for a broader discussion of the implications.

\section{Higher-order moments}
\label{app:kurtosis}

\subsection{The fourth-order moment Jeans equations}
\label{app:fourth_order_jeans}
The fourth-order moment Jeans equations can be expressed as \citep{1990AJ.....99.1548M}:
\begin{align}
    &\frac{1}{\nu}\frac{\partial}{\partial r}(\nu \vrfour) - \frac{3}{r}\vrvtfour + \frac{2}{r}\vrfour = -3 \sigma_r^2 \frac{GM}{r^2}, \\
    &\frac{1}{\nu}\frac{\partial}{\partial r}(\nu \vrvtfour) - \frac{1}{r}\vtfour + \frac{4}{r}\vrvtfour = - \sigma_t^2 \frac{GM}{r^2},
\end{align}
where $\braket{\cdot}$ denotes an expectation over the velocity distribution, so that $\braket{v_r^2} \equiv \sigma_r^2$ is the radial velocity dispersion and $\braket{v_t^2} \equiv \sigma_t^2$ the tangential one (assuming negligible mean motion).

As shown in \cite{2013MNRAS.432.3361R}, the first equation can be simplified to:
\begin{equation}
    \frac{1}{\nu}\frac{\partial}{\partial r}(\nu \vrfour) + \frac{2 \betaprime}{r} \vrfour = -3 \sigma^2_r \frac{GM}{r^2},
\end{equation}
by introducing
\begin{equation}
    \betaprime(r) \equiv 1 - \frac{3}{2}\frac{\braket{v_r^2 v_t^2}}{\vrfour},
\end{equation}
which plays a role analogous to the second-order anisotropy $\beta(r) = 1 - \sigma_t^2/(2\sigma_r^2)$ (Equation~\ref{eq:beta_ani}), parameterizing the deviation of the fourth-order velocity ellipsoid from isotropy.
We can also define
\begin{equation}
    \gprime(r) \equiv \exp\left(2\int\frac{\betaprime(s)}{s}\mathrm{d}s\right),
\end{equation}
which serves as the integrating factor for the fourth-order Jeans equation, analogous to the second-order integrating factor $g(r)$ (Equation~\ref{eq:gint})\footnote{Following convention in \cite{2013MNRAS.432.3361R}, $\betaprime(r)$ and $\gprime(r)$ denote the fourth-order anisotropy equivalent of $\beta(r)$ and $g(r)$ in Eqs.~\ref{eq:beta_ani} and \ref{eq:gint}, and \textit{not} the derivative of $\beta(r)$ and $g(r)$.}.

The solution for the fourth-order radial velocity moment is then
\begin{equation}
    \vrfour(r) = \frac{1}{\nu(r)\gprime(r)} \int_r^\infty 3 \gprime(s) \frac{GM(<s)\, \sigma^2_r(s)\,\nu(s)}{s^2}\,\mathrm{d}s.
\end{equation}
Applying the Abel transform yields the projected LOSV fourth moment:
\begin{equation}
    \braket{\vlos^4}(R) = \frac{2}{\Sigma_\star(R)} \int_R^\infty F_\mathrm{los}(r, R) \frac{\nu(r)\, r}{\sqrt{r^2 - R^2}}\,\mathrm{d}r,
\end{equation}
where
\begin{align}
    F_\mathrm{los}(r, R) \equiv & \left(1 -  2 \betaprime \frac{R^2}{r^2} + \frac{1}{2}\betaprime(1+\betaprime)\frac{R^4}{r^4} - \frac{1}{4}\frac{\partial \betaprime}{\partial r} \frac{R^4}{r^3}\right) \vrfour \nonumber \\
    & + \frac{3}{4}(\betaprime - \beta)\, \sigma_r^2\, GM\,\frac{R^4}{r^5}.
\end{align}

The kurtosis profile is then defined as the dimensionless ratio
\begin{equation}
    \kappa_\mathrm{los}(R) \equiv \frac{\braket{\vlos^4}(R)}{\sigma_\mathrm{los}^4(R)},
\end{equation}
where $\sigma_\mathrm{los}^2(R)$ is the projected LOSV dispersion obtained from the second-order Jeans equation.
For a Gaussian LOSV  distribution, $\kappa_\mathrm{los} = 3$, so deviations from this value quantify the non-Gaussianity of the velocity distribution at projected radius $R$.
Values $\kappa_\mathrm{los} > 3$ indicate a leptokurtic (heavy-tailed) distribution, characteristic of radially anisotropic orbits, while $\kappa_\mathrm{los} < 3$ indicates a platykurtic distribution, characteristic of tangentially anisotropic orbits \citep{2002MNRAS.333..697L}.

In general, the fourth-order anisotropy does not equal to the second-order anisotropy, i.e. $\betaprime \neq \beta$.
In practice, evaluating $\kappalos(R)$ from the fourth-order Jeans equations requires specifying a functional form for $\betaprime(r)$, which in general is independent of $\beta(r)$ and introduces additional degrees of freedom into the mass modeling.
For example, methods such as those in \citet{2013MNRAS.432.3361R} and \citet{2026A&A...705A.212B} treat
$\betaprime(r)$ as a free parameter independent of $\beta(r)$,

In this work, we assume the COM distribution function \citep{1991MNRAS.253..414C} of the form:
\begin{equation}
    f = L^{-2\beta_0}g(Q),
\end{equation}
where $L$ is the specific angular momentum, $Q \equiv E + L^2/2r_a^2$ is the augmented energy, $E$ is the specific orbital energy, $r_a$ is the anisotropy radius, and $g(Q)$ is an arbitrary non-negative function of $Q$ that vanishes for $Q \geq 0$.
This DF belongs to the separable augmented density class \citep{2011ApJ...736..151A, 2013MNRAS.432.3361R}, for which one can show by explicit moment integration that $\betaprime(r) = \beta(r)$ exactly, regardless of the specific form of $g(Q)$.
We therefore adopt the closure $\betaprime = \beta$ when computing the kurtosis profiles in Sections~\ref{section:mock} and \ref{section:result}.

\subsection{Fitting the kurtosis}
\label{app:kurtosis_fit}

We measure the projected LOSV kurtosis profile $\kappalos(R)$ from the discrete stellar kinematics by fitting a generalized Gaussian velocity distribution to the observed velocities in radial bins.
The generalized Gaussian (also known as the stretched exponential) is given by
\begin{equation}
    f(v) = \frac{\beta}{2\alpha\,\Gamma(1/\beta)}
    \exp\!\left(-\left|\frac{v - \mu}{\alpha}\right|^\beta\right),
\end{equation}
where $\mu$ is the systemic velocity, $\alpha$ is a scale parameter, $\beta$ controls the shape of the distribution, and $\Gamma$ is the Gamma function.
The intrinsic velocity dispersion and kurtosis are related to $\alpha$ and $\beta$ by
\begin{equation}
    \sigma^2 = \alpha^2\,\frac{\Gamma(3/\beta)}{\Gamma(1/\beta)},
    \qquad
    \kappa = \frac{\Gamma(5/\beta)\,\Gamma(1/\beta)}{\Gamma(3/\beta)^2},
\end{equation}
so that $\beta = 2$ recovers a Gaussian with $\kappa = 3$, values $\beta < 2$ give leptokurtic distributions ($\kappa > 3$), and values $\beta > 2$ give platykurtic distributions ($\kappa < 3$).

Measurement errors $\Delta_i$ are incorporated via the analytic convolution of the generalized Gaussian with a Gaussian error distribution $\mathcal{N}(0, \Delta_i^2)$.
The observed velocity of each star is modeled as $v_i = v_\mathrm{int} + \epsilon_i$,
where $\epsilon_i \sim \mathcal{N}(0, \Delta_i^2)$, so that the likelihood
contribution of each star is
\begin{equation}
    \mathcal{L}_i = \int_{-\infty}^{\infty}
    f_\mathrm{int}(v_\mathrm{int} \mid \mu, \alpha, \beta)\,
    \mathcal{N}(v_i \mid v_\mathrm{int}, \Delta_i^2)\,
    \mathrm{d}v_\mathrm{int},
\end{equation}
which is evaluated numerically for each star.
Unlike the fast approximation of folding errors into an effective scale parameter, this approach makes no assumption on the relative size of $\Delta_i$ and $\sigma$, and correctly propagates measurement uncertainties into the inferred kurtosis at the cost of increased computational expense.

We sample the posterior over $(\mu,\, \log\alpha,\, \log\beta)$ using the affine-invariant ensemble sampler \textsc{emcee} \citep{2013PASP..125..306F}.
We adopt flat priors $\mu \in (-100, 100)\, \kms$,  $\alpha \in (0.1, 100)\, \kms$ and $\beta \in (0.5, 4.0)$.
The velocity dispersion $\sigma$ and kurtosis $\kappa$ are then derived from the posterior samples via the relations above.
For each dataset, we first construct radial bins with equal number of tracers and perform the above fit independently for each bin.
The median and 68\% percentile intervals of the resulting $\sigma$ and $\kappa$ posterior samples are used as the binned LOSV dispersion and kurtosis profiles and their uncertainties.

\bsp	
\label{lastpage}
\end{document}